\documentclass[aps,prx,reprint,groupedaddress,floatfix]{revtex4-1}
\pdfoutput=1

\usepackage{siunitx} 

\usepackage{amsmath} 
\usepackage{amssymb}
\usepackage{mathrsfs}
\usepackage{mathtools}
\usepackage{stmaryrd}
\usepackage{esint}

\usepackage[shortlabels]{enumitem}
\usepackage{graphicx} 

\usepackage{multirow}
\usepackage{booktabs} 
\usepackage{hyperref} 

\DeclareMathOperator{\Tr}{Tr}
\DeclareMathAlphabet{\mathpzc}{OT1}{pzc}{m}{it}

\begin{document}


\title{Morphogenesis and self-organization of persistent filaments confined within flexible biopolymeric shells}
\date{\today}
\author{Maxime M. C. Tortora}
\thanks{Correspondence to \href{mailto:maxime.tortora@ens-lyon.fr}{maxime.tortora@ens-lyon.fr}}
\author{Daniel Jost}
\affiliation{Université de Lyon, ENS de Lyon, Univ Claude Bernard, CNRS, Laboratoire de Biologie et Modélisation de la Cellule, Lyon, France}



\begin{abstract}

We systematically explore the self-assembly of semi-flexible polymers in deformable spherical confinement across a wide regime of chain stiffness, contour lengths and packing fractions by means of coarse-grained molecular dynamics simulations. Compliant, DNA-like filaments are found to undergo a continuous crossover from two distinct surface-ordered quadrupolar states, both characterized by tetrahedral patterns of topological defects, to either longitudinal or latitudinal bipolar structures with increasing polymer concentrations. These transitions, along with the intermediary arrangements that they involve, may be attributed to the combination of an orientational wetting phenomenon with subtle density- and contour-length-dependent variations in the elastic anisotropies of the corresponding liquid crystal phases. Conversely, the organization of rigid, microtubule-like polymers evidences a progressive breakdown of continuum elasticity theory as chain dimensions become comparable to the equilibrium radius of the encapsulating membrane. In this case, we observe a gradual shift from prolate, tactoid-like morphologies to oblate, erythrocyte-like structures with increasing contour lengths, which is shown to arise from the interplay between nematic ordering, polymer and membrane buckling. We further provide numerical evidence of a number of yet-unidentified, self-organized states in such confined systems of stiff achiral filaments, including spontaneous spiral smectic assemblies, faceted polyhedral and twisted bundle-like arrangements. Our results are quantified through the introduction of several order parameters and an unsupervised learning scheme for the localization of surface topological defects, and are in excellent agreement with field-theoretical predictions as well as classical elastic theories of thin rods and spherical shells.

\end{abstract}


\maketitle 


\section{Introduction} \label{sec:Introduction}

The tight packing of macromolecules within confined, crowded domains is commonly observed in a vast array of biological contexts~\cite{ellis2001macromolecular}. Examples range from DNA arrangements in viral capsids~\cite{marenduzzo2010biopolymer} and chromosome organization in eukaryotic cells~\cite{sinden1994organization} to collagen fibrillar assemblies in connective tissues~\cite{saedi2012molecular}, and from protein (mis)folding~\cite{chiti2006protein} to the formation of actin bundles and networks in the cellular cytoskeleton~\cite{silva2011self}. Geometrical confinement, which may arise from the presence of physical boundaries such as membrane walls or fluid interfaces, generally reduces the entropy of individual molecules by restricting the ensemble of accessible conformations that they may adopt. These constraints are generally supplemented by additional steric forces emerging from the dense bulk environment, which may drastically affect organization even in the absence of significant specific interactions between macromolecules and their surroundings~\cite{zhou2008macromolecular}. Such excluded-volume contributions have been suggested to largely control the conformational stability of endogenous proteins in the crowded intra-cellular medium~\cite{cheung2005molecular,lucent2007protein}, and are crucial to numerous technological applications in such fields as microfluidics~\cite{squires2005microfluidics}, mechanical filtration~\cite{dlamini2019critical} and drug delivery~\cite{tianmeng2014engineered}.
\par
Beyond these purely entropic considerations, the folding of macromolecules may be further constrained by the finite compliance of their constituent intra-molecular bonds. In polymer physics, a canonical example of such systems is the so-called worm-like chain (WLC) model of semi-flexible polymers, which describes the elastic response of fiber-like structures in terms of a bending rigidity penalizing local deviations away from a straight, linear state~\cite{rubinstein2003polymer}. Instances of macromolecules whose conformational statistics may be well-captured by WLCs include many common biopolymers such as double-stranded DNA (dsDNA), microtubules, F-actin and intermediate filaments~\cite{broedersz2014modeling}.
\par
In self-avoiding WLCs, the local shape anisotropy imparted by bending stiffness, combined with the steric repulsion resulting from the impenetrability of the macromolecular backbone, is often sufficient to induce liquid-crystalline (LC) behavior at high-enough polymer concentrations~\cite{binder2020understanding}. The simplest instance of such structures, known as the \textit{nematic phase}, is characterized by the presence of long-ranged orientational order driven by the spontaneous alignment of neighboring chain segments, and is ubiquitously observed \textit{in vivo} and \textit{in vitro} in dense solutions of biopolymers~\cite{hamley2010liquid,mitov2017cholesteric}. This higher level of organization is associated with an additional form of elasticity, which emerges at the scale of the LC phase to describe the thermodynamic costs of distorting the spatial pattern of chain orientations away from its stable equilibrium arrangement~\cite{deGennes1993physics}.
\par
The distinction between intrinsic (molecular) and emergent (LC) elasticity has given rise to two orthogonal and somewhat complementary theoretical descriptions of WLCs in confinement. The first, based on self-consistent field theory (SCFT), aims to predict the equilibrium state under given assembly conditions directly from the structural and mechanical properties of individual polymers~\cite{chen2016theory}. This approach is, however, largely intractable to analytical methods, and relies rather fundamentally on the second-virial approximation~\cite{onsager1949effects} --- which is not obviously suited to the high packing densities relevant in many biological settings. The second proceeds from a continuum-level representation of the polymeric phase in terms of a mesoscopic \textit{director field}, corresponding to the locally-averaged direction of alignment of chain backbone segments~\cite{svensek2010confined,shin2011filling}. This framework conversely leads to the coarse-graining of all salient thermodynamic quantities into a handful of scalar parameters, known as the Oseen-Frank (OF) elastic moduli~\cite{deGennes1993physics}, which are linked in a non-trivial fashion to the underlying molecular features of the polymer~\cite{revignas2020interplay} --- and are often understood as purely phenomenological, mean-field quantities. 
\par
Molecular simulations have thus retained a crucial role in exploring the quantitative interplay between self-organization, polymer mechanics and geometrical constraints in confined macromolecular systems. In particular, the case of WLCs confined within spherical cavities has been extensively studied as an idealized model for macromolecular packing in biological and microfluidic compartments~\cite{micheletti2011polymers,shaebani2017compaction,curk2019spontaneous}. In this context, a further source of frustration commonly arises when the director field at the cavity boundary lies coplanar to the sphere surface, as is typically favored by simple steric interactions between polymers and the confining wall or fluid interface. This additional constraint implies that the preferred local alignment cannot be maintained throughout space, and thus drives the formation of so-called \textit{topological defects} --- corresponding to localized ``bald spots'' in the surface orientational order --- whose presence is imposed by topology to provide a total \textit{topological charge} matching the Euler characteristic $s=2$ of the sphere, as first shown by Poincar\'e and Hopf~\cite{frankel2003geometry}. Besides broad fundamental interest~\cite{bowick1994cosmological,alexander2012colloquium}, such topological defects have gathered considerable attention from the point of view of materials science due to their ease of functionalization~\cite{wang2016topological}, which holds significant promise for the synthesis of patchy colloids with tunable anisotropic interactions~\cite{nelson2002toward,lopez2011frustrated,ravnik2020topological}.
\par
Despite this rich theoretical and experimental backdrop, the link between molecular properties and defect morphology in spherically-confined polymers has remained largely unresolved. The vast majority of numerical and analytical studies to date have focused on the case of a single long polymer chain, and generally predicted a spool-like folded structure as the system ground state~\cite{spakowitz2005dna,katzav2006statistical,petrov2008packaging,reith2012effective,liang2019orientationally}. These arrangements typically bear two surface topological defects of charge $s=1$ located along the spool symmetry axis, and resemble the organization of packaged dsDNA within small viral capsids~\cite{petrov2008packaging}. Contrastingly, investigations of low-molecular-weight LCs in spherical cavities and shells have revealed a rich phenomenology involving various combinations and patterns of bulk and surface topological defects in the ground state, depending on shell thickness, anchoring conditions and relative values of the OF elastic moduli~\cite{urbanski2017liquid}. Such structures have remained highly elusive in macromolecular systems~\cite{nikoubashman2017semiflexible,milchev2018densely,khadilkar2018self}, with recent reported shortcomings of continuum elasticity for strongly-confined polymers raising further questions about their practical relevance and stability~\cite{garlea2016finite}.
\par
Another largely unexplored question regards the potential interplay between compartment flexibility and macromolecular organization. With few exceptions~\cite{marenduzzo2007dynamics,fosnaric2013montecarlo,vetter2014morphogenesis,zou2018packing}, a significant limitation shared by most previous investigations is their assumption of a perfectly rigid confining cavity, which may be inappropriate for many biophysical systems. Indeed, spontaneous deviations from ideal sphericity in phospholipid vesicles and biopolymeric shells may commonly result from thermal fluctuations or polymer-membrane interactions, as observed in microtubules and actin/filamin networks within liposomes and red blood cells (or \textit{erythrocytes})~\cite{elbaum1996buckling,fygenson1997mechanics,limozin2002polymorphism,tsai2015shape}. The morphology of eukaryotic nuclei has been similarly shown to be intimately linked to the structure and mechanics of the encapsulated chromatin fibers~\cite{heo2016differentiation,zhu2017comprehensive,stephens2019chromatin}, suggesting that the mutual feedback between polymer folding and membrane conformation may play an essential role in genome organization and thus in the regulation of cellular function. From a practical standpoint, the ability to control the coiling of elastic nanowires through their confinement within swelling polymer shells~\cite{xu2010mechanical} or liquid droplets~\cite{chen2012general} has also been demonstrated experimentally, and could be exploited as the basis of a variety of applications for nanomechanical energy storage~\cite{chen2011controlling}.
\par
To shed light on these issues, we here combine a generic model of self-avoiding WLCs with a minimal elastic description of thin amorphous shells. This numerical framework, introduced in Sec.~\ref{sec:Numerical model}, allows for the direct control of the chain packing fraction in the case of pure steric interactions between macromolecules and their encapsulating membrane, and is found to provide a practical kinetic pathway to equilibrium in such hysteresis-prone systems of dense, confined polymers. By mapping the model mechanical parameters to experimental measurements, we illustrate its application to two biologically-relevant scenarios --- namely, flexible, dsDNA-like filaments confined within the nuclear envelope (Secs.~\ref{sec:From quadrupolar to bipolar order}-\ref{sec:Splay- versus bend-dominated regimes}), and rigid, tubulin-like chains enclosed within the red blood cell plasma membrane (Secs.~\ref{sec:Tactoid-like and faceted morphologies}-\ref{sec:Buckled, sickled and toroidal states}). We finally summarize our results in Sec.~\ref{sec:Discussion}, and discuss their potential physical and biological implications.


\section{Numerical model} \label{sec:Numerical model}

\subsection{Discretized elastic description of polymerized membranes}
 
Biological membranes may be broadly classified into two categories, depending on the nature of their underlying microscopic interactions. Fluid membranes, such as vesicles and liposomes, are typically stabilized by weak hydrophobic forces involving the hydrocarbon tail of amphiphilic chains, which generally lead to a dynamic, liquid-like form of molecular organization in physiological temperature conditions~\cite{nelson1989statistical,seifert1997configurations}. Polymerized (or ``tethered'') membranes, on the other hand, are characterized by fixed internal connectivity, which arises from the presence of stable cross-links between membrane molecular units, as in the case of the spectrin network of erythrocyte plasma membranes and of the lamina network of eukaryotic nuclear envelopes~\cite{nelson1989statistical}.
\par
Despite their superficial similarity, these two types of surfaces belong to different universality classes, and therefore display markedly distinct physical behavior~\cite{bowick2001statistical}. The vanishing shear modulus imposed by the lack of in-plane structure allows fluid membranes to continuously switch between different geometric shapes at relatively low energetic cost, as illustrated by the budding and filopodial projections induced by protein-lipid interactions in the cell plasma membrane~\cite{hurley2010membrane}. Conversely, the response of polymerized shells to distributed mechanical or osmotic stresses usually involves a non-linear buckling bifurcation, corresponding to a sudden collapse of the shell away from its reference elastic state, upon raising the magnitude of the applied constraints above a certain critical threshold~\cite{hutchinson2016buckling}. Although the importance of buckling as a morphogenetic process has long been recognized~\cite{nelson2016buckling,lin2018shell}, the role of encapsulated macromolecules on its onset and on post-buckling structure in systems of polymerized membranes remains largely unexplored.
\par
Following previous investigations of erythrocyte and nuclear envelopes~\cite{li2005spectrin,banigan2017mechanics}, we focus on the case of \textit{amorphous} polymerized shells, in which elastic properties are taken to be homogeneous. To that end, we construct a discretized mesh representation of a spherical membrane by solving a numerical formulation of the classical Thomson problem~\cite{saff1997distributing}. We randomly distribute a number $N_v$ of vertices on the surface of a sphere of radius $R$, and endow each vertex with an elementary point charge $q=e$. Relaxation then amounts to minimizing the cost function $C(N_v)=\sum_{i<j} r_{ij}^{-1}$, with $r_{ij}$ the vertex pairwise distance, which is proportional to the total vertex electrostatic repulsion energy. Note that the local hexagonal order of the corresponding ground state is generally associated with $N_5 \geq 12$ minimal fivefold disclinations imposed by spherical topology~\cite{kamien2002geometry}, thus leading to elastic instabilities that have been shown to drive the faceting of crystalline membranes such as viral capsids~\cite{lidmar2003virus}. However, for the large values of $N_v$ considered here, the strains induced by these disclinations are largely screened by complex patterns of dislocation defects~\cite{perez1997influence} --- which is found to provide a good quantitative approximation of uniform elastic shells for our purposes, as discussed further below. 
\par
We allow the particles to interact via a truncated Coulomb potential with cutoff radius $r_{\rm cut}=0.5\,R$, which we verified to be large enough to bear a negligible effect on the final structure. We evolve the system towards equilibrium through the fast inertial relaxation engine~\cite{bitzek2006structural}, using a Lagrange multiplier approach to constrain vertex positions to the sphere surface. Upon convergence, we discard all electrostatic interactions and construct the membrane topology by computing the Delaunay triangulation of the relaxed configuration via the QuickHull algorithm~\cite{barber1996quickhull}. The adjacency matrix describing the connected neighborhood $\mathscr{V}(i)$ of any vertex $i$ is then kept fixed throughout the simulations, which mirrors the static character of membrane cross-links~\cite{nelson1989statistical,gompper1997network}. Each generated bond is represented by a harmonic potential of stiffness $k_m$, yielding a total stretching energy~\cite{kantor1987phase}
\begin{equation}
  \label{eq:memb_stretch}
  \mathscr{H}_{\rm memb}^{\rm stretch} = \frac{k_m}{4} \sum_{i=1}^{N_v} \sum_{j\in \mathscr{V}(i)} \Big(r_{ij} - r_{ij}^0\Big)^2,
\end{equation}
where the sum runs over all pairs of linked vertices $i$ and $j$, denoting by $r_{ij}^0$ their separation distance in the relaxed spherical reference state. In practice, for all systems considered here, we find $r_{ij}^0 \simeq r_0 = \big(8\pi R^2/\sqrt{3}N_v\big)^{0.5}$, indicating that the vertex distributions produced by our relaxation procedure are locally indistinguishable from uniform hexagonal packing. In this case, the membrane bending energy may be cast in the convenient form~\cite{kantor1987phase}
\begin{equation}
    \label{eq:memb_bend}
  \mathscr{H}_{\rm memb}^{\rm bend} = \kappa_m \sum_{\langle k,l\rangle}\Big[1-\cos\big(\phi_{kl}-\phi_0\big)\Big],
\end{equation}
where $\phi_{kl}\equiv \arccos\big(\widehat{\mathbf{n}}_k\cdot\widehat{\mathbf{n}}_l\big)$ is the dihedral angle between the normals $\widehat{\mathbf{n}}_k$ and $\widehat{\mathbf{n}}_l$ of each unique pair of adjacent triangles $\langle k,l \rangle$ sharing a common edge, and the hat notation indicates normalized vectors. In Eq.~\eqref{eq:memb_bend}, the offset $\phi_0=r_0/\sqrt{3}R$ accounts for spontaneous spherical curvature. We finally enforce membrane self-avoidance by incorporating a short-ranged steric repulsive contribution between all non-bonded pairs of vertices,
\begin{equation}
  \label{eq:memb_exc}
  \mathscr{H}_{\rm memb}^{\rm exc} = \frac{1}{2}\sum_{i=1}^{N_v} \sum_{j\notin\mathscr{V}(i)} u^{\rm WCA}_{r_0}\big(r_{ij}\big),
\end{equation}
in which the Weeks-Chandler-Andersen (WCA) term $u^{\rm WCA}_{r_0}$ represents excluded-volume interactions with effective diameter $r_0$ through a truncated and shifted Lennard-Jones potential~\cite{weeks1971role},
\begin{equation*}
  u^{\rm WCA}_{r_0}(r)= 
  \begin{dcases}
    4\epsilon \Bigg [ \bigg(\frac{r_0}{r} \bigg)^{12} - \bigg( \frac{r_0}{r}\bigg)^6 + \frac{1}{4} \Bigg] & \text{\!if } r < 2^{1/6}r_0 \\
    0 &\text{\!if } r \geq 2^{1/6}r_0
  \end{dcases},
\end{equation*}
where $\epsilon$ defines the model unit of energy.
\par
In the following, we make use of $N_v=\num[group-separator={,},group-minimum-digits={3}]{9800}$ vertices, in which case the shell topology is found to bear $N_7\simeq 250$ sevenfold defects associated with $N_5=N_7+12$ fivefold disclinations~\cite{kamien2002geometry}. This arbitrarily-large value of $N_v$ was chosen to allow for a fine-mesh discretization of the membrane surface, rather than to emulate any specific details of its underlying molecular structure. In the continuum limit, the parameters $k_m$ and $\kappa_m$ may be respectively linked to the Young's modulus~\cite{kantor1987phase} $Y_0=2k_m/\sqrt{3}$ and bare bending modulus~\cite{schmidt2012universal} $\kappa_0=\sqrt{3}\kappa_m/2$ of the two-dimensional (2D) material. To assess the quantitative accuracy of the model, we include in Supplemental Sec.~SI a comparison of its computed force-extension behavior against the classical analytical predictions of Reissner for thin uniform elastic shells~\cite{reissner1946stresses}. The excellent agreement observed in the linear response regime (see Supplemental Fig.~S1) suggest that the system is largely unaffected by the reported shortcomings of Eq.~\eqref{eq:memb_bend} for irregular surface meshes~\cite{guckenberger2017theory}, at least in the case of weak membrane deformations. Hence, the triangular tessellation obtained by Thomson relaxation provides an appropriate middle ground between ordered and randomized network descriptions of spherical shells~\cite{li2005spectrin}, and yields a tractable expression for the Hamiltonian $\mathscr{H}_{\rm memb} = \mathscr{H}_{\rm memb}^{\rm stretch}+\mathscr{H}_{\rm memb}^{\rm bend}+\mathscr{H}_{\rm memb}^{\rm exc}$ suitable for molecular dynamics (MD) simulations.

\subsection{The Kremer-Grest chain model} \label{subsec:kg}

Similarly, since we restrict our focus to polymer chains characterized by a persistence length ($l_p$) significantly larger than the length of chemical bonds, we describe the encapsulated macromolecules through the Kremer-Grest model~\cite{grest1986molecular,kremer1990dynamics}, which corresponds to a discretized realization of the classical Kratky-Porod WLC~\cite{kratky1949diffuse,kratky1949rontgenuntersuchung} with excluded volume constraints. In this framework, each polymer chain $\mathpzc{C}$ is represented as a linear assembly of $N_m$ monomeric units indexed by $k$, interacting via intra- and inter-molecular WCA repulsion,
\begin{equation}
  \label{eq:poly_exc}
   \mathscr{H}_{\rm poly}^{\rm exc}=\sum_{\mathpzc{C}} \sum_{k,l\in \mathpzc{C}} u^{\rm WCA}_{\sigma}\big(r_{kl}\big) + \sum_{\mathpzc{C},\mathpzc{C}'}\sum_{\substack{k\in\mathpzc{C} \\ k' \in\mathpzc{C}'}} u^{\rm WCA}_{\sigma}\big(r_{kk'}\big),
\end{equation}
where comma-separated indices imply summation over unique pairs of distinct elements and the effective chain diameter $\sigma$ sets the model unit of length. Chain connectivity is enforced by linking adjacent monomers through finitely-extensible non-linear elastic (FENE) springs,
\begin{equation*}
  u^{\rm FENE}_{\rm poly}\big(\Delta_k\big)= 
  \begin{dcases}
    -\frac{k_c r_c^2}{2} \log\Bigg[1-\bigg(\frac{\Delta_k}{r_c}\bigg)^2\Bigg] & \text{\!if } \Delta_k < r_c \\
    0 &\text{\!if } \Delta_k \geq r_c
  \end{dcases},
\end{equation*}
in which $\Delta_k\equiv \lVert \mathbf{r}_k-\mathbf{r}_{k-1}\rVert$ is the corresponding bond length, with $\mathbf{r}_k$ the center-of-mass position of the $k$-th bead. Chain stiffness is then governed by a simple angular potential of a similar form to Eq.~\eqref{eq:memb_bend},
\begin{equation*}
  u^{\rm bend}_{\rm poly}(\Theta_k) = \kappa_c \big(1-\cos\Theta_k\big),
\end{equation*}
where $\Theta_k \equiv \arccos\big(\widehat{\mathbf{t}}_k\cdot\widehat{\mathbf{t}}_{k+1}\big)$ is the inter-bond angle associated with the triplet of consecutive monomers $\big(k-1,k,k+1\big)$, with $\widehat{\mathbf{t}}_k\equiv \big(\mathbf{r}_{k}-\mathbf{r}_{k-1}\big)/\Delta_k$ the bond unit vector. We set $r_c=1.5\,\sigma$ and $k_c=30\,\epsilon/\sigma^2$, so that the individual bond length resulting from the competition between WCA repulsion and FENE stretching remains effectively constant throughout the simulations, and is given by $\Delta_k =  l_b \simeq 0.97\,\sigma$~\cite{egorov2016insight}. In this case, the polymer persistence length is simply related to the bending modulus $\kappa_c$ via $l_p\simeq \beta \kappa_c l_b$, with $\beta \equiv 1/k_BT$ the inverse temperature, which holds for stiff chains ($\beta \kappa_c \gtrsim 2$)~\cite{hsu2010standard}.
\par
Lastly, we introduce the coupling between macromolecules and confining membrane through a steric repulsion term involving each heterogeneous pair of vertices $i\in\llbracket1,N_v\rrbracket$ and monomers $k$,
\begin{equation}
  \label{eq:coupling}
  \mathscr{H}^{\rm poly}_{\rm memb} = \sum_{i=1}^{N_v}\Bigg(\sum_{\mathpzc{C}} \sum_{k\in \mathpzc{C}} u^{\rm WCA}_{\Sigma}\big(r_{ik}\big)\Bigg),
\end{equation}
where the effective excluded diameter $\Sigma = \big(r_0+\sigma\big)/2$ is chosen based on Lorentz-Berthelot combining rules so as to ensure that the shell surface may not be crossed by the encapsulated chains in any conformation. The total Hamiltonian then reads as $\mathscr{H}=\mathscr{H}_{\rm memb}+\mathscr{H}_{\rm poly}+\mathscr{H}^{\rm poly}_{\rm memb}$, with $\mathscr{H}_{\rm poly}=\mathscr{H}_{\rm poly}^{\rm FENE}+\mathscr{H}_{\rm poly}^{\rm bend}+\mathscr{H}_{\rm poly}^{\rm exc}$, which fully determines the mechanics of the polymer-membrane system. Simulations were conducted in the canonical ensemble as realized by a Langevin thermostat at fixed temperature $T=\epsilon/k_B$, and were evolved via a standard velocity-Verlet integration scheme~\cite{frenkel2002understanding}. Calculations were performed on multiple graphics processing units (GPUs) using the HOOMD-blue software package~\cite{anderson2008general,glaser2015strong}.
\par
Unless stated otherwise, a fixed number $N=\num[group-separator={,},group-minimum-digits={3}]{32768}$ of monomers was used in most simulations, although some larger systems with up to $N=\num[group-separator={,},group-minimum-digits={3}]{131072}$ were also employed to assess the influence of finite-size effects, as further discussed in Sec.~\ref{sec:Discussion}. The monomers were evenly distributed among a number $N_c = N/N_m$ of identical chains with contour length $l_c=\big(N_m-1\big)\,l_b+\sigma\simeq N_m\, \sigma$ and volume $v_c\simeq \pi l_c \sigma^2/4$, which were initially arranged in an ideal crystalline configuration. The initial membrane radius was set to $R=4\,l_c$ for all chains considered, which ensures that the starting state of all simulations lied deep in the stability range of the disordered (isotropic) phase and in conditions of weak confinement. 
\par
Relaxation runs of $\mathcal{O}\big(10^9\big)$ MD steps were performed to achieve full decorrelation from the initial orientational and positional order. Production runs then consisted in a slow isothermal compression carried out by gradually reducing the reference length $r_0$ of the membrane springs in $\mathcal{O}\big(10^2\big)$ increments. A stationary state was typically reached in $\mathcal{O}\big(10^8\big)$ MD steps following each membrane volume move, after which measurements were performed over a further $\mathcal{O}\big(10^8\big)$ MD steps. The density regimes thus explored by our simulations span over 4 orders of magnitudes for the longest chains studied ($l_c=256\,\sigma$), which was made computationally feasible by the use of state-of-the-art tree-based neighbor list calculations~\cite{howard2016efficient}.
\par
To evaluate the equilibration of the system, additional ``quenching'' simulations were performed by progressively increasing or lowering the polymer persistence length between the two extremal values investigated ($l_p=25\,\sigma$ to $1000\,\sigma$) at fixed $r_0$. This procedure was found to allow for a reversible switching between the densest ordered structures obtained in these two cases for all considered contour lengths. Hysteretic effects may nonetheless be occasionally observed at lower densities close to the isotropic-nematic (I-N) transition region, and are discussed in the following sections where relevant. The robustness of the results was finally ascertained by averaging measurements over 2 to 3 independent runs.


\section{From quadrupolar to bipolar order} \label{sec:From quadrupolar to bipolar order}

\subsection{Describing non-uniform orientational alignment} \label{subsec:describing}

Before discussing the results of the model, let us briefly recall some general considerations relevant to the quantification of orientational order in confinement. To describe nematic organization in assemblies of apolar macromolecules, it is customary to introduce the so-called Landau-de Gennes $\mathcal{Q}$-tensor~\cite{deGennes1993physics},
\begin{equation}
  \label{eq:deGennes}
  \mathcal{Q}_k^{\alpha\beta} = \frac{3\,t_k^\alpha t_k^\beta - \delta_{\alpha\beta}}{2},
\end{equation}
where $\alpha,\beta\in \{x,y,z\}$ denote the components of a bond vector $\widehat{\mathbf{t}}_k$ in the fixed laboratory frame and $\delta$ is the Kronecker delta. In the following, we omit the hat notation from unit vectors when no confusion can arise. In spatially-uniform systems at equilibrium, one may probe the potential collective anisotropy of the structure from the ascending eigenvalues $\lambda_1\leq \lambda_2 \leq \lambda_3$ of the tensor $\mathcal{Q}\equiv\big\langle\mathcal{Q}_k\big\rangle$, with $\big\langle\cdot\big\rangle$ a time and ensemble average over all constituent bonds $k$. In this case, the eigenvector $\mathbf{e}_3$ associated with $\lambda_3$ may be identified as the preferred direction of alignment $\mathbf{n}_0$ of the chains, referred to as the \textit{nematic director}, while $\lambda_3$ is related to the angular dispersion $\theta_k \equiv \arccos \big(\mathbf{t}_k \cdot \mathbf{n}_0\big)$ of bonds $\mathbf{t}_k$,
\begin{equation}
  \label{eq:legendre}
  \lambda_3 = \frac{3\big\langle\cos^2 \theta_k\big\rangle-1}{2}\equiv \big\langle P_2\big(\cos\theta_k\big)\big\rangle,
\end{equation}
with $P_2$ the second Legendre polynomial.
\par
However, in confined systems, the loss of homogeneity imposed by the presence of physical boundaries implies that both the director and the local degree of orientational order are generally position-dependent. Furthermore, the geometry of the confining walls may induce additional breakings of local phase symmetries. For instance, in the isotropic phase, spherical invariance dictates $\lambda_1=\lambda_2=\lambda_3\,(=0)$ in the bulk, while the inhibition of chain fluctuations along one direction --- which may arise (e.g.) from the vicinity of a planar interface --- typically yields $\lambda_1 < \lambda_2=\lambda_3\,(=-\lambda_1/2)$ near a flat boundary, where the eigenvector $\mathbf{e}_1$ is borne by the plane normal. Conversely, one expects $\lambda_3 > \lambda_1=\lambda_2\,(=-\lambda_3/2)$ in the bulk nematic phase of uniaxial molecules such as WLCs, reflecting the cylindrical symmetry of the equilibrium chain distribution about $\mathbf{e}_3=\mathbf{n}_0$ --- while the existence of a preferred direction of alignment lifts the degeneracy in $\lambda_2$ and $\lambda_3$ at the confining walls, thus leading to a biaxial surface structure ($\lambda_1 < \lambda_2< \lambda_3$).
\par
Hence, the interpretation of such tensorial order parameters (OPs) is necessarily ambiguous in finite-size samples. At the local level, it nonetheless follows from the previous discussion that a generic alignment parameter $\alpha \in [0,1]$ may be introduced to measure the degree of nematic order,
\begin{equation}
  \label{eq:alpha}
  \alpha \equiv \frac{2\big(\lambda_3-\lambda_2\big)}{3}.
\end{equation}
Eq.~\eqref{eq:alpha} satisfies $\alpha=0$ in the absence of local alignment ($\lambda_2=\lambda_3$) and $\alpha=1$ in the limit of perfect orientational order ($\lambda_3=1$, $\lambda_2=-1/2$), both in the bulk and near the boundary, and simply reduces to $\alpha=\lambda_3$ in the case where the order is purely uniaxial ($\lambda_2=-\lambda_3/2)$. More generally, since $\mathcal{Q}$ is a traceless tensor, only 2 of the 3 eigenvalues $\lambda_i$ are linearly independent, implying that a minimal set of 2 OPs may be constructed by considering diverse combinations of the $\lambda_i$. However, due to the variety of such definitions used in the literature~\cite{groh1999fluids,vanRoij2000orientational,trukhina2008computer,ivanov2013wall}, we here report the 3 relevant eigenvalues wherever possible to limit any possible confusion.
\par
To transparently quantify orientational order in confined, inhomogeneous LCs, it is necessary to devise a hierarchy of OPs capable of distinguishing the \textit{global} symmetries of the system, as described by the long-ranged spatial patterns of the director field $\mathbf{n}$, from the \textit{local} structure arising from the detailed arrangements of the polymers about $\mathbf{n}$. For this purpose, we partition the volume $V$ enclosed by the membrane into $\mathcal{O}\big(10^2\big)$ mesoscopic elements $\Xi_{\mathbf{r}}$ centered on $\mathbf{r}\in V$ with identical volumes, and introduce the locally-averaged tensor
\begin{equation}
  \label{eq:q_loc}
  \mathcal{Q}_{\rm loc}^{\alpha\beta}(\mathbf{r}) = \Big\langle\mathcal{Q}_k^{\alpha\beta}\Big\rangle_{\Xi_{\mathbf{r}}},
\end{equation}
in which $\big\langle\cdot\big\rangle_{\Xi_{\mathbf{r}}}$ is an ensemble average over all bonds $k$ whose center of mass lies within $\Xi_{\mathbf{r}}$. We may then quantify local nematic order at position $\mathbf{r}$ through the eigenvalues $\lambda_i^{\rm loc}(\mathbf{r})$ of $\mathcal{Q}_{\rm loc}$, and define the spatially-resolved director field $\mathbf{n}(\mathbf{r})\equiv \mathbf{e}_3^{\rm loc}(\mathbf{r})$ as the eigenvector associated with $\lambda_3^{\rm loc}$. Note that to improve statistical sampling, Eq.~\eqref{eq:q_loc} may be further averaged over multiple equilibrated configurations prior to spectral analysis~\cite{trukhina2008computer}. Finally, we may quantify long-wavelength fluctuations of the director field over any collection $\Xi$ of volume elements $\Xi_{\mathbf{r}}$ via the eigenvalues $\Lambda_i(\Xi)$ of the tensor
\begin{equation}
  \label{eq:q_dir}
  \mathcal{Q}_{\rm dir}^{\alpha\beta}(\Xi) = \int_{\mathbf{r}\in\Xi} d^3\mathbf{r} \, w_\Xi(\mathbf{r}) \frac{3\,n^\alpha(\mathbf{r})n^\beta(\mathbf{r})-\delta_{\alpha\beta}}{2},
\end{equation}
where the weight $w_\Xi(\mathbf{r}) \equiv \big[\lambda_3^{\rm loc}-\lambda_2^{\rm loc}\big](\mathbf{r}) / \int_\Xi \big(\lambda_3^{\rm loc}-\lambda_2^{\rm loc}\big)$ effectively subdues contributions to the integral Eq.~\eqref{eq:q_dir} from regions of low orientational order, in which the local director and its vector components $n^{\alpha,\beta}(\mathbf{r})$ are ill-defined.

\begin{figure*}[htpb]
  \includegraphics[width=2\columnwidth]{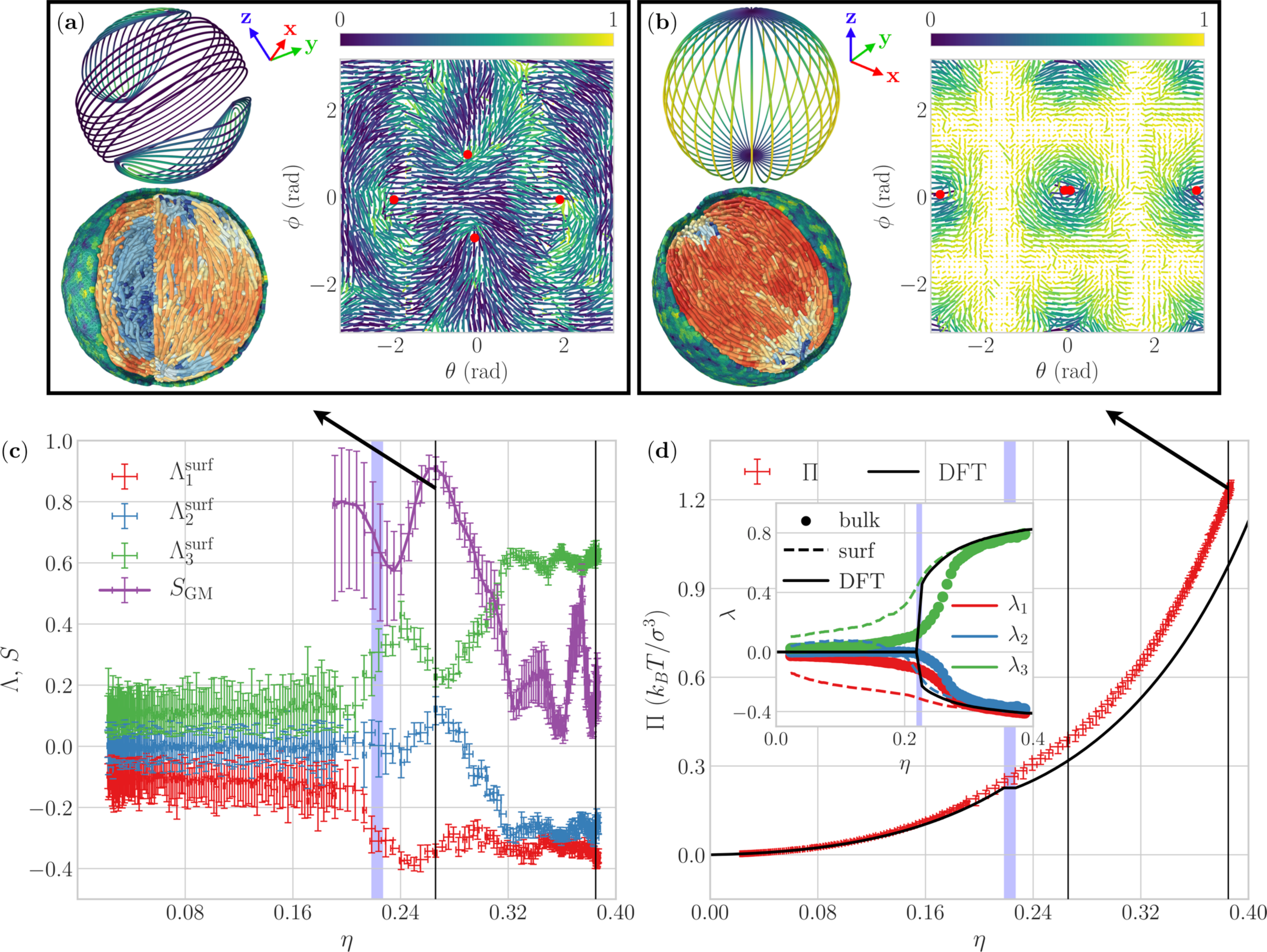}
  \caption{\label{fig1}Self-organization of short DNA-like chains ($l_c=16\,\sigma$, $l_p=25\,\sigma$) confined within a nuclear-like envelope. (a) Top left: Sketch of the idealized ``cricket-ball'' (CB) surface texture. The axis tripod represents the eigenvectors of $\mathcal{Q}_{\rm dir}(\Xi_{\rm surf})$ (Eq.~\eqref{eq:q_dir}), with eigenvalues $\Lambda_x^{\rm surf} = \Lambda_y^{\rm surf}=\Lambda/2 > 0$, $\Lambda_z^{\rm surf}=-\Lambda$, where $\Lambda$ generally depends on the arbitrary thickness $r_s$ of $\Xi_{\rm surf}$ ($\Lambda(r_s=0)=1/3$ in the limit of an infinitely-thin shell). Bottom left: Cut and peeled view of a simulation snapshot taken at volume fraction $\eta\simeq 0.27$, displaying the CB structure. Chains are colored according to the local order parameter (OP) $\lambda_3^{\rm loc}(\mathbf{r})$, with red (blue) regions being associated with strong (weak) orientational alignment. Right: Stereographic map of the surface director field~\cite{vitelli2006nematic}, obtained from simulations as described in the text. $\phi$ and $\theta$ represent the azimuthal and polar angles relative to the $\mathbf{x}$ and $\mathbf{y}$ axes, respectively. Headless arrows represent the director projection onto the $x$-$y$ plane at position $(\theta,\phi)$, and colors denote the $z$-component of the director. Red dots mark the computed positions of topological defects. (b) Same as (a) for the longitudinal bipolar configuration at $\eta\simeq 0.38$, with eigenvalues $\Lambda_z^{\rm surf}\equiv \Lambda$, $\Lambda_x^{\rm surf} = \Lambda_y^{\rm surf} = -\Lambda/2$ [$\Lambda(r_s=0)=1/4$]. (c) Surface director global OPs ($\Lambda_i^{\rm surf}$) and defect Glassmeier parameter ($S_{\rm GM}$, Eq.~\eqref{eq:glassmeier}) as a function of polymer density ($\eta$). (d) Osmotic pressure ($\Pi$) as a function of $\eta$, compared against the bulk predictions of DFT (black line, Eq.~\eqref{eq:p_dft}). Inset: Local OPs $\lambda_i \equiv \mu_\Xi \big(\lambda_i^{\rm loc}\big)$, where $\mu_\Xi$ denotes a uniform volume average over $\Xi \in \{\Xi_{\rm surf}, \Xi_{\rm bulk}\}$. The I-N bulk coexistence region calculated by DFT is highlighted in blue. Error bars are obtained as standard deviations from 2 independent simulations.}
\end{figure*}

\subsection{Orientational wetting of DNA-like chains in membrane confinement}

Let us now consider the case of generic dsDNA-like WLCs confined within a deformable membrane mimicking the nuclear envelope. The geometric diameter $\sigma$ and persistence $l_p$ of the chains may be respectively approximated as $\sigma \simeq \SI{2}{\nano\meter}$ and $l_p \simeq \SI{50}{\nano\meter} \simeq 25\,\sigma$ in usual solvent conditions~\cite{herrero2013mechanical}. The bare Young's modulus of the envelope lamina network has been measured as $Y_0\simeq\SI{25}{\milli\newton\per\meter}$ by micropipette aspiration experiments on swollen nuclei~\cite{dahl2004nuclear}, which yields $k_m \simeq 20 \, k_BT/\sigma^2$ in reduced model units at $T=\SI{300}{\kelvin}$. The bending modulus of the nuclear lamina has similarly been estimated as $\kappa_0 \simeq \SI{3.5e-19}{\newton\meter}$~\cite{vaziri2007mechanics}, which leads to $\kappa_m \simeq 100\,k_BT$.
\par
We summarize in Fig.~\ref{fig1} the equilibrium phase behavior observed for short chains with contour length $l_c = 16\,\sigma$ ($\sim \SI{32}{\nano\meter}$) as a function of the packing fraction $\eta \equiv N_c v_c/V = N \pi \sigma^3/4 V$, where the volume $V$ is evaluated using standard algorithms for closed and oriented triangular surface meshes~\cite{zhang2001efficient}. In this case, we first remark that the confining membrane retains a stable spherical conformation throughout the density range studied, although polymer-membrane interactions lead to an increasing degree of isotropic swelling of the shell with rising polymer concentrations, as evidenced by growing deviations of the mean inter-vertex distance $r_{ij}$ away from the reference-state value $r_0$ (see Supplemental Fig.~S2). This inflation effect may be related to the osmotic pressure $\Pi = \Tr{\mathcal{P}}/3$ of the encapsulated polymer solution, with $\Tr{\cdot}$ the trace operator and $\mathcal{P}$ the classical pressure tensor obtained from the virial theorem~\cite{gray1984theory},
\begin{equation*}
  \mathcal{P}^{\alpha\beta} = \rho_0 k_B T\, \delta_{\alpha\beta}+ \frac{1}{3V}\Bigg\langle\sum_{k} r_{k}^\alpha F_{k}^\beta\Bigg\rangle,
\end{equation*}
where $\rho_0\equiv N/V=4\eta/\pi\sigma^3$ is the monomer density and $\mathbf{F}_{k}$ is the total instantaneous force exerted on any given monomer $k$ (Fig.~\ref{fig1}d). 
\par
Let us denote by $R_0$ the average radius of the thermalized membrane at vanishing internal pressure, defined as the mean separation distance between each constituent vertex and the shell center of mass in the absence of polymer, and $R_\eta \simeq \sigma \,(3N/16\eta)^{1/3}$ its counterpart at finite macromolecular concentration $\eta$. In the limit of weak osmotic swelling ($\Delta R \equiv R_\eta-R_0 \ll R_0$), the radial force resisting expansion may be linked to the resulting tangential elastic stresses within the shell via~\cite{landau1986theory}
\begin{equation}
  \label{eq:swelling}
 \frac{\Delta R}{R_0} = \frac{\Pi}{2Y_0}R_0(1-\nu),
\end{equation}
with $\nu=1/3$ the membrane Poisson's ratio~\cite{seung1988defects}. Eq.~\eqref{eq:swelling} is found to hold quantitatively for $\Delta R/R_0 \lesssim 0.04$ (see Supplemental Fig.~S2). It is shown in Sec.~\ref{sec:Tactoid-like and faceted morphologies} that the stability of this spherical shell morphology may be attributed to the relative compliance of the encapsulated DNA chains and their emergent LC elastic moduli, while the osmotic inflation effect generally leads to an increase in the effective surface tension of the membrane, which mirrors the reported role of chromatin in the regulation of nuclear stiffness~\cite{stephens2019chromatin} --- as further discussed in Sec.~\ref{sec:Discussion}.
\par
In the rest of the paper, we set the origin of the frame to the system center of mass. The spherical symmetry of the membrane enables us to represent the discrete $\Xi_{\mathbf{r}}$ (c.f.~Sec.~\ref{subsec:describing}) as standard spherical volume elements, and to separately probe the onset of orientational order near the surface and in the bulk by discerning the outer shell $\Xi_{\rm surf} = \big\{\Xi_{\mathbf{r}} \mid r > R_\eta-r_s\big\}$ of the polymer solution from the complementary central core $\Xi_{\rm bulk}$, where $\mathbf{r} \equiv (r,\theta,\phi)$ are spherical coordinates. Following Refs.~\cite{nikoubashman2017semiflexible,milchev2018densely}, we set the thickness of the shell to $r_s\equiv 0.15\,R_\eta$, although the implications of this arbitrary choice are qualitatively discussed in Sec.~\ref{subsec:quantifying}. 
\par
\begin{figure*}[htpb]
  \includegraphics[width=1.5\columnwidth]{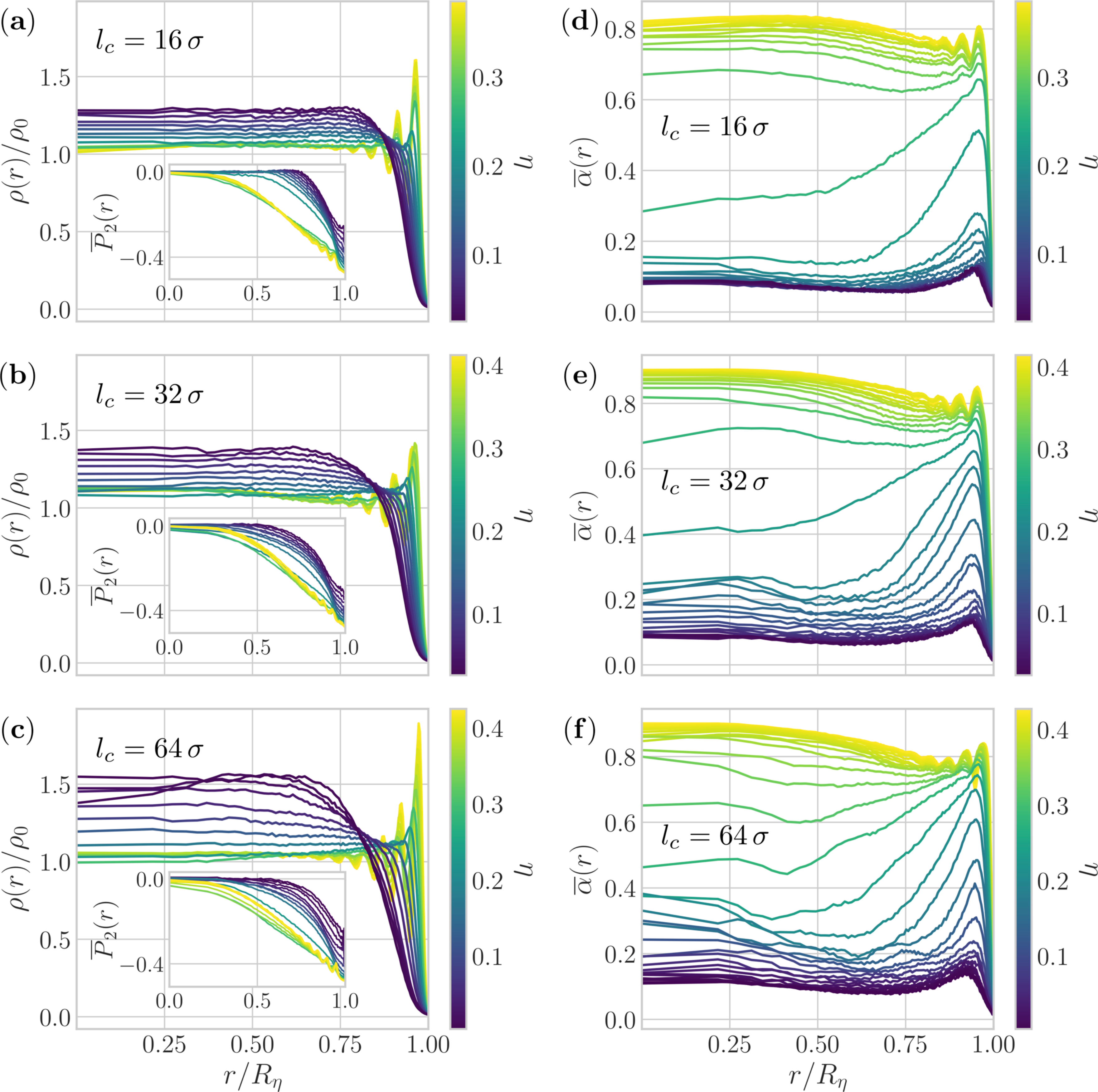}
  \caption{\label{fig2}Orientational wetting of DNA-like chains ($l_p=25\,\sigma$) in membrane confinement. (a-c) Radial monomer density ($\rho/\rho_0$) as a function of distance $r$ from the membrane center for various polymer volume fractions $\eta$ and contour lengths $l_c$. Inset: Mean projection $\overline{P}_2(r) \equiv \big\langle P_2\big(\mathbf{t}_k \cdot \widehat{\mathbf{r}}_k\big)\big\rangle_r$ of bond vectors $\mathbf{t}_k$ on the unit radial vector $\widehat{\mathbf{r}}_k$, with $P_2$ the second Legendre polynomial (Eq.~\eqref{eq:legendre}) and $\big\langle\cdot\big\rangle_r$ a volume average over a thin shell of radius $r$. $\overline{P}_2(r)=-1/2$ signifies that all chain segments at radial distance $r$ lie in the local tangent plane of the membrane ($\mathbf{t}_k \cdot \widehat{\mathbf{r}}_k=0$), while $\overline{P}_2(r)=0$ indicates that chain orientations are uncorrelated with the membrane surface. (d-f) Alignment parameter $\overline{\alpha}(r) \equiv \big\langle 2\big(\lambda_3^{\rm loc}-\lambda_2^{\rm loc}\big)/3\big\rangle_r$ (Eq.~\eqref{eq:alpha}) as a function of $r$. $\overline{\alpha}(r)\simeq 0$ evidences a lack of preferred local direction of alignment, implying either a 3D isotropic arrangement ($\overline{P}_2(r)=0$) or a uniform distribution of chains in the membrane tangent plane ($\overline{P}_2(r)< 0$). $\overline{\alpha}(r)=1$ in the limit of perfect orientational order.}
\end{figure*}

In Fig.~\ref{fig1}d, we report that the surface layer displays a near-uniaxial disordered structure at low densities ($\lambda_1^{\rm surf} < \lambda_2^{\rm surf}\simeq \lambda_3^{\rm surf}$). The weak residual biaxiality observed in this regime may be attributed to the finite curvature of the membrane, which breaks the symmetry in $\lambda_2^{\rm surf}$ and $\lambda_3^{\rm surf}$ expected from the proximity of a flat wall in the absence of in-plane alignment~\cite{groh1999fluids}. The system is then found to undergo a surface transition to a biaxial nematic state around $\eta\simeq 0.24$, in which the growing disparity between $\lambda_2^{\rm surf}$ and $\lambda_3^{\rm surf}$ indicates an increase in local orientational order at the surface (Eq.~\eqref{eq:alpha}). This observation may be attributed to an orientational ``wetting'' phenomenon~\cite{vanRoij2000orientational,ivanov2013wall}, characterized by the formation of a thin nematic film covering the interior of the shell at polymer concentrations such that the bulk solution still lies in the isotropic phase (Fig.~\ref{fig1}a). This wetting effect is evidenced by a continuous increase in the local monomer density $\rho$ near the membrane, which exhibits radial oscillations characteristic of a layered structure for $\eta \gtrsim 0.24$ (Fig.~\ref{fig2}a) --- as well as a growing peak in the radial variations of the local alignment parameter $\overline{\alpha}$ located at a distance $\sim 0.05\,R_\eta$ from the shell surface, associated with a low degree of orientational order $\overline{\alpha} \lesssim0.2$ in the core region $r \lesssim 0.75\,R_\eta$ for $\eta\in[0.24,0.26]$ (Fig.~\ref{fig2}d).
\par
The surface arrangement of the corresponding director field displays large fluctuations in the vicinity of the uniaxial-biaxial (UB) transition, and gradually converges to the quadrupolar configuration depicted in Fig.~\ref{fig1}a around $\eta\simeq 0.26$. This structure may be represented as a collection of 4 topological defects of half-integer charge $s=1/2$ lying on the vertices of a regular tetrahedron, and connected in pairs by the nematic field through the shortest (geodetic) segment of great-circle arcs. The resulting texture consists of two mirror-symmetric domains lying at an angle $\pi/2$, joined together by field lines tracking the seams of a cricket ball near the membrane equatorial plane. Upon further increase of the concentration, we observe that these linked pairs of surface defects progressively move apart to reach antipodal positions, while the associated nematic domains simultaneously rotate to realign along a common axis (Fig.~\ref{fig1}c). This process leads to the growth and eventual coalescence of the two domains through the fusion of unconnected defect pairs, which yields a longitudinal bipolar pattern with two $s=1$ ``hedgehog'' defects at the poles for $\eta \gtrsim 0.32$ (Fig.~\ref{fig1}b).

Interestingly, a CB texture similar to that in Fig.~\ref{fig1}a was predicted theoretically nearly 15 years ago by Vitelli and Nelson~\cite{vitelli2006nematic} for thin nematic shells in the so-called \textit{one-constant approximation} of OF elasticities (c.f.~Sec.~\ref{subsec:density}). In this case, the director patterns derived using the method of conformal mappings~\cite{vitelli2006nematic} are in remarkable agreement with the stereographic projections obtained from our simulations near the UB transition. A closer inspection of the density variations in the radial structure of the phase further reveals that the destabilization of this tetrahedral surface arrangement is associated with a rapid divergence in the thickness of the wetting nematic film, which gradually takes over the entire cavity with increasing densities in the range $\eta \in [0.26,0.30]$, as evidenced by the broadening of the peak in $\overline{\alpha}$ in Fig.~\ref{fig2}d. Hence, the quadrupolar-to-bipolar transition observed in this regime is also qualitatively consistent with the Vitelli-Nelson theory, in which the lowest-energy state was found to progressively switch from a tetrahedral to a \mbox{(quasi-)bipolar} configuration by increasing the thickness of the nematic shell, as the two bulk disclination lines associated with the four $s=1/2$ surface defects are supplanted by more favorable 3D ``escaped'' arrangements~\cite{urbanski2017liquid,vitelli2006nematic}.

\subsection{Quantifying surface topological transitions} \label{subsec:quantifying}

Qualitatively, the crossover from quadrupolar to bipolar surface order may be characterized by the eigenvalues $\Lambda_i^{\rm surf}$ of $\mathcal{Q}_{\rm dir}(\Xi_{\rm surf})$ (Eq.~\eqref{eq:q_dir}). Elementary symmetry considerations lead to $\Lambda_2^{\rm surf}=\Lambda_3^{\rm surf}=-\Lambda_1^{\rm surf}/2$ for the ``cricket-ball'' (CB) texture and $\Lambda_1^{\rm surf}=\Lambda_2^{\rm surf}=-\Lambda_3^{\rm surf}/2$ for the longitudinal state, although the actual values of the $\Lambda_i$ generally depend on both thermal fluctuations and the finite thickness $r_s$ of the surface shell $\Xi_{\rm surf}$ (Fig.~\ref{fig1}c). Several studies~\cite{nikoubashman2017semiflexible,milchev2018densely,khadilkar2018self} have attempted to provide a more quantitative description of such topological crossovers by introducing a tensor involving pairs of adjacent bonds,
\begin{equation*}
  \mathcal{N}_k \equiv \Big(\widehat{\mathbf{t}_k\times \mathbf{t}_{k+1}}\Big) \otimes \Big(\widehat{\mathbf{t}_k\times \mathbf{t}_{k+1}}\Big).
\end{equation*}
We show in Supplemental Sec.~SII that in the case of semi-flexible chains, the equipartition theorem imposes
\begin{equation*}
  \big\langle \mathcal{N}_k\big\rangle = \frac{\mathcal{I}- \big\langle \mathcal{Q}_k\big\rangle}{3} + \mathcal{O}\bigg(\frac{l_b}{l_p}\bigg),
\end{equation*}
with $\mathcal{I}$ the identity tensor and $\mathcal{Q}_k$ the standard Landau-de Gennes tensor (Eq.~\eqref{eq:deGennes}), which is found to be quantitatively satisfied for all systems considered here (see Supplemental Fig.~S3). Hence, $\big\langle\mathcal{N}_k\big\rangle$ is effectively equivalent to $\big\langle \mathcal{Q}_k\big\rangle$, and is similarly affected by fluctuations as well as the arbitrary choice of $r_s$ --- suggesting that it may not reliably yield any additional insights as to the nature of these transitions. 

\begin{figure*}[htpb]
  \includegraphics[width=1.5\columnwidth]{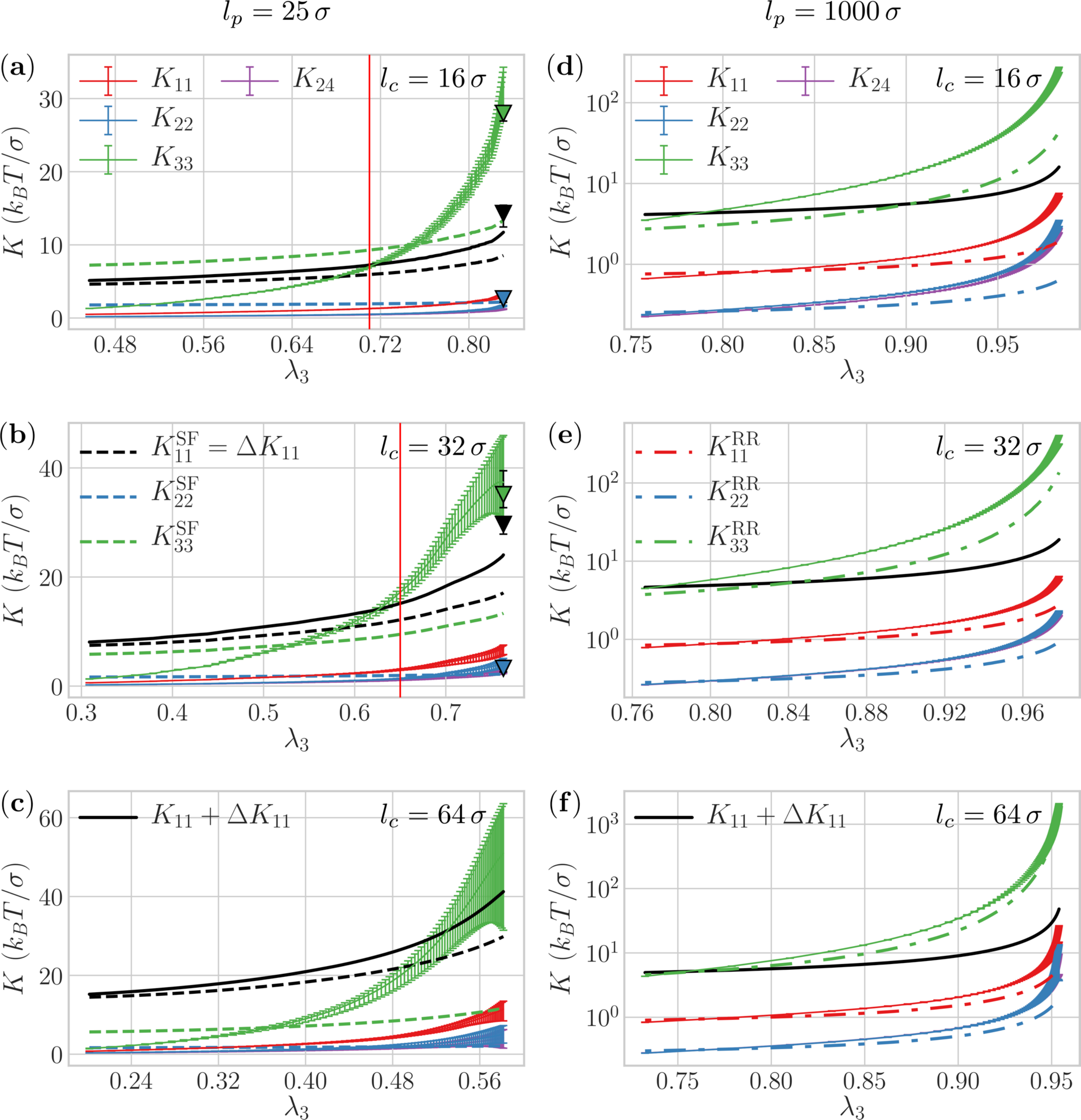}
  \caption{\label{fig3}Oseen-Frank elastic moduli of DNA- (a-c) and microtubule-like (d-f) chains as a function of the degree $\lambda_3$ of nematic order (Eqs.~\eqref{eq:legendre},~\eqref{eq:op_dft} and~\eqref{eq:op_bond}). Solid lines represent DFT predictions (Eqs.~\eqref{eq:K1}--\eqref{eq:delta_K1} and~\eqref{eq:saupe}). Error bars were estimated via a bootstrapping procedure~\cite{tortora2020chiral}, and are omitted in the case of $K_{11}+\Delta K_{11}$ for readability. Markers denote the simulation results of Ref.~\onlinecite{milchev2018nematic}, obtained for the same molecular model using $l_p=16\,\sigma$. Dashed and dash-dotted lines respectively correspond to the asymptotic analytical expressions of Ref.~\cite{odijk1986elastic} obtained for semi-flexible chains in the limit $l_c \gg l_p \gg \sigma$ ($K_{ii}^{\rm SF}$) and for fully-rigid rods with $\l_c \gg \sigma$ ($K_{ii}^{\rm RR}$). The lowest reported values of $\lambda_3$ correspond to the DFT-predicted nematic binodal point for the different systems. Red vertical lines denote the estimated value of the local surface OP $\lambda_3^{\rm surf}$ at the onset of tetrahedral CB order for dsDNA-like polymers with $l_c=16\,\sigma$ (Fig.~\ref{fig1}) and $l_c = 32\,\sigma$ (Fig.~\ref{fig4}).}
\end{figure*}

More accurately, we here implement a simple and robust algorithm to directly compute the positions of surface topological defects, which may be summarized as follows. We pick a random point $\mathbf{r}_{\rm trial}$ on the membrane surface, and consider its arbitrarily-large neighborhood $\Xi_{\rm trial} \equiv \big\{\mathbf{r}\in V \mid \lVert\mathbf{r}-\mathbf{r}_{\rm trial}\rVert < r_{\rm probe}\big\}$. We identify all polymer bonds with centers of mass $\mathbf{r}_k \in \Xi_{\rm trial}$, and compute the tensor $\mathcal{Q}_{\rm trial} \equiv \big\langle \mathcal{Q}_k\big\rangle_{\Xi_{\rm trial}}$ as in Eq.~\eqref{eq:q_loc}, from which the local degree of surface alignment may be estimated as $\alpha_{\rm trial} \equiv 2\big(\lambda_3^{\rm trial}-\lambda_2^{\rm trial}\big)/3$ (Eq.~\eqref{eq:alpha}). Since points associated with higher values of $\alpha_{\rm trial}$ are less likely to be located in the immediate vicinity of a defect, we may construct a loose collection of potential candidates by recording the position $\mathbf{r}_{\rm trial}$ with a probability $1-\alpha_{\rm trial}$, and repeating the process until a number $\mathcal{O}\big(10^4\big)$ of observations is reached. In the following, we set $r_{\rm probe}=5\,\sigma$, although the results were verified to be largely insensitive to any choice of value in the range $\sigma \ll r_{\rm probe} \ll R_\eta$.
\par
The two-fold (``head-tail'') local symmetry of the surface director field, which results from the strong tangential anchoring of the system at densities $\eta \gtrsim 0.24$ beyond the UB transition (Fig.~\ref{fig2}a, inset)~\cite{nelson2002toward,vitelli2006nematic}, implies that the number of stable topological defects may not exceed 4 in the ground state~\cite{lubensky1992orientational}. We may thus infer the most likely surface arrangements of defects in equilibrated simulations by spherical $k$-means clustering analysis~\cite{banerjee2005clustering} of the previous set of trial points, using a fixed number $k=4$ of centroids. The geometry of the defects may then be quantified through the Glassmeier parameter~\cite{robert1998tetrahedron},
\begin{equation}
  \label{eq:glassmeier}
  S_{\rm GM} = \frac{1}{2} \bigg(\frac{V_{\rm tet}}{V_{\rm reg}}+\frac{A_{\rm tet}}{A_{\rm reg}}\bigg),
\end{equation}
with $V_{\rm tet}$ and $A_{\rm tet}$ the respective volume and surface area of the tetrahedron defined by the 4 defects, rescaled by those $V_{\rm reg}$ and $A_{\rm reg}$ of the regular tetrahedron with identical circumsphere. $S_{\rm GM}=1$ characterizes regular tetrahedral order, while $S_{\rm GM}=0$ if all 4 defects are colinear --- indicating a degenerate bipolar configuration with two antipodal pairs of adjacent defects. $S_{\rm GM}=1/2$ in the intermediary case where all defects lie coplanar in the membrane equatorial plane, which identifies the so-called ``great-circle'' configuration (c.f.~Sec.~\ref{subsec:circumventing}). 
\par
Hence, the CB texture in Fig.~\ref{fig1}a may be further evidenced by the peak $S_{\rm GM}\simeq 0.9$ in Fig.~\ref{fig1}c at $\eta \simeq 0.26$, indicating a near-ideal tetrahedral distribution of defects, followed by a rapid decay in the range $\eta \in[0.26, 0.32]$, which reflects the gradual migration of defects towards the poles (Fig.~\ref{fig1}d). In this context, the fluctuations of $S_{\rm GM}$ for $\eta \gtrsim 0.32$ reveal the transient splitting of the two resulting $s=1$ hedgehogs into four ``half-hedgehogs'', corresponding to two antipodal pairs of close $s=1/2$ defects, through an elastic instability that has been similarly discussed by Nelson~\cite{nelson2002toward}.


\section{Splay- versus bend-dominated regimes} \label{sec:Splay- versus bend-dominated regimes}

However, in the Vitelli-Nelson description, a ground-state degeneracy was predicted in the thin-shell limit between the CB arrangement and a competing structure, obtained by linking the regular tetrahedral defect pattern pairwise through the longest (complementary) section of the same great-circle arcs. This latter configuration conversely leads to the partition of the membrane surface into two congruent nematic domains, joined together along a track matching the seams of a tennis ball (c.f.~Sec.~\ref{subsec:circumventing})~\cite{vitelli2006nematic}. To quantitatively understand the absence of stable ``tennis ball'' (TB) order for DNA-like chains with $l_c=16\,\sigma$, we write the free energy of the confined nematic system in the standard OF form~\cite{deGennes1993physics},
\begin{equation}
  \label{eq:f_tot}
   \mathscr{F} = \mathscr{F}_0 + \int_V d\mathbf{r}\, \mathpzc{f}_d(\mathbf{r}),
\end{equation}
where $\mathscr{F}_0$ is the free energy of a nematic phase with uniform director $\mathbf{n}_0$, and $\mathpzc{f}_d(\mathbf{r})$ is the elastic free energy density associated with director distortions~\cite{deGennes1993physics},
\begin{multline}
  \label{eq:f_frank}
  \mathpzc{f}_d= \frac{K_{11}}{2} \big(\nabla\cdot\mathbf{n}\big)^2 + \frac{K_{22}}{2} \big(\mathbf{n} \cdot [\nabla\times\mathbf{n}]\big)^2+ \frac{K_{33}}{2} \big(\mathbf{n}\times[\nabla\times\mathbf{n}]\big)^2 \\ - K_{24} \big(\nabla \cdot \big[\mathbf{n}\{\nabla\cdot \mathbf{n}\} + \mathbf{n}\times\{\nabla\times \mathbf{n}\} \big]\big).
\end{multline}
\par
In Eq.~\eqref{eq:f_frank}, $K_{11}$, $K_{22}$ and $K_{33}$ --- collectively referred to as the OF elastic moduli --- respectively quantify the thermodynamic penalties incurred by splay, twist and bend deformations, and generally depend on both the molecular system considered and the local structure $\lambda_i^{\rm loc}(\mathbf{r})$ of the orientational order~\cite{deGennes1993physics}. The $K_{24}$ term, known as the \textit{saddle-splay} elasticity, takes the form of a total divergence, and therefore integrates out to a boundary contribution in the limit of homogeneous orientational order, which is typically neglected in studies of bulk phases. This approximation is, however, generally inappropriate in confined geometries or in the presence of topological defects~\cite{kos2016relevance}, implying that $K_{24}$ may not be discarded \textit{a priori} in our case. Note that due to the strong tangential anchoring of the chains (Fig.~\ref{fig2}a, inset), we do not include an explicit surface coupling term in Eq.~\eqref{eq:f_frank}, which may be regarded as an external constraint for the purpose of variational analysis as long as the membrane does not significantly deviate from a spherical conformation~\cite{williams1986two}.
\par
To proceed, we remark that the local OPs $\lambda_i^{\rm loc}\equiv \lambda_i^{\rm surf}$ at the onset of the UB transition does not appreciably depend on position within the wetting nematic film, outside of the microscopic core of radius $\sim \sigma$ of the defects (Fig.~\ref{fig1}a), and may therefore be taken as uniform throughout the thin shell. Furthermore, we observe that the local surface biaxiality $\lambda_2^{\rm surf}-\lambda_1^{\rm surf}\ll \lambda_3^{\rm surf}$ is relatively weak for $\eta \gtrsim 0.26$ (Fig.~\ref{fig1}d), and may be neglected as a first approximation. Hence, the degree of molecular alignment within the shell may be fully described by the single scalar $\lambda_3^{\rm surf}\simeq \alpha_{\rm surf}$ (Eq.~\eqref{eq:alpha}). In this limit, the saddle-splay contribution simply reduces to an irrelevant additive constant~\cite{williams1986two}, and the degeneracy of the CB and TB textures may be lifted by relaxing the one-constant approximation $K_{11}^{\rm surf}=K_{33}^{\rm surf}(=K_{22}^{\rm surf})$, with $K_{ii}^{\rm surf} \equiv K_{ii}(\lambda_3^{\rm surf})$ the homogeneous OF moduli. 

\subsection{Density functional theory of LC elasticities} \label{subsec:density}

Under these assumptions, a convenient theoretical route is to treat Eq.~\eqref{eq:f_tot} as a perturbative expansion with respect to the uniform and uniaxial reference nematic state with director $\mathbf{n}_0$~\cite{tortora2017perturbative}. A microscopic expression for the corresponding reference free energy $\mathscr{F}_0$ may be obtained in the general functional form~\cite{woodward1991density}
\begin{equation}
  \label{eq:f}
  \mathscr{F}_0[\psi] = \mathscr{F}_{\rm id}[\psi] + \mathscr{F}_{\rm ex}[\psi],
\end{equation}
with $\psi \equiv \psi\big(\mathbf{u}\cdot\mathbf{n}_0\big)$ the molecular distribution quantifying the probability of finding a chain with arbitrary orientation $\mathbf{u}$ at any point in space. Note that the dependence of this orientation distribution function on the sole quantity $\mathbf{u}\cdot\mathbf{n}_0$ results from the assumed cylindrical symmetry of the phase about $\mathbf{n}_0$. In this case, the \textit{ideal} component $\mathscr{F}_{\rm id}$ of the free energy reads as~\cite{onsager1949effects}
\begin{multline}
  \label{eq:f_id}
  \frac{\beta \mathscr{F}_{\rm id}[\psi]}{V} = \rho_m(\log\rho_m-1)\\+\rho_m\oint d\mathbf{u}\,\log\Big\{\psi\big(\mathbf{u}\cdot\mathbf{n}_0\big)\Big\}\psi\big(\mathbf{u}\cdot\mathbf{n}_0\big),
\end{multline}
with $\rho_m \equiv N_c/V$ the molecular density, which corresponds to the entropy of an anisotropic ideal gas of macromolecules~\cite{woodward1991density}. 

\begin{figure*}[htpb]
  \includegraphics[width=2\columnwidth]{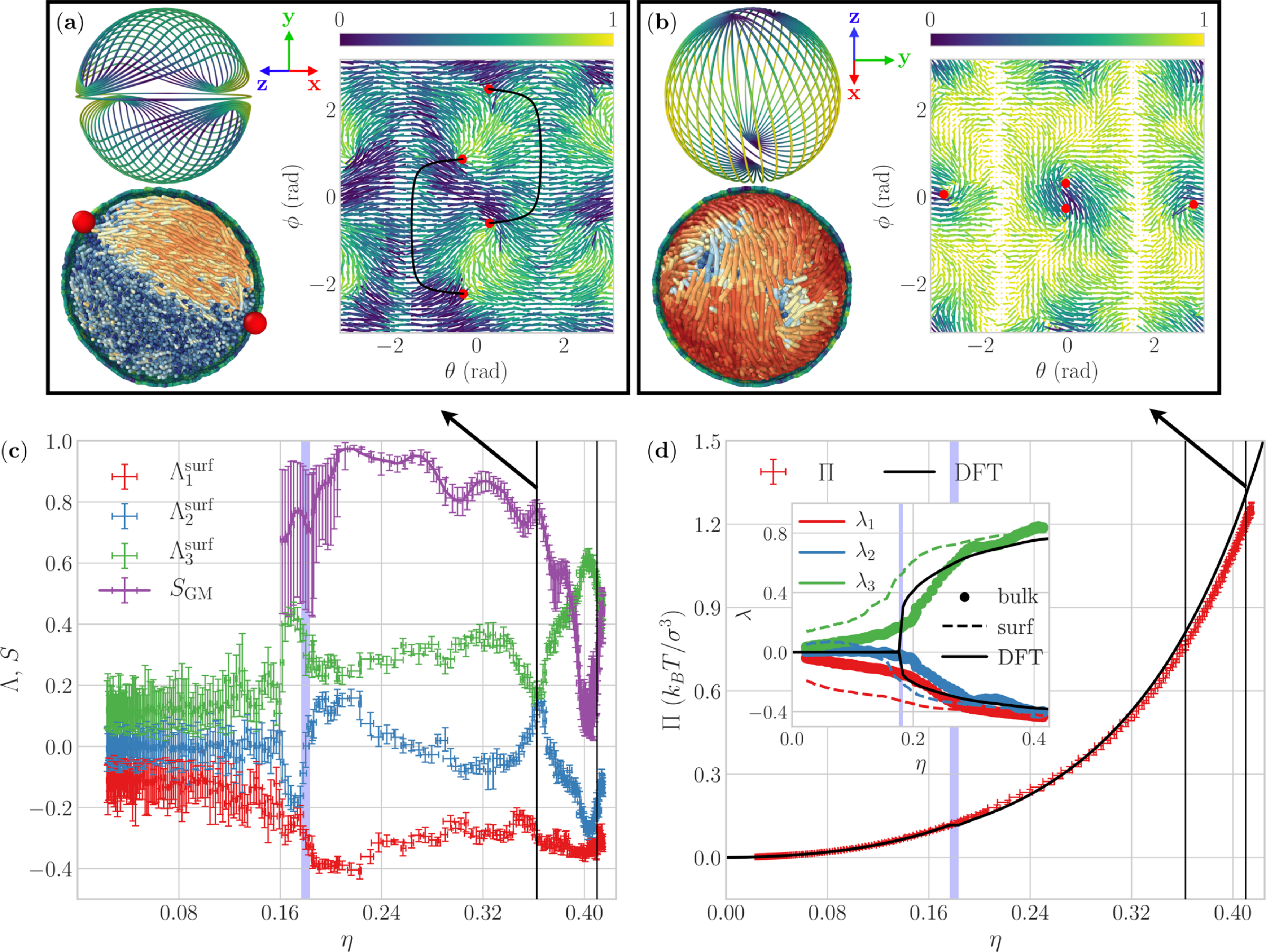}
  \caption{\label{fig4}Same as Fig.~\ref{fig1} for $l_c = 32\,\sigma$. (a) The ``great-circle'' state, with $\Lambda_x^{\rm surf}=\Lambda_z^{\rm surf}=\Lambda/2$, $\Lambda_y^{\rm surf}=-\Lambda$ [$\Lambda(r_s=0)=1/8$]. All defects approximately lie in the $\theta=0$ ($x$-$z$) equatorial plane. Bottom left: Equatorial cut of a simulation snapshot through the $y$-$z$ plane. Chains are colored according to the projection of bond orientations onto $\mathbf{z}$, and red spheres mark the computed locations of surface topological defects. Right: Stereographic map of the simulated surface director field. Black lines represent the two circular arcs connecting pairs of defects within the same nematic domain through the poles. Note that the $x$- and $z$-axes used for the projection are rotated by \SI{45}{\degree} around $\mathbf{y}$ to facilitate visualization. (b) The twisted bipolar configuration. The splitting of $s=1$ antipodal defects into two $s=1/2$ half-hedgehogs gives rise to a biaxial structure $\Lambda_z^{\rm surf} > \Lambda_y^{\rm surf} > \Lambda_x^{\rm surf}$, with the asymmetry in $\Lambda_x^{\rm surf}$ and $\Lambda_y^{\rm surf}$ reflecting the spontaneous chirality of the system. (c) and (d) are as in Fig.~\ref{fig1}.}
\end{figure*}

Conversely, the exact functional form of the excess contribution $\mathscr{F}_{\rm ex}$, which accounts for the presence of inter-molecular interactions, is generally not known. A successful ansatz due to Fynewever and Yethiraj~\cite{fynewever1998phase} yields $\mathscr{F}_{\rm ex}$ as a direct extension of the Onsager second-virial expression~\cite{onsager1949effects},
\begin{multline}
  \label{eq:f_exc}
  \frac{\beta \mathscr{F}_{\rm ex}[\psi]}{V} = -\frac{\rho_m^2 G(\eta)}{2} \int_V d\mathbf{r} \oiint d\mathbf{u}_1d\mathbf{u}_2  \\ \times \overline{f}\big(\mathbf{r},\mathbf{u}_1,\mathbf{u}_2\big) \psi\big(\mathbf{u}_1\cdot \mathbf{n}_0\big) \psi\big(\mathbf{u}_2\cdot \mathbf{n}_0\big),
\end{multline}
where the prefactor $G(\eta)=(1-0.75\,\eta)/(1-\eta)^2$ is a rescaling correction to approximate the effects of higher-order virial terms based on the Carnahan-Starling equation of state~\cite{parsons1979nematic,lee1987numerical}. In Eq.~\eqref{eq:f_exc}, $\overline{f}\equiv \big\langle\big\langle f_{\mathpzc{C}_1\mathpzc{C}_2}\big\rangle\big\rangle$ is the conformational average of the Mayer $f$-function, which involves the inter-molecular component of Eq.~\eqref{eq:poly_exc} for two arbitrary chain conformations $\mathpzc{C}_{1,2}$ with center-of-mass separation $\mathbf{r}$ and respective orientations $\mathbf{u}_{1,2}$~\cite{tortora2018incorporating},
\begin{equation}
  \label{eq:mayer}
  f_{\mathpzc{C}_1\mathpzc{C}_2}(\mathbf{r},\mathbf{u}_1,\mathbf{u}_2)=\exp\Bigg\{-\beta\sum_{k\in\mathpzc{C}_1} \sum_{k'\in\mathpzc{C}_2} u^{\rm WCA}_{\sigma}\big(r_{kk'}\big)\Bigg\}-1.
\end{equation}
\par
At thermodynamic equilibrium, the most probable distribution $\psi_\eta^{\rm eq}$ proceeds from the functional minimization of Eqs.~\eqref{eq:f}--\eqref{eq:f_exc} at fixed density $\eta$, and may be obtained using standard numerical methods detailed elsewhere~\cite{tortora2018incorporating}. For now, we simply note that unlike previous investigations, which had to resort to approximate empirical expressions for the generalized excluded volume $\overline{f}$ in Eq.~\eqref{eq:f_exc}~\cite{binder2020understanding}, we here evaluate $\overline{f}$ directly from Eq.~\eqref{eq:mayer} using representative ensembles of polymer conformations generated from our simulations. Hence, we are able to accurately apply this density functional theory (DFT) framework to the exact molecular model described in Sec.~\ref{sec:Numerical model} in the limit of a bulk, uniform nematic system. To assess the validity of the theory, we report in Fig.~\ref{fig1}d the osmotic pressure predicted by DFT,
\begin{equation}
  \label{eq:p_dft}
  \Pi = -\frac{\partial \mathscr{F}\big[\psi_\eta^{\rm eq}\big]}{\partial V}\bigg|_{N,T},
\end{equation}
as well as the theoretical nematic OPs~\cite{tortora2018incorporating},
\begin{equation}
\begin{gathered}
  \label{eq:op_dft}
  \lambda_3 = S_{\rm mol} \times S_{\rm bond}, \\
  \lambda_1 = \lambda_2 = -\frac{\lambda_3}{2}, \\
\end{gathered}
\end{equation}
where $S_{\rm mol}$ and $S_{\rm bond}$ respectively quantify the alignment of macromolecular axes $\mathbf{u}$ and internal bonds $\mathbf{t}_k$~\cite{tortora2018incorporating},
\begin{equation}
\begin{gathered}
  \label{eq:op_bond}
  S_{\rm mol} = \oint d\mathbf{u} \, \psi_\eta^{\rm eq}\big(\mathbf{u}\cdot\mathbf{n}_0\big) \times P_2\big(\mathbf{u}\cdot \mathbf{n}_0\big), \\
  S_{\rm bond} = \big\langle P_2\big(\mathbf{t}_k\cdot \mathbf{u}\big)\big\rangle.
  \end{gathered}
\end{equation}
The good agreement with simulations obtained outside of the orientational wetting regime ($\eta\in[0.24,0.30]$) suggests that the local structure of the nematic phase is remarkably unaltered by the effects of confinement, even for the limited system sizes considered here, thus validating the premise of the perturbative treatment of director distortions underlying Eqs.~\eqref{eq:K1}--\eqref{eq:K3}~\cite{tortora2017perturbative}. Moderate underestimations of osmotic pressure may nonetheless be observed at higher densities, which may reflect potential shortcomings of the rescaled virial approximation (Eq.~\eqref{eq:f_exc})~\cite{egorov2016insight}.
\par
Let us set the $z$-axis of the laboratory frame to the homogeneous director $\mathbf{n}_0$. Using Eqs.~\eqref{eq:f_tot}--\eqref{eq:f_exc}, lengthy but straightforward manipulations yield the OF elastic moduli in the form~\cite{poniewierski1979statistical,somoza1989frank,yokoyama1997density,tortora2017perturbative,tortora2020chiral}
\begin{equation}
\label{eq:K1}
\begin{split}
  \beta K_{11} &= \frac{\rho^2_m G(\eta)}{2} \int_V d\mathbf{r} \oiint d\mathbf{u}_1 d\mathbf{u}_2 \\&\quad\times \overline{f}\big(\mathbf{r},\mathbf{u}_1,\mathbf{u}_2\big) \dot{\psi}_\eta^{\rm eq}(u_{1z}) \dot{\psi}_\eta^{\rm eq}(u_{2z}) r_x^2 u_{1x} u_{2x},
\end{split}
\end{equation}
\begin{equation}
  \label{eq:K2}
  \begin{split}
  \beta K_{22} &= \frac{\rho^2_m G(\eta)}{2} \int_V d\mathbf{r} \oiint d\mathbf{u}_1 d\mathbf{u}_2 \\&\quad\times \overline{f}\big(\mathbf{r},\mathbf{u}_1,\mathbf{u}_2\big) \dot{\psi}_\eta^{\rm eq}(u_{1z}) \dot{\psi}_\eta^{\rm eq}(u_{2z}) r_x^2 u_{1y} u_{2y},
  \end{split}
  \end{equation}
  \begin{equation}
  \begin{split}
  \label{eq:K3}
  \beta K_{33} &=\frac{\rho^2_m G(\eta)}{2} \int_V d\mathbf{r} \oiint d\mathbf{u}_1 d\mathbf{u}_2 \\&\quad\times \overline{f}\big(\mathbf{r},\mathbf{u}_1,\mathbf{u}_2\big) \dot{\psi}_\eta^{\rm eq}(u_{1z}) \dot{\psi}_\eta^{\rm eq}(u_{2z}) r_z^2 u_{1x} u_{2x},
  \end{split}
  \end{equation}
where $\dot{\psi}$ denotes the derivative of $\psi$. Note that Eqs.~\eqref{eq:K1}--\eqref{eq:K3} further rely on the so-called \textit{quasi-homogeneous approximation}~\cite{somoza1989frank,yokoyama1997density}, which postulates that the local molecular density $\rho_m$ remains unaffected by orientational fluctuations. This assumption is generally inadequate in the case of long macromolecules, for which splay deformations necessarily incur an additional entropic penalty due to the local accumulation of chain extremities~\cite{meyer1982macroscopic}. This effect gives rise to an effective renormalization of the splay modulus $K_{11}^{\rm eff} = K_{11}+\Delta K_{11}$, where the compressibility correction $\Delta K_{11}$ may be approximated as~\cite{meyer1982macroscopic,milchev2018nematic}
\begin{equation}
  \label{eq:delta_K1}
  \beta \Delta K_{11} = \frac{4\eta}{\pi} \frac{l_c}{\sigma^2}.
\end{equation}
\par 
The corresponding OF moduli are reproduced in Fig.~\ref{fig3}, and are found to be entirely consistent with the limited simulation results of Ref.~\onlinecite{milchev2018nematic} for Kremer-Grest chains with $l_p=16\,\sigma$ (Fig.~\ref{fig3}a). This observation contrasts with the significant discrepancies of the DFT predictions reported therein~\cite{milchev2018nematic}, which highlights the importance of accurately representing the interaction kernel $\overline{f}$ for the computation of emergent LC elasticities~\cite{tortora2017perturbative,tortora2017hierarchical}. Some slight underestimations of the splay elastic modulus $K_{11}^{\rm eff}$ may nonetheless be observed, which may suggest limited deficiencies of the compressibility contribution in Eq.~\eqref{eq:delta_K1}~\cite{romani2018elastic}. In the case of DNA-like chains with $l_c=16\,\sigma$, we observe that the regime of stable tetrahedral order is characterized by a weak stiffening of the bending rigidity relative to the splay modulus ($K_{11}^{\rm eff} \lesssim K_{33}$). This subtle anisotropy may therefore account for the stability of the splay-rich CB over the bend-rich TB texture, in full agreement with both our simulations and the arguments of Ref.~\onlinecite{vitelli2006nematic}.

\subsection{Circumventing splay: The twisted \& ``tennis-ball'' states} \label{subsec:circumventing}

We now turn our focus to longer chains with $l_c=32\,\sigma$ (Fig.~\ref{fig4}). Similarly, we find that the UB transition, occurring now at $\eta \simeq 0.20$, induces a stable CB surface structure ($\Lambda_2^{\rm surf}=\Lambda_3^{\rm surf}=-\Lambda_1^{\rm surf}/2$, $S_{\rm GM} \simeq 1$, Fig.~\ref{fig4}c), which may also be attributed to a slight asymmetry $K_{11}^{\rm eff} \lesssim K_{33}$ in the splay and bend elastic modes (Fig.~\ref{fig3}b). The fluctuations of the defect positions --- quantified by the width of the error bars in $S_{\rm GM}$ --- are found to be substantially weaker than in the case of shorter chains (Fig.~\ref{fig1}c), which may reflect the larger magnitude of the mean OF modulus $K \equiv \sqrt{K_{11}^{\rm eff} K_{33}}$, relative to the thermal stability threshold $K_0 = 16\,k_BT/3\pi\sigma$~\cite{nelson2002toward,vitelli2006nematic}, at the onset of CB order ($K \simeq 2\,K_0$ and $4\,K_0$ for $l_c=16\,\sigma$ and $32\,\sigma$, respectively). 

Interestingly, we observe that longer polymers display partial orientational wetting across a wider concentration range than their shorter counterparts. Indeed, increasing densities in the range $\eta\in [0.20,0.25]$ for $l_c = 32\,\sigma$ significantly enhances the degree of alignment within the wetting layer, as quantified by the height of the peaks in $\alpha$ near the membrane (Fig.~\ref{fig2}e), but leads to a slower increase in the layer thickness --- measured by the limited increments in the corresponding peak widths. This effect may reflect the increasingly-weak first-order character of the I-N transition for chains with lower $l_p/l_c$~\cite{debraaf2017self}, which results in a more gradual growth of the surface-nucleated nematic domains as the density is raised beyond the bulk binodal region predicted by DFT (Fig.~\ref{fig4}d). 
\par
Consequently, the local order parameter $\lambda_3^{\rm surf}$ within the stable wetting layer rises from $\lambda_3^{\rm surf} \simeq 0.65$ at the onset of CB order ($\eta \simeq 0.20$) to $\lambda_3^{\rm surf} \simeq 0.75$ for $\eta\simeq 0.25$, which induces a rapid shift in the relative OF moduli from $K_{33}/ K_{11}^{\rm eff}\simeq 1$ to $K_{33} / K_{11}^{\rm eff} \gtrsim 1.5$ (Fig.~\ref{fig3}b). This increasing elastic anisotropy is associated with a migration of the four $s=1/2$ defects towards a common equatorial plane, accompanied by the azimuthal expansion of the two corresponding CB domains to fully occupy each hemisphere (Fig.~\ref{fig4}a). This so-called ``great-circle'' (GC) configuration has also been predicted by theory for thin nematic shells in the limit $K_{33} \gg K_{11}$~\cite{shin2008topological}, and may be obtained from the bipolar surface pattern in Fig.~\ref{fig1}b by cutting the system through any plane containing the two poles, followed by a rotation of one of the hemispheres by an angle $\pi/2$. Remarkably, the rapid increase in layer thickness for $\eta \gtrsim 0.25$ then gives rise to a bulk-ordered state retaining the GC symmetry, which may be similarly realized by applying the previous cut-and-rotate surgery through the bulk of the bipolar state~\cite{shin2008topological} (Fig.~\ref{fig4}a). This arrangement bears a grain boundary wall at the equatorial junction between the two nematic domains, and could not be spontaneously recovered from quenching simulations (c.f.~Sec.~\ref{subsec:kg}, data not shown). Hence, it is likely to be only metastable, although its apparent lifetime lies well beyond the simulation timescale.
\par
Further increases of the density lead to the gradual annealing of the grain boundary by progressive realignment of the two hemispheric regions. However, unlike in Fig.~\ref{fig1}b, this mutual rotation does not lead to the full coalescence of unconnected defect pairs, but rather yields four stable ``half-hedgehogs'' (c.f.~Sec.~\ref{subsec:quantifying}), associated with a finite angular mismatch between the two corresponding surface nematic domains (Fig.~\ref{fig4}b). This arrangement is strongly reminiscent of the twisted bipolar texture observed in tangentially-anchored droplets of low-molecular-weight LCs~\cite{volovik1983topological}, whose stability has been explained in terms of the Williams inequality~\cite{williams1986two}
\begin{equation}
  \label{eq:williams}
  K_{33} < 2.32\,\big(K_{11}-K_{22}\big).
\end{equation}
We show in Supplemental Fig.~S4 that Eq.~\eqref{eq:williams} is violated in the high-density regime of DNA-like chains with $l_c=16\,\sigma$, but holds for $l_c \geq 32\,\sigma$. This crossover may be attributed to the stiffening of the splay compressibility contribution $\Delta K_{11}$ for longer chains (Eq.~\eqref{eq:delta_K1}), which results in the increasing favorability of a twisted bulk structure over the large splay distortion imposed by $s=1$ hedgehog defects~\cite{williams1986two}. For $l_c=32\,\sigma$, Fig.~\ref{fig3}b yields $K_{22}/ K_{11}^{\rm eff}\simeq 0.2$ and $K_{33}/ K_{11}^{\rm eff}\simeq 1.7$ in the high density regime, from which the Williams theory predicts a twist angle at the surface of about $\SI{30}{\degree}~$\cite{williams1986two} --- in close agreement with our simulated value of $\sim\SI{25}{\degree}$ (Fig.~\ref{fig4}b).

\begin{figure*}[htpb]
  \includegraphics[width=2\columnwidth]{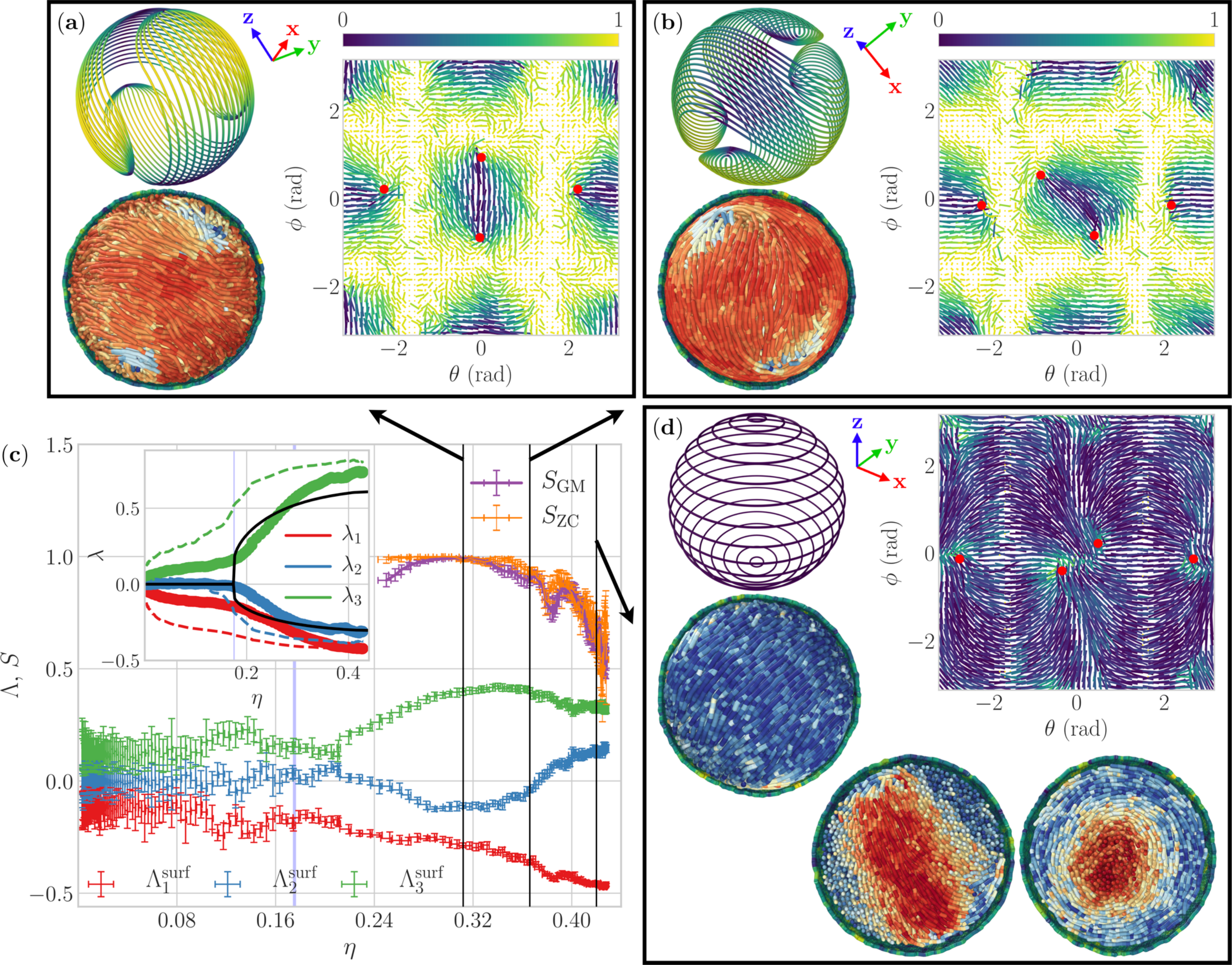}
  \caption{\label{fig5}Same as Fig.~\ref{fig1} for $l_c = 64\,\sigma$. (a) The ``tennis-ball'' (TB) state, with $\Lambda_z^{\rm surf}=\Lambda$, $\Lambda_x^{\rm surf}=\Lambda_y^{\rm surf}=-\Lambda/2$ [$\Lambda(r_s=0)=1/3$]. The TB surface texture matches the equipotential lines of the CB director field (Fig.~\ref{fig1}a)~\cite{vitelli2006nematic}. (b) The buckled TB arrangement, characterized by the migration of $s=1/2$ defects towards the $\phi=0$ ($y$-$z$) meridional plane. Note the simultaneous rotation of the domain central regions towards the $x$-$z$ equatorial lines. (c) Zhang-Chen axial OP~\cite{zhang2011tennis} ($S_{\rm ZC}$), which quantifies the degree of TB order, as a function of polymer packing fraction ($\eta$). Since $S_{\rm ZC}$ does not rely on the determination of defect positions~\cite{zhang2011tennis}, the close quantitative agreement observed with $S_{\rm GM}$ (Eq.~\eqref{eq:glassmeier}) may be interpreted as evidence of the accuracy of our defect localization algorithm. Other notations and symbols are as in Fig.~\ref{fig1}. (d) The latitudinal bipolar configuration, with $\Lambda_x^{\rm surf}=\Lambda_y^{\rm surf}=\Lambda/2$, $\Lambda_z^{\rm surf}=-\Lambda$ [$\Lambda(r_s=0)=1/2$]. Bottom: Equatorial cuts of the simulated structure through the $y$-$z$ (left) and $x$-$y$ (right) planes. Chains are colored according to the projection of bond orientations onto $\mathbf{z}$.}
\end{figure*}

Lastly, we report in Fig.~\ref{fig5} the self-assembling behavior obtained for $l_c=64\,\sigma$. In this case, we find that the splay-rich CB texture is replaced by the bend-rich tennis-ball (TB) state post the UB transition (Fig.~\ref{fig5}a). Note that the DFT approach underlying Eqs.~\eqref{eq:f}--\eqref{eq:f_exc} implicitly assumes that the polymer conformational fluctuations may be decoupled from the degree of nematic order~\cite{fynewever1998phase}, which leads to increasing underestimations of nematic OPs for chains with $l_c \gg l_p$~\cite{tortora2018incorporating}. Hence, we are now unable to accurately estimate the OF moduli in the strong alignment regime $\lambda_3^{\rm surf} \simeq 0.75$ associated with the TB structure (Fig.~\ref{fig5}c), although a naive extrapolation of the trends in Fig.~\ref{fig3}c suggests that $K_{11}^{\rm eff}$ and $K_{33}$ are likely of comparable magnitude. Qualitatively, $K_{33}$ is expected to be asymptotically independent of contour length for $l_c \gg l_p$~\cite{odijk1986elastic}, and we therefore presumably impute the stability of the TB structure to $K_{11}^{\rm eff} \gtrsim K_{33}$~\cite{vitelli2006nematic}, based on the stiffening of $K_{11}^{\rm eff}$ with increasing $l_c$ (Sec.~\ref{subsec:density}).
\par
Similarly to Fig.~\ref{fig4}, we observe that increasing the polymer concentration induces a progressive migration of the 4 defects from a regular tetrahedral configuration ($S_{\rm GM} = 1$) towards a common meridional plane ($S_{\rm GM} \simeq 0.5$, Fig.~\ref{fig5}c). However, this transition is in this case accompanied by the buckling of the two nematic domains (Fig.~\ref{fig5}b), which yields a bend-dominated analog of the GC state (Fig.~\ref{fig4}a). This buckling allows for the gradual realignment of the central sections of the two domains along the equatorial lines, and eventually leads to a latitudinally-ordered configuration bearing two pairs of near-coplanar defects near the poles (Fig.~\ref{fig5}d). Note that this kinetic pathway results in the juxtaposition of $s=1/2$ defect pairs with opposite orientations in the bipolar latitudinal state (Fig.~\ref{fig5}b), which precludes their coalescence into the $s=1$ hyperbolic defects expected from the idealized depiction in Fig.~\ref{fig5}d.
\par
\begin{figure}[htpb]
  \includegraphics[width=\columnwidth]{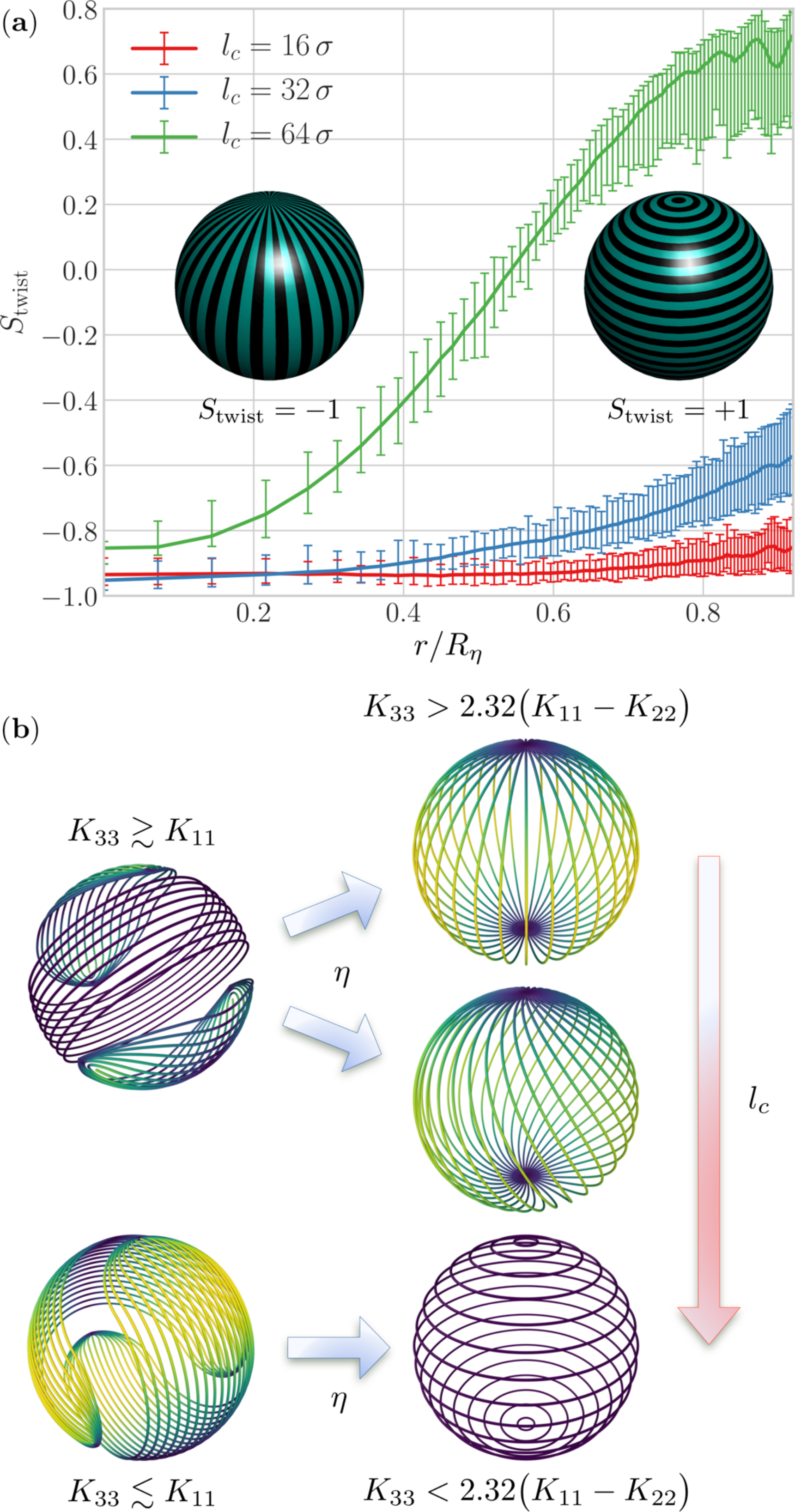}
  \caption{\label{fig6}(a) Helicoidal OP~\cite{zhang2011tennis} $S_{\rm twist}(r) \equiv \big\langle\cos(2\Phi_k)\big\rangle_r$ for DNA-like chains at $\eta \simeq 0.40$, where $\Phi_k$ is the angle between a bond vector and the local latitude circle, and we used the notations of Fig.~\ref{fig2}. $S_{\rm twist}(r)=-1$ implies that chains at radial distance $r$ point along the polar axis, and $S_{\rm twist}=1$ indicates latitudinal order. Intermediary values specify a finite mean twist angle. (b) Schematic summary of the self-assembling behavior of DNA-like chains in membrane confinement.}
\end{figure}

The director field is found to adopt a radially-twisted structure in the bulk, in which chains located close to the polar axis point along the north-south direction, and undergo a continuous rotation towards the latitude lines near the membrane surface with increasing axial distance (Figs.~\ref{fig5}d-\ref{fig6}a). This arrangement, variously referred to as an escaped concentric configuration~\cite{fernandez2007topological} or condensed Hopf fibration~\cite{liang2019orientationally}, is typically only metastable in low-molecular-weight LCs~\cite{fernandez2007topological}. Here, we simply identify this texture as a special case of the twisted bipolar state (Fig.~\ref{fig4}b), in which the surface twist angle reaches a full \SI{90}{\degree} (Fig.~\ref{fig6}), and thus attribute its stability to the Williams criterion (Eq.~\eqref{eq:williams}) --- which we expect to be quite generally satisfied for $l_c \gg \sigma$ due to the splay compressibility correction in Eq.~\eqref{eq:delta_K1}. Hence, the latitudinal structure in Fig.~\ref{fig5}d may provide a robust and generic template for the dense packing of long DNA-like chains, as further discussed in Sec.~\ref{sec:Discussion}.

\section{Tactoid-like \& faceted morphologies} \label{sec:Tactoid-like and faceted morphologies}

\begin{figure*}[htpb]
  \includegraphics[width=2\columnwidth]{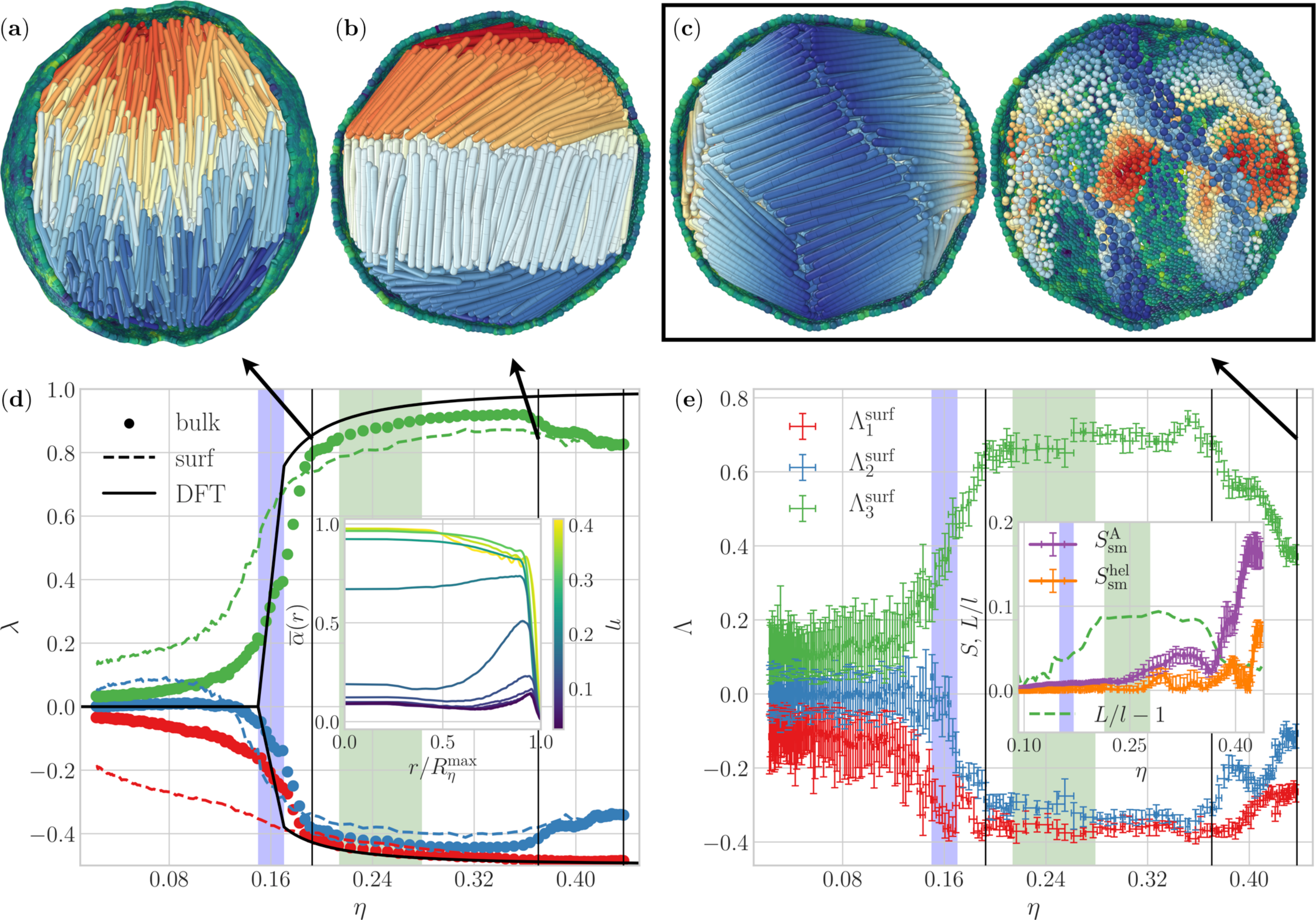}
  \caption{\label{fig7}Self-organization of short tubulin-like chains ($l_c=16\,\sigma$, $l_p=1000\,\sigma$) confined within an erythrocyte-like spectrin shell. (a) Nematic and (b) smectic states in cut and peeled view, respectively. Colors reflect the projection of the filament center-of-mass positions onto the polar axis, in order to facilitate the visualization of potential smectic order. (c) Peeled view of the spiral smectic phase, displaying the full polymers (left) and the chain extremities (right). Monomers are colored according to their radial distance from the helical axis. (d) Local nematic OPs ($\lambda_i$) as a function of polymer density ($\eta$). Inset: Radial variations of the local alignment parameter $\overline{\alpha}$, defined as in Fig.~\ref{fig2}. $R_\eta^{\rm max}$ denotes the maximum radial extent of the membrane. (e) Global surface director OPs ($\Lambda_i^{\rm surf}$) as a function of $\eta$. Inset: Membrane anisotropy ($L/l$), smectic ($S_{\rm sm}^{\rm A}$, Eq.~\eqref{eq:smectic}) and helical smectic OPs ($S_{\rm sm}^{\rm hel}$, Eq.~\eqref{eq:helical_smectic}). Blue regions delimit the exact I-N coexistence range calculated by DFT in the bulk, while the approximate predictions of Ref.~\cite{odijk1986theory} for WLCs in the limit $l_p \gg \l_c \gg \sigma$ are highlighted in green (Eq.~\eqref{eq:odijk_bin}).}
\end{figure*}

Let us now consider the case of stiff, tubulin-like filaments confined within an elastic shell matching the mechanical properties of the red blood cell plasma membrane. The model unit of length $\sigma$, defined as the approximate outer diameter of a microtubule, now reads as $\sigma \simeq \SI{25}{\nano\meter}$. The flexural rigidity of microtubules is generally dependent on both contour length~\cite{pampaloni2006thermal,vandenHeuvel2008microtubule} and the magnitude of the applied strains~\cite{memet2018microtubules}, due to the combination of their finite shear modulus with a cross-sectional buckling instability. For simplicity, we here set the polymer persistence length to $l_p = 1000\,\sigma\simeq\SI{25}{\micro\meter}$, which likely provides a reasonable order of magnitude for the regime of limited chain deformations and contour lengths considered here ($l_c \in [\SI{0.4}{\micro\meter},\SI{6.4}{\micro\meter}]$)~\cite{pampaloni2006thermal,vandenHeuvel2008microtubule,memet2018microtubules}, although significantly larger values $l_p \in[\SI{1}{\milli\meter},\SI{10}{\milli\meter}]$ may be obtained for longer filaments~\cite{pampaloni2006thermal}. The area modulus $K_A \equiv \lambda+\mu$ of the membrane spectrin network has been measured as $K_A \simeq \SI{50}{\micro\newton\per\meter}$~\cite{park2010measurement}, where the 2D Lam\'e coefficients $(\lambda,\mu)$ are related to the stiffness parameter $k_m$ via $\lambda=\mu=\sqrt{3}k_m/4$~\cite{seung1988defects}. Hence, $k_m \simeq 10 \,k_BT/\sigma^2$, and the corresponding bending modulus is similarly set to $\kappa_m \simeq 15\,k_B T$~\cite{park2010measurement}.

\subsection{Ellipsoidal, smectic \& spiral smectic states} \label{subsec:ellipsoidal}
 
We report in Fig.~\ref{fig7} the phase behavior of short chains with $l_c=16\,\sigma\simeq\SI{0.4}{\micro\meter}$. We first remark that orientational wetting is substantially inhibited, compared to the more compliant DNA-like chains with similar reduced dimensions $l_c/\sigma$ (Fig.~\ref{fig1}). Indeed, although a weak UB surface transition --- characterized by a moderate peak in the radial alignment parameter $\overline{\alpha}$ near the membrane (Fig.~\ref{fig7}d, inset) --- is still observable at densities $\eta \simeq 0.12$ slightly below the bulk I-N binodal range predicted by DFT ($0.15\lesssim\eta\lesssim 0.17$), the degree of order within the nematic surface layer is found to be limited in the bulk-disordered regime ($\overline{\alpha} \lesssim 0.5$ for $\eta \lesssim 0.17$, Fig.~\ref{fig7}d). Consequently, we find no tangible evidence of a discernible quadrupolar surface texture, and the system directly transitions from a near-isotropic state to a longitudinal bipolar structure with $\Lambda_1^{\rm surf} = \Lambda_2^{\rm surf}=-\Lambda_3^{\rm surf}/2$ around $\eta\simeq 0.17$ (Fig.~\ref{fig7}e). 
\par
Furthermore, increasing the concentration beyond $\eta\simeq 0.17$ leads to a rapid shift in the radial distribution of orientational order from a surface- ($\lambda_3^{\rm surf} > \lambda_3^{\rm bulk}$) to a bulk-dominated arrangement ($\lambda_3^{\rm surf} < \lambda_3^{\rm bulk}$, Fig.~\ref{fig7}d), which is also evidenced by the monotonic decrease in $\alpha$ with increasing radial distance in the bulk nematic stability range $\eta \gtrsim 0.17$ (Fig.~\ref{fig7}d, inset). We attribute this effect to the incompatibility of the large chain bending rigidity with the finite curvature of the membrane for $R_\eta \ll l_p$. Indeed, as discussed further in Sec.~\ref{subsec:tetrahedral}, geometrical frustration generally imposes a local drop in polymer density as one approaches the interface, if the membrane radius $R_\eta$ and the chain contour length $l_c$ are of comparable magnitude~\cite{groh1999fluids}. These observations suggest a conflicting dual role of the membrane on the ordering behavior of stiff chains, whereby the induced restrictions in the range of accessible orientations favor the nucleation of a nematic wetting layer, while the lower adsorption of polymers at the interface limits the subsequent degree of surface nematic alignment --- in agreement with previous numerical studies of spherically-confined rigid rods~\cite{trukhina2008computer}.
\par
The onset of nematic order is found to induce a significant directional elongation of the shell along the polar axis, borne by the eigenvector $\mathbf{e}_3^{\rm surf}$ of $\mathcal{Q}_{\rm dir}(\Xi_{\rm surf})$, which yields an ellipsoidal overall shape (Fig.~\ref{fig7}a). Using the cylindrical symmetry of the system about $\mathbf{e}_3^{\rm surf}$, the anisotropy of the membrane may be evaluated from the dimensions $L$ and $l$ of its semi-major and semi-minor axes, respectively defined as the extremal projections of vertex positions parallel and perpendicular to $\mathbf{e}_3^{\rm surf}$. Thus, we find that the sample displays an axial aspect ratio $L/l\simeq 1.1$, which is largely independent of polymer density in the range $\eta\in[0.20,0.35]$ (Fig.~\ref{fig7}e, inset).
\par
In this context, a rigorous quantitative theoretical analysis of the system would be significantly more involved, since the simultaneous determination of the optimal director configuration and its associated membrane morphology generally amounts to a complex free-boundary variational problem. However, some useful insights may be gleaned from phenomenological studies of finite-size LC droplets, referred to as \textit{tactoids}, which commonly occur in bulk systems near the I-N transition as a consequence of the nucleation of the nematic phase from the surrounding isotropic fluid~\cite{prinsen2003shape}. Assuming a uniaxial longitudinal symmetry, the droplet aspect ratio may be shown to be governed by the combination of a dimensionless surface anchoring strength $\omega$ and splay-to-surface-tension ratio $\mathpzc{K}$~\cite{prinsen2003shape},
\begin{equation}
  \label{eq:kappa_tactoid}
  \mathpzc{K} = \frac{K_{11}^{\rm eff}-K_{24}}{\gamma V^{1/3}},
\end{equation}
with $\gamma$ the interfacial tension. Note that due to the presence of potential surface shape fluctuations and/or deviations from strong tangential anchoring, the saddle-splay term $K_{24}$ must now be retained --- although its contribution to Eq.~\eqref{eq:kappa_tactoid} reduces to an apparent softening of the splay rigidity $K_{11}^{\rm eff}$ for a bipolar director field. Under the assumptions of Eqs.~\eqref{eq:K1}-\eqref{eq:K3}, a simple microscopic expression for $K_{24}$ may be obtained in the Cauchy-Nehring-Saupe form~\cite{yokoyama1997density},
\begin{equation}
  \label{eq:saupe}
  K_{24} = \frac{K_{11}+K_{22}}{4}.
\end{equation}
\par
In the case of polymerized shells, an effective surface tension $\gamma_{\rm eff}$ may be derived in the small-strain limit~\cite{landau1986theory},
\begin{equation}
  \label{eq:surf_tens}
 \gamma_{\rm eff} = \frac{Y_0}{2(1-\nu)}\frac{\Delta A}{A_0},
\end{equation}
where $\Delta A \equiv A_\eta - A_0$ denotes the excess surface area of the membrane mesh at finite density $\eta$ relative to that $A_0$ in the reference elastic state. Using Eq.~\eqref{eq:surf_tens} and Fig.~\ref{fig3}d, it is shown in Supplemental Fig.~S5 that $\mathpzc{K}$ varies in the range $[0.02,0.04]$ throughout the stable nematic regime. Assuming a typical value $\omega \simeq 1$, as commonly accepted for rod-like particles~\cite{prinsen2003shape}, the theory of Ref.~\onlinecite{prinsen2003shape} would predict the equilibrium morphology of the system to display a strongly-anchored bipolar director arrangement, associated with an aspect ratio $L/l\in [1.1,1.2]$. These results are fully consistent with our findings, although the moderate overestimations of $L/l$ could reflect the additional costs of membrane bending deformations in our simulations, which have no obvious equivalent in phase-separated tactoids. In this framework, the absence of directional elongation observed in the case of dsDNA may be attributed to a low reduced modulus $\mathpzc{K} < 0.01$ (see Supplemental Fig.~S5), due to the larger effective tension $\gamma_{\rm eff}$ of the nuclear envelope relative to the splay modulus $K_{11}^{\rm eff}$ of the more compliant chains at fixed density $\eta$ (Fig.~\ref{fig3}a), which would lead to a predicted relative anisotropy $(L-l)/l \lesssim 0.05$~\cite{prinsen2003shape}.
\par
Upon raising the density around $\eta\simeq 0.35$, we observe the emergence of smectic layers, which retain the approximate uniaxial symmetry of the longitudinal nematic state (Fig.~\ref{fig7}b). This partial positional ordering may be quantified through the standard smectic OP~\cite{deGennes1993physics},
\begin{equation}
  \label{eq:smectic}
  S_{\rm sm}^{\rm A} = \max_{\Delta\in\mathbb{R}} \Bigg | \frac{1}{N_c}\sum_{m=1}^{N_c} \exp\bigg(-\frac{2\mathrm{i}\pi}{\Delta} \mathbf{r}_m \cdot \mathbf{e}_3^{\rm surf}\bigg) \Bigg |^2,
\end{equation}
with $|\cdot|$ the complex modulus and $\mathbf{r}_m$ the center-of-mass position of the $m$-th filament. $S_{\rm sm}^{\rm A}$ is thus found to rapidly increase towards a plateau value $S_{\rm sm}^{\rm A}\simeq 0.2$ for $\eta \gtrsim 0.35$ (Fig.~\ref{fig7}e, inset). Interestingly, although the smectic phase is generally known to be destabilized by finite chain flexibility~\cite{debraaf2017self,milchev2018nematic}, the nematic-smectic A transition of fully-rigid rods with $l_c/\sigma \simeq 15$ has been predicted to occur at significantly higher densities $\eta \simeq 0.45$ in the bulk~\cite{bolhuis1997tracing}. In our case, the onset of smectic organization around $\eta\simeq 0.35$ is further associated with a reduction in the anisotropy of the membrane, which reverts back to a near-spherical conformation with diameter $2R_\eta \simeq \sigma \,(3N/2\eta)^{1/3} \simeq 50\,\sigma \simeq 3\,l_c$ for $N=\num[group-separator={,},group-minimum-digits={3}]{32768}$ monomers (Fig.~\ref{fig7}b). Hence, we attribute the stability of the smectic state to a ``magic number'' effect, similar to the layering transitions reported for WLCs confined within spherical cavities whose dimensions approach an integer multiple of $l_c$~\cite{nikoubashman2017semiflexible,milchev2018densely}. This conclusion is supported by simulations of larger systems with $N=\num[group-separator={,},group-minimum-digits={3}]{98304}$, for which no stable smectic phase could be observed for $\eta < 0.45$ (data not shown).
\par
Further increasing the density beyond $\eta \simeq 0.42$ eventually induces a buckling of the smectic layers, which gives rise to a striking double-helical structure (Fig.~\ref{fig7}c). By analogy with Eq.~\eqref{eq:smectic}, a helicoidal smectic OP may be derived based on the Fourier transform $\widetilde{\mathbf{r}}$ of the center-of-mass distribution $\mathbf{r}_m$ relative to the smectic axis $\mathbf{e}_3^{\rm surf}$,
\begin{equation*}
  \widetilde{\mathbf{r}}(s) = \sum_{m=1}^{N_c} \mathbf{r}_m \exp\bigg(-\frac{2\mathrm{i}\pi}{s} \mathbf{r}_m \cdot \mathbf{e}_3^{\rm surf}\bigg).
\end{equation*}
Using the shorthand $\widetilde{r}_i \equiv \widetilde{\mathbf{r}} \cdot \mathbf{e}_i^{\rm surf}$, the cross-correlation function $\widetilde{c}_{12}$ of the transverse components $r_{1,2}$ reads as
\begin{equation*}
  \widetilde{c}_{12}(s) = \widetilde{r}_1(s) \times \widetilde{r}_2^{\,*}(s),
\end{equation*}
where $\widetilde{r}_2^{\,*}$ is the complex conjugate of $\widetilde{r}_2$, from which the helicity parameter $\mathpzc{H}$ may be defined~\cite{tortora2020chiral},
\begin{equation*}
  \mathpzc{H}(s) = \frac{2 \times\Im\big\{\widetilde{c}_{12}(s) \big\}}{\widetilde{c}_{11}(s)+\widetilde{c}_{22}(s)},
\end{equation*}
with $\Im\big\{\widetilde{c}_{12}\big\}$ the imaginary part of $\widetilde{c}_{12}$. One may verify that $\mathpzc{H}(s)\in [-1,1]$ --- with $\mathpzc{H}(s) = \pm 1$ if and only if the chain centers of mass adopt an ideal helical configuration of pitch $s$ about $\mathbf{e}_3^{\rm surf}$, whose handedness is determined by the sign of $\mathpzc{H}$~\cite{tortora2020chiral}. The degree of spiral smectic order may thus be quantified through
\begin{equation}
  \label{eq:helical_smectic}
  S_{\rm sm}^{\rm hel} = S_{\rm sm}^{\rm A}  \times \big|\mathpzc{H}\big(s_{\rm max}\big)\big|,
\end{equation}
which is accordingly found to jump to non-vanishing values $S_{\rm sm}^{\rm hel} \simeq 0.08$ for $\eta \gtrsim 0.42$ (Fig.~\ref{fig7}e, inset).
\par
This ``spiral'' configuration has also been predicted by SCFT in the lamellar phase of block copolymers confined to the surface of a sphere, which bears a strong analogy to the smectic A state~\cite{chantawansri2007self}. In this case, the crossover between uniaxial and spiral smectic arrangements was attributed to a Helfrich-Hurault transition, whereby the elastic costs of lamellar compression may be offset by bending distortions of the layers when the sphere circumference significantly differs from an integer number of lamellar periods~\cite{chantawansri2007self}. This interpretation is fully consistent with the previous discussion, and suggests that a similar instability could serve to promote a spiral smectic texture in membrane-confined systems as the shell diameter deviates away from a value compatible with a regular stacking of layers.

\subsection{Tetrahedral \& sphericonical morphologies} \label{subsec:tetrahedral}

\begin{figure*}[htpb]
  \includegraphics[width=2\columnwidth]{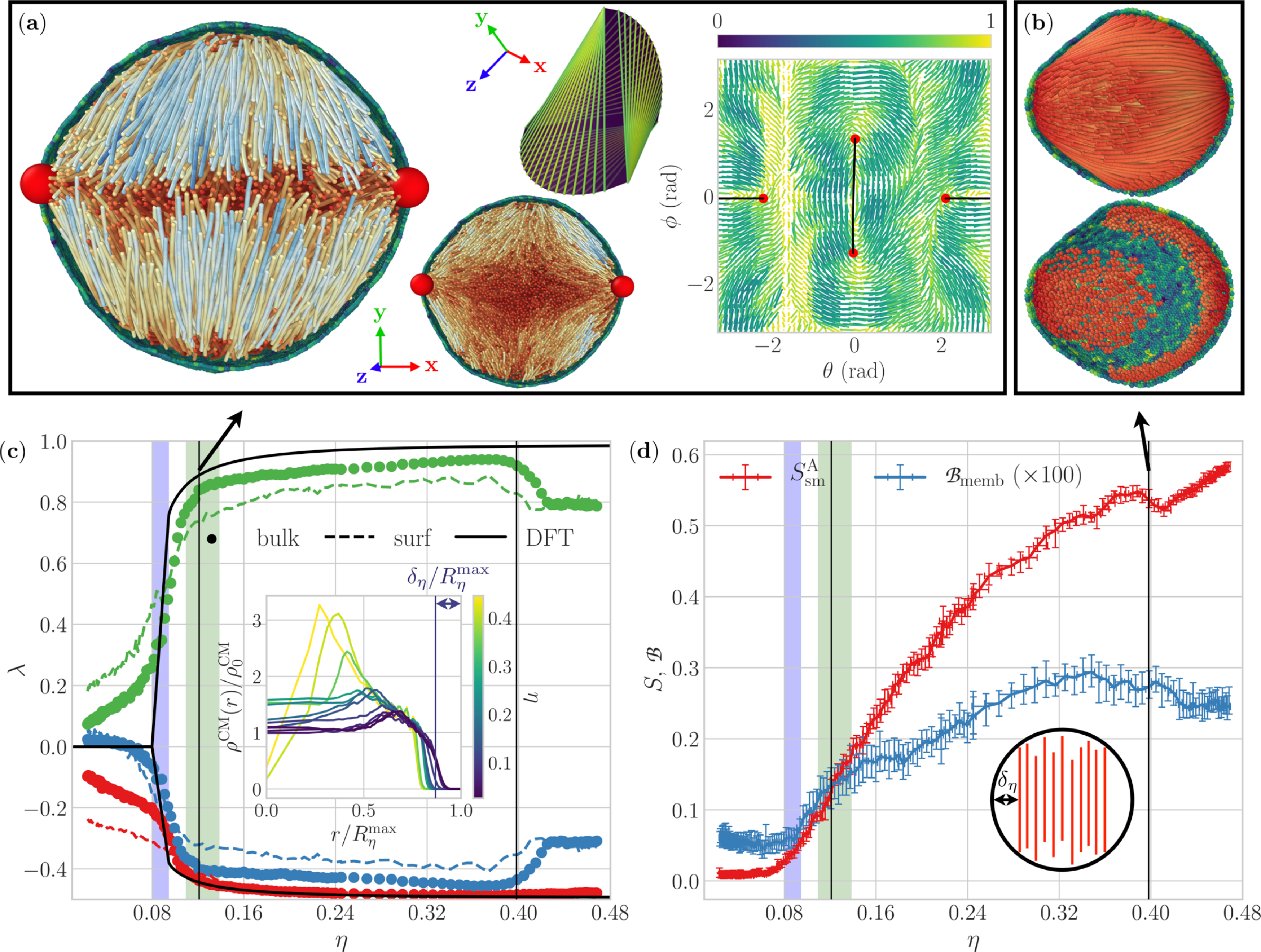}
  \caption{\label{fig8}Same as Fig.~\ref{fig7} for $l_c = 32\,\sigma$. (a) The faceted ``tennis ball'' (TB) state. Left: Peeled view of the simulated structure, revealing the accumulation of chain ends in the two symmetry planes of the TB surface texture ($x$-$z$ and $y$-$z$, c.f.~Fig.~\ref{fig5}a), associated with $s=1/2$ bulk disclination lines. Chains are colored according to the projection of bond orientations onto $\mathbf{z}\equiv \mathbf{e}_3^{\rm surf}$, and red spheres mark the computed locations of surface topological defects. Bottom center: Cut view of the same through a plane containing the two highlighted topological defects. Top center: Schematic representation of the membrane conformation and corresponding surface director field. Right: Stereographic map of the same, using the conventions of Fig.~\ref{fig1}a. Black lines represent the two $s=1/2$ disclination lines. Note the regular tetrahedral arrangement of surface defects. (b) The ``sphericonical'' state, displaying the full polymers (left) and the chain ends (right). Note that all surface defects now lie in the same plane. (c) Same as Fig.~\ref{fig7}d. Inset: Radial variations of the normalized center-of-mass density ($\rho^{\rm CM}/\rho^{\rm CM}_0$) of the filaments. Notations are as in Fig.~\ref{fig7}d. (d) Smectic OP ($S_{\rm sm}^{\rm A}$, Eq.~\eqref{eq:smectic}) and membrane asphericity ($\mathpzc{B}_{\rm memb}$, Eq.~\eqref{eq:asphericity}) as a function of $\eta$. Inset: Illustration of the lateral depletion effect in the nematic phase of chains with $l_p \gg l_c \simeq R_\eta$. This phenomenon is largely irrelevant for isotropic systems, in which the excluded region $\delta_\eta$ may be penetrated by the extremal sections of chains with different orientations, implying that the faceting instability requires the presence of underlying nematic order.}
\end{figure*}

For longer chains with $l_c =32\,\sigma$ ($\sim \SI{0.8}{\micro\meter}$), we observe a direct crossover from the isotropic state to a bulk-ordered nematic arrangement near the DFT-predicted binodal region $\eta \in[0.08, 0.10]$ (Fig.~\ref{fig8}c). Remarkably, although this transition does similarly not involve any appreciable orientational wetting of the membrane, we find that the resulting structure displays a quadrupolar surface pattern strongly reminiscent of the TB texture (Fig.~\ref{fig5}a). More accurately, this assembly may be identified as a TB arrangement bearing a strong accumulation of chain ends along the symmetry planes of the two nematic domains, associated with a significant faceting of the 4 corresponding half-domains (Fig.~\ref{fig8}a). The ensuing positional order is also captured by the smectic OP $S_{\rm sm}^{\rm A}$ (Fig.~\ref{fig8}d), and induces a morphological transition of the shell from a spherical to a rounded tetrahedral shape, which may be quantified through the asphericity parameter~\cite{xing2012morphology}
\begin{equation}
  \label{eq:asphericity}
  \mathpzc{B}_{\rm memb} = \frac{\big\langle \lVert \mathbf{r}_i\rVert ^2 \big\rangle_v-\big\langle \lVert \mathbf{r}_i\rVert  \big\rangle_v^2}{\big\langle \lVert \mathbf{r}_i \rVert  \big\rangle_v^2},
\end{equation}
with $\big\langle\cdot\big\rangle_v$ an ensemble average over all membrane vertices $i\in[1,N_v]$ (Fig.~\ref{fig8}d).
\par
We attribute this faceting to the increasingly-large volume inaccessible to the filaments due to the presence of the spherical membrane. Indeed, elementary geometry shows that the center of mass of rigid chains with $l_p\gg l_c$ may not penetrate a layer of thickness $\delta_\eta = R_\eta - \sqrt{R_\eta^2-l^2_c/4}$ in the vicinity of the curved confining walls (Fig.~\ref{fig8}d), which is confirmed by inspection of the radial center-of-mass density $\rho^{\rm CM}$ (Fig.~\ref{fig8}c, inset). In the nematic regime, this effect leads to a growing depletion of polymers near the shell's equator as $l_c/R_\eta$ reaches order unity ($l_c/R_\eta \simeq 1$ at the nematic binodal concentration $\eta \simeq 0.10$ for $l_c=32\,\sigma$ and $N=\num[group-separator={,},group-minimum-digits={3}]{32768}$), which may be offset by a reduction in local membrane curvature through the formation of flat polyhedral facets.
\par
Since a non-degenerate polyhedron must span at least 4 distinct vertices, the minimal candidate shape is necessarily a tetrahedron, in which the Gaussian curvature is concentrated at each vertex --- and which therefore provide ideal locations for the surface topological defects of the underlying LC structure~\cite{xing2012morphology}. However, due to the presence of nematic order in the bulk, these vertices are associated with $s=1/2$ disclination lines joining pairs of defects along the edges of the tetrahedron (Fig.~\ref{fig8}a). This effect, combined with the energy penalty of the sharp membrane bends localized at the edges, leads to the stabilization of a rounded regular tetrahedral conformation in the ground state --- which minimizes the total edge length at fixed surface area~\cite{xing2012morphology}. Hence, despite its similarity with the TB arrangement, this faceted TB configuration has no equivalent in rigid spherical confinement, and may not be accounted for by continuum (OF) descriptions of LC elasticities, which are agnostic to the additional inhomogeneities imposed by depletion effects for $l_p \gg l_c \simeq R_\eta$.

\begin{figure*}[htpb]
  \includegraphics[width=2\columnwidth]{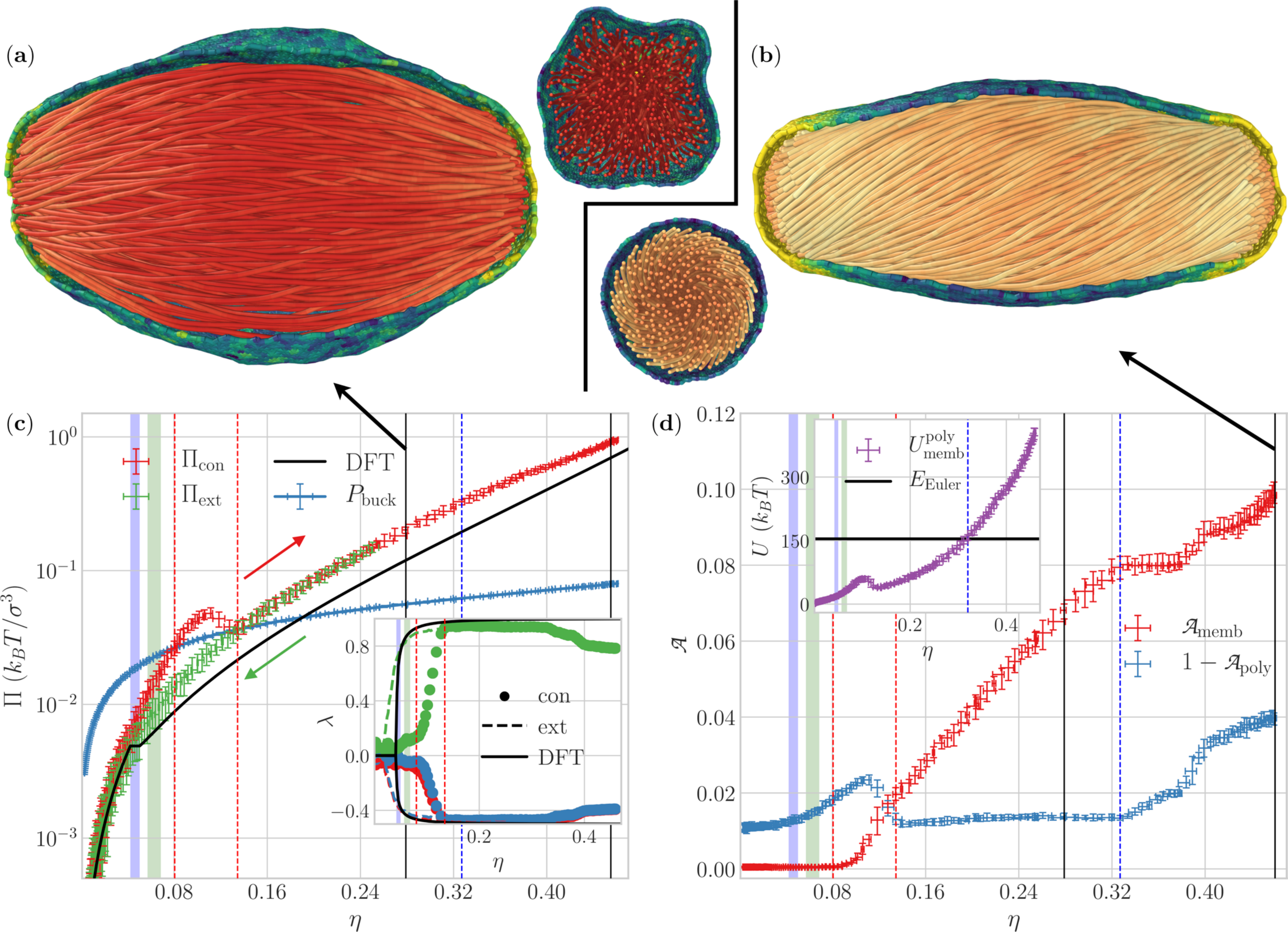}
  \caption{\label{fig9}Same as Fig.~\ref{fig7} for $l_c = 64\,\sigma$. (a) Bundle and (b) twisted-bundle states in longitudinal (peeled) and transverse (cut) views. Chains are colored according to the projection of bond orientations onto the system-averaged director $\mathbf{e}_3$ (Eq.~\eqref{eq:deGennes}), and membrane vertices are colored according to local strain. (c) Classical buckling pressure ($P_{\rm buck}$, Eq.~\eqref{eq:p_buck}) and osmotic pressures obtained via membrane contraction ($\Pi_{\rm con}$) and extension ($\Pi_{\rm ext}$) routes. Inset: Nematic OPs ($\lambda_i$) for the same. Note that due to the strong anisotropy of the shell, the $\lambda_i$ are now obtained by simple averaging of Eq.~\eqref{eq:deGennes} over all polymer bonds, without spatial discretization. (d) Anisotropy parameter ($\mathpzc{A}$, Eq.~\eqref{eq:aniso}) of the membrane and individual chains. $\mathpzc{A}_{\rm memb}>0$ characterizes anisotropic (oblate or prolate) shell conformations, while $\mathpzc{A}_{\rm poly}<1$ indicates deviations of the filaments from the linear ground state. Inset: Polymer-membrane interaction energy ($U^{\rm poly}_{\rm memb}$, Eq.~\eqref{eq:coupling}) and Euler critical buckling load ($E_{\rm Euler}$, Eq.~\eqref{eq:euler}).}
\end{figure*}

Let us denote by $R_\eta^{\rm max}$ the maximum vertex distance from the membrane center of mass, such that $R_\eta^{\rm max} \simeq R_\eta$ in the case of a spherical conformation. Further increases of the density induced by the contraction of the shell lead to a reduction in the length of the nematic half-domains, corresponding to the approximate height of each triangular facet, which eventually approaches the dimension $l_c$ of individual chains around $R_\eta^{\rm max} \simeq l_c/\sqrt{2}$ at $\eta\simeq 0.40$. Beyond this point, the chains in the bulk adopt an increasingly layered arrangement, evidenced by the increasing peak in $\rho^{\rm CM}$ around $r\simeq 0.4\, R_\eta^{\rm max}$ (Fig.~\ref{fig8}c, inset), while the 4 defects gradually migrate towards a common plane to allow for the dense packing of a single-filament layer at the surface (Fig.~\ref{fig8}b) --- which finally yields a geometrical shape sometimes referred to as the \textit{sphericon}~\cite{hirsch2020polycons}.


\section{Buckled, sickled \& toroidal states} \label{sec:Buckled, sickled and toroidal states}

It follows from the discussions of Sec.~\ref{subsec:tetrahedral} that surface faceting requires the typical membrane radius $R_\eta$ to be such that $1/\sqrt{2} \lesssim R_\eta/l_c$ ($\lesssim 1$) in the stable nematic regime, so that each of the 4 minimal facets may accommodate at least one monolayer of fully-extended polymers. For longer chains with $l_c=64\,\sigma$, this inequality is violated at the nematic binodal density ($\eta \simeq 0.05$) for $N=\num[group-separator={,},group-minimum-digits={3}]{32768}$, implying that the excluded region $\delta_\eta$ associated with nematic order may no longer be mitigated by purely-local shell deformations. In this case, we find that the onset of the I-N transition is pushed to significantly higher concentrations ($\eta \simeq 0.08$), and is associated with an abrupt collapse of the membrane into a strongly-elongated, multi-lobed structure bearing long-wavelength lateral buckles (Fig.~\ref{fig9}a). This transition may be captured by the eigenvalues $\Gamma_i$ of the membrane gyration tensor,
\begin{equation*}
  \mathcal{G}_{\rm memb}^{\alpha\beta} = \Big\langle r_i^\alpha r_i^\beta\Big\rangle_v,
\end{equation*}
which yield the dimensionless anisotropy parameter~\cite{theodorou1985shape},
\begin{equation}
  \label{eq:aniso}
  \mathpzc{A}_{\rm memb} = \frac{3}{2}\frac{\Gamma_1^4+\Gamma_2^4+\Gamma_3^4}{\big(\Gamma_1^2+\Gamma_2^2+\Gamma_3^2\big)^2}-\frac{1}{2}.
\end{equation}
One readily checks that $0\leq \mathpzc{A}_{\rm memb}\leq1$, with $\mathpzc{A}_{\rm memb}=1$ indicating an extended linear membrane conformation, and $\mathpzc{A}_{\rm memb}=0$ if the vertex distribution is identical along the 3 corresponding eigenvectors. 
\par
The drastic spontaneous breaking of spherical symmetry around $\eta \simeq 0.08$ (Fig.~\ref{fig9}d) is somewhat reminiscent of the classical buckling instability observed in spherical elastic shells subject to an inward external pressure $P_{\rm ext}$, which undergo a sudden collapse when the compressive load exceeds a critical value $P_{\rm buck}<P_{\rm ext}$~\cite{hutchinson2016buckling},
\begin{equation}
  \label{eq:p_buck}
  P_{\rm buck} = \frac{4\sqrt{\kappa_0Y_0}}{R_\eta^2}.
\end{equation}
Interestingly, we find that the internal osmotic pressure $\Pi$ precisely reaches the buckling threshold predicted by Eq.~\eqref{eq:p_buck} for $\eta = 0.08$ (Fig.~\ref{fig9}c). Hence, we attribute the transition of the membrane into a collapsed, prolate state to the anisotropic stress induced by the chain exclusion effect (Fig.~\ref{fig8}d), which yields an effective depletion pressure of magnitude $\sim \Pi$ acting on the lateral sections of the shell to reduce the inaccessible region $\delta_\eta$. 
\par
In this context, the delayed onset of nematic order may be imputed to the fact that this depletion volume vanishes for isotropic systems (Fig.~\ref{fig8}d), and therefore imposes an additional entropic cost for nematic organization within the spherical membrane. This penalty may be reduced by the large anisotropy of the post-buckling membrane conformation, which is compatible with denser arrangements of encapsulated chains (Fig.~\ref{fig9}a). In this case, the stability of the nematic phase is thus contingent on the inequality $\Pi > P_{\rm buck}$, to allow for the shell to spontaneously snap into the buckled state post the I-N transition. The resulting interplay between membrane buckling and LC assembly gives rise to a significant hysteresis in the vicinity of the transition region, which may be evidenced by allowing the system to cycle back to the isotropic regime by subsequent expansion of the shell. Through this route, it is found that the buckled configuration remains metastable down to the bulk binodal density $\eta \simeq 0.05$, at which point the underlying nematic order gradually disappears and the membrane eventually recovers its spherical reference shape (Fig.~\ref{fig9}c).
\par
The axial dimension of the collapsed shell approaches the chain contour length $l_c$ at densities $\eta \simeq 0.27$, beyond which the further contraction of the membrane leads to the disappearance of the lateral lobes to achieve a more compact, convex structure bearing a bundle-like arrangement of fibers (Fig.~\ref{fig9}a-b). A second transition is then found to take place around $\eta \simeq 0.32$, characterized by the onset of buckling of the polymers due to the increasing compressive strain exerted by the shell. This transition may be similarly quantified through the anisotropy parameter $\mathpzc{A}_{\rm poly}$ of the chains (Fig.~\ref{fig9}d), and is shown to occur when the membrane-polymer coupling energy (Eq.~\eqref{eq:coupling}) approaches the theoretical threshold known as the Euler critical load~\cite{landau1986theory},
\begin{equation}
  \label{eq:euler}
  E_{\rm Euler} = \frac{\pi^2 l_p k_B T}{l_c},
\end{equation}
which corresponds to the maximum stress that simply-supported filaments may sustain without undergoing lateral deflections (Fig.~\ref{fig9}d, inset). In the post-buckling state ($\eta \gtrsim 0.32$), peripheral chains are found to adopt increasingly-curved, C-like conformations resembling the Euler fundamental mode~\cite{landau1986theory}, which wrap around the axis of the bundle to yield a spontaneous twisted structure (Fig.~\ref{fig9}b).
\par
The critical load in Eq.~\eqref{eq:euler} vanishes as $1/l_c$, which leads to an increasingly-extensive range of post-buckling behavior for longer chains with $l_c\geq 128\,\sigma$. This effect is further compounded by the softening of the buckling transition due to thermal fluctuations. Indeed, at finite temperature, the onset of buckling is generally expected to be shifted to a lower compressive load $E_{\rm resc}<E_{\rm Euler}$ than the classical, ``zero-temperature'' limit (Eq.~\eqref{eq:euler}), which approximately scales as~\cite{odijk1998microfibrillar,emanuel2007buckling}
\begin{equation}
  \label{eq:euler_resc}
  E_{\rm resc} \simeq E_{\rm Euler}\bigg(1-\frac{l_c}{l_p}\bigg).
\end{equation} 
Hence, Eq.~\eqref{eq:euler_resc} gives rise to increasing deviations from Eq.~\eqref{eq:euler} as $l_c$ reaches comparable magnitudes to $l_p$. Analogous considerations have also been suggested to affect the buckling of large spherical elastic shells, provided that to following inequality is satisfied~\cite{paulose2012fluctuating,kosmrlj2017statistical},
\begin{equation*}
  \frac{k_B T}{\kappa_0} \sqrt{\gamma_{\rm FvK}} \gg 1,
\end{equation*}
where $\gamma_{\rm FvK} \equiv Y_0 R_\eta^2/\kappa_0$ is the dimensionless F\"oppl-von K\'arm\'an number. In this case, thermal fluctuations may be shown to lead to an effective renormalization of the applied pressure $\Pi$~\cite{paulose2012fluctuating},
\begin{equation}
  \label{eq:p_resc}
  \Pi_{\rm resc} \simeq \Pi + \frac{1}{24\pi}\frac{k_BT}{\kappa_0}\sqrt{\gamma_{\rm FvK}}\bigg(P_{\rm buck}+\frac{63\pi}{128}\Pi\bigg),
\end{equation}
which may similarly induce buckling at bare osmotic pressures $\Pi < \Pi_{\rm resc}=P_{\rm buck}$ well below the expected classical value (Eq.~\eqref{eq:p_buck}) for sufficiently-large membrane radii $R_\eta$ at fixed $Y_0$ and $\kappa_0$~\cite{paulose2012fluctuating,kosmrlj2017statistical}. This effect is of limited relevance for chains with $l_c \leq 64\,\sigma$, for which $k_B T \sqrt{\gamma_{\rm FvK}}/\kappa_0\lesssim 3$ in the stable nematic regime for the system sizes considered here --- but becomes consequential for longer chains, which may display stable orientational order at increasingly-low densities (i.e., increasingly-large $R_\eta$) for a fixed total number of monomers $N$, as described in the next paragraphs.
\par
\begin{figure*}[htpb]
  \includegraphics[width=2\columnwidth]{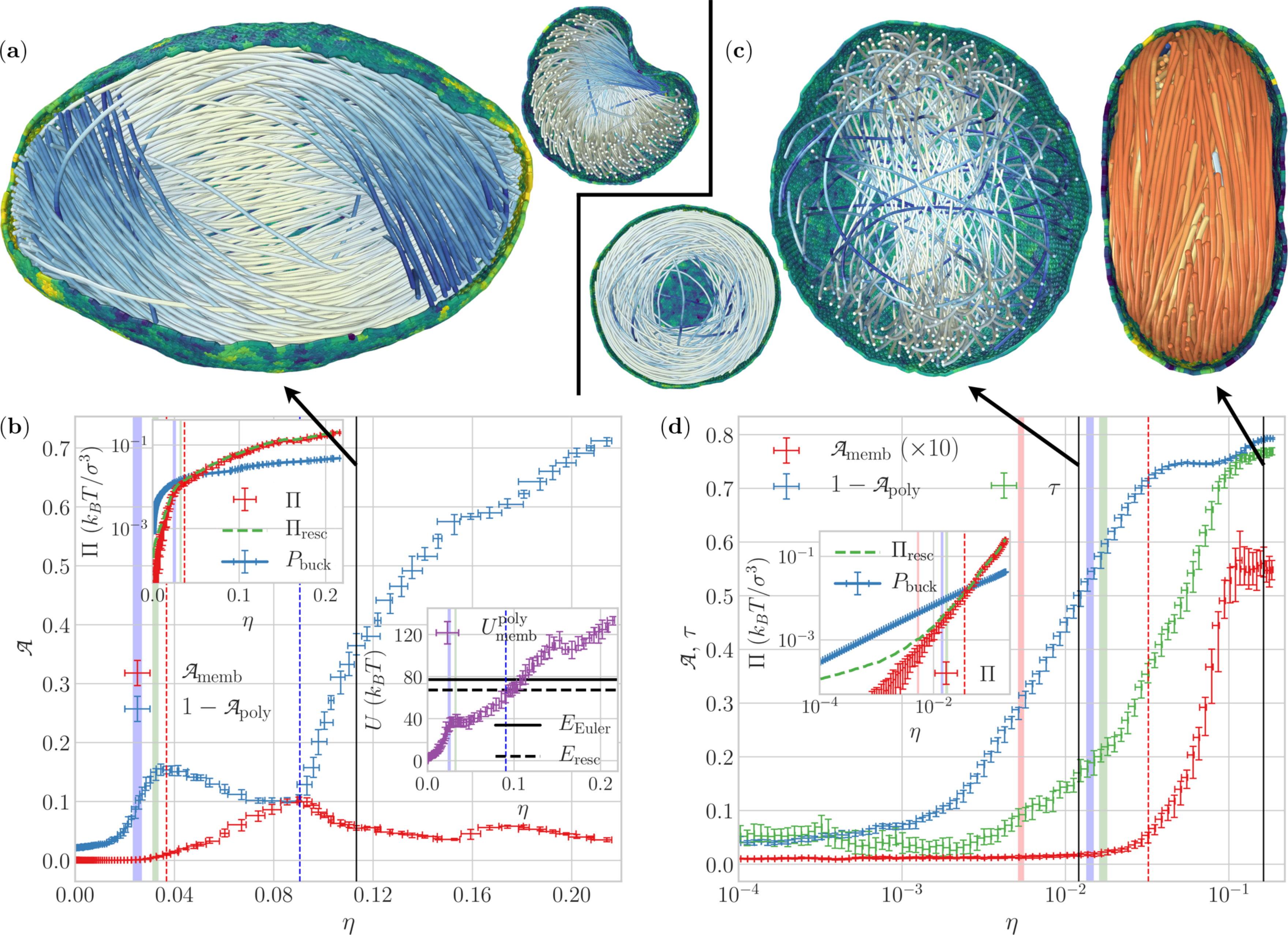}
  \caption{\label{fig10}Same as Fig.~\ref{fig7} for $l_c = 128\,\sigma$ (a-b) and $256\,\sigma$ (c-d). (a) The ``sickled'' state. As in Fig.~\ref{fig9}b, each filament is buckled into a C-like circular segment, whose height now exceeds the typical diameter $d\propto \sqrt{N_c}\sigma$ of the bundle. This stronger curvature precludes a simple axisymmetric arrangement, and leads to a crescent-like post-buckling morphology similar to that of sickled red blood cells. Colors are as in Fig.~\ref{fig9}. (b) Same as Fig.~\ref{fig9}c-d, with $E_{\rm resc}$ and $\Pi_{\rm resc}$ the respective rescaled filament buckling load (Eq.~\eqref{eq:euler_resc}) and osmotic pressure (Eq.~\eqref{eq:p_resc}) due to thermal fluctuations. (c) The ``toroidal'' state. Chains are colored according to their projection in the plane normal to the approximate symmetry axis $\mathbf{z}\equiv \mathbf{e}_1$ (Fig.~\ref{fig5}d). (d) Toroidal OP ($\tau$, Eq.~\eqref{eq:tau}) as a function of $\eta$. The red highlighted region delimits the I-N coexistence range predicted by DFT, rescaled to account for the non-uniform distribution of chains within the shell (Eqs.~\eqref{eq:odijk_bin}-\eqref{eq:eta_eff}). Other notations and symbols are as in (b).}
\end{figure*}

For $l_c=128\,\sigma$ ($\sim \SI{3.2}{\micro\meter}$), we also observe that the onset of nematic order --- and the associated collapse of the membrane --- occurs at densities $\eta\simeq 0.04$ near the shell buckling threshold ($\Pi_{\rm resc} \simeq P_{\rm buck}$, Fig.~\ref{fig10}b), followed by the buckling of the encapsulated filaments around $\eta \simeq 0.09$ ($U_{\rm memb}^{\rm poly} \simeq E_{\rm resc}$). However, the post-buckling state is in this case strongly non-axisymmetric, and exhibits a concave, crescent-like structure (Fig.~\ref{fig10}a). This morphology interestingly resembles the pathological shape of sickled red blood cells~\cite{lu2019quantitative}, which may be attributed to the abnormal growth of hemoglobin chains within the cells, whose typical diameter $\sigma \simeq \SI{21}{\nano\meter}$ and persistence length $l_p\simeq \SI{240}{\micro\meter}$~\cite{wang2002micromechanics} are both comparable in magnitude to our filaments --- as further discussed in Sec.~\ref{sec:Discussion}.

Finally, for longer chains with $l_c=256\,\sigma$, the buckling of the filaments is no longer associated with a sharp transition, but is rather found to occur gradually across a wide range of concentrations $\eta \in [10^{-3},10^{-1}]$, broadly corresponding to the confinement regime $2R_\eta \lesssim l_c$, as evidenced by the slow saturation in the anisotropy parameter $1-\mathpzc{A}_{\rm poly}$ (Fig.~\ref{fig10}d). This behavior is consistent with theoretical predictions for stiff polymers of increasing lengths at finite temperature~\cite{emanuel2007buckling}, for which thermal fluctuations may lead to significant deviations from a straight conformation ($\mathpzc{A}_{\rm poly}=1$) even in the absence of axial compressive forces. 
\par
In this case, the buckled chains undergo a transition to a partially-ordered state at densities $\eta \simeq 5\times 10^{-3}$ (Fig.~\ref{fig10}c, left \& center), far below the I-N binodal region predicted by DFT in the bulk for $l_p\gg l_c \gg \sigma$~\cite{odijk1986theory}, 
\begin{equation}
  \begin{gathered}
    \label{eq:odijk_bin}
    \eta_{\rm iso}^{\rm bin} = \frac{\sigma}{l_p} \bigg(3.340 \times \frac{l_p}{l_c} + 4.990\bigg), \\
    \eta_{\rm nem}^{\rm bin} = \frac{\sigma}{l_p} \bigg(4.486 \times \frac{l_p}{l_c} - 1.458\bigg),
  \end{gathered}
\end{equation}
which is delimited by the green shaded area in Fig.~\ref{fig10}d. This discrepancy is due to the fact that the distribution of filaments in this regime is now strongly non-uniform, with most monomers being concentrated within a spherical shell of thickness $r_{\rm shell} \propto R^2_\eta /l_p$ near the inner surface of the membrane~\cite{gao2014free} (Fig.~\ref{fig10}c, center). The effective chain density within this peripheral layer hence reads as
\begin{equation}
  \label{eq:eta_eff}
  \eta_{\rm eff} \simeq \frac{N_c v_c}{4\pi R^2_\eta r_{\rm shell}} \propto \frac{N l_p \sigma^3}{16R_\eta^4} \simeq \frac{l_p}{\sigma N^{1/3}}\bigg(\frac{2\eta}{3}\bigg)^{4/3}.
\end{equation}
Equating Eqs.~\eqref{eq:odijk_bin} and~\eqref{eq:eta_eff}, the predicted transition region is found to be shifted to lower overall concentrations $\eta \simeq 5\times 10^{-3}$ by this local accumulation effect, in good agreement with the simulation results (Fig.~\ref{fig10}d).
\par
The alignment of buckled filaments within the thin surface shell leads to a toroidal structure reminiscent of the latitudinal nematic arrangement of long DNA-like chains (Fig.~\ref{fig5}d), associated with a strong depletion of monomers along the polar axis (Fig.~\ref{fig10}c, left). By analogy with Eq.~\eqref{eq:alpha}, a toroidal (oblate) OP may be introduced based on the eigenvalues $\lambda_i$ of the system-averaged $\mathcal{Q}$-tensor,
\begin{equation}
\label{eq:tau}
  \tau \equiv \frac{4\big(\lambda_2-\lambda_1)}{3}.
    \vspace{0mm}
\end{equation}
One readily checks that $\tau=0$ in the absence of orientational organization, while in the limit of perfect toroidal order, the chain bond vectors $\mathbf{t}_k$ must lie entirely in the plane normal to the cylindrical symmetry axis $\mathbf{z}\equiv\mathbf{e}_1$ --- thus yielding $\tau = 1$ ($\lambda_1=-1/2$, $\lambda_2=1/4$). Accordingly, we report in Fig.~\ref{fig10}d that $\tau$ increases from $\sim0.05$ to $0.75$ over a broad density range $\eta \in [ 5\times 10^{-3},10^{-1}]$, indicating that the toroidal transition predicted by Eqs.~\eqref{eq:odijk_bin}-\eqref{eq:eta_eff} is in this case likely only weakly of the first order.
\par
This structure is akin to the peripheral ring of microtubules known as the \textit{marginal band}, typically observed in non-mammalian erythrocytes and thrombocytes~\cite{lee2004shape} as well as mammalian erythroblasts and platelets~\cite{patel2008vizualization}. Interestingly, it is similarly found to induce the buckling of the membrane at a critical density $\eta \simeq 3\times 10^{-2}$ such that $\Pi_{\rm resc} \simeq P_{\rm buck}$ (Fig.~\ref{fig10}d, inset), which now results in an oblate, disk-like post-buckling morphology resembling that of a healthy red blood cells or unactivated platelet~\cite{dmitrieff2017balance} (Fig.~\ref{fig10}c, right). This observation is consistent with the reported role of the marginal band in stabilizing the equilibrium shape of the cells~\cite{dmitrieff2017balance}, and further highlights the potential biological relevance of such buckling transitions as a general and versatile morphogenetic mechanism.


\section{Discussion \& Conclusion} \label{sec:Discussion}

We have extensively explored the self-assembling behavior of semi-flexible, self-avoiding chains encapsulated within elastic biopolymeric shells. We report that the ordering transitions of dsDNA-like filaments involve an orientational wetting phenomenon at densities close to the bulk I-N coexistence region, provided that their persistence length $l_p\simeq \SI{50}{\nano\meter}$ is smaller or comparable in magnitude to the mean radius $R\equiv (3V/4\pi)^{1/3}$ of the confining membrane. This effect is found to extend across an increasingly-wide concentration range for polymers with larger contour lengths $l_c$ (Fig.~\ref{fig2}d-f), which we impute to the increasingly-weak first-order character of the I-N transition reported in bulk phases of chains with lower effective rigidities $l_p/l_c$~\cite{debraaf2017self}.
\par
The nematic surface layer is found to be associated with tetrahedral patterns of $s=1/2$ topological defects, which evolve towards ``escaped'' 3D arrangements bearing 2 antipodal $s=1$ defects at higher densities, as the orientational order gradually extends to the entire cavity. This observation is in agreement with the theoretical predictions of Vitelli and Nelson~\cite{vitelli2006nematic} for spherical nematic shells of increasing thickness in the one-constant approximation of nematic elasticities, which we show by means of density functional theory to be consistent with the limited degree of local alignment within the wetting layer for the various chains studied (Fig.~\ref{fig3}a-c). More precisely, we report that the surface director field concomitant with the four $s=1/2$ defects adopts a splay-rich cricket-ball texture for short filaments with $l_c \lesssim 32 \,\sigma$ ($\sim 200$ base pairs), associated with a weak elastic anisotropy $K_{11}^{\rm eff} \lesssim K_{33}$ --- while longer chains with $l_c = 64 \,\sigma$ display a bend-rich tennis-ball configuration, which we attribute to a relative stiffening of the splay elasticity $K_{11}^{\rm eff} \gtrsim K_{33}$ with increasing polymer contour lengths (Fig.~\ref{fig6}b).
\par
This rigidification effect would eventually lead to the stabilization of a splay-free state in the wetting regime of very long chains, for which one expects $K_{11}^{\rm eff} \gg K_{33}$ throughout the nematic stability range~\cite{meyer1982macroscopic}. A suitable candidate would be the latitudinal bipolar surface texture depicted in Fig.~\ref{fig5}d, which features a pure bend distortion localized at the two $s=1$ hyperbolic defects~\cite{fernandez2007topological}. Such a configuration --- characterized by the presence of a disordered, isotropic core region surrounded by a spool-like arrangement of filaments near the membrane --- has indeed been recently predicted by SCFT for long WLCs such that $l_p \leq R$~\cite{liang2019orientationally}. This assembly was found to be stable at densities just above the bulk I-N transition point, and could therefore be similarly interpreted as an orientational wetting phenomenon --- while its absence from the phase diagram of chains with $l_p > R$~\cite{liang2019orientationally} may reflect the inhibition of wetting observed in our simulations in the case of stiffer filaments, as discussed further below.
\par
At higher concentrations, this splay stiffening process analogously induces a transition from a splay- (longitudinal) to a bend-dominated (latitudinal) bipolar surface texture in the bulk-ordered state as one increases $l_c$ (Fig.~\ref{fig6}b). This crossover gives rise to an growing degree of radial twist through the bulk of the phase, which may be quantitatively captured by the classical theory of Williams~\cite{williams1986two} for tangentially-anchored spherical nematic droplets (Eq.~\eqref{eq:williams}). The twisted, spool-like arrangement thus obtained for $l_c=64\,\sigma$ (Fig.~\ref{fig5}d) is consistent with the folded genome structures reported in mature spherical and icosahedral bacteriophages~\cite{petrov2008packaging}, and mirrors the so-called \textit{condensed Hopf fibration} predicted as the general ground state in the high-density regime of spherically-confined long chains~\cite{liang2019orientationally}. These results would suggest the Williams theory as a simple framework to interpret the chiral assemblies observed in small viral capsids at various genome packing fractions~\cite{leforestier2010bacteriophage} --- and more broadly demonstrate a universal mechanism for the establishment of spontaneous twist in such systems based on the interplay between confinement and LC elasticities, rather than specific cholesteric interactions involving the detailed molecular structure of DNA~\cite{tortora2018incorporating}.
\par
Setting the elastic parameters of the polymerized shell to typical values for the lamina network of the nuclear envelope, we find that these LC assemblies are not associated with a discernable directional elongation of the confining membrane. In the case of simple longitudinal bipolar order, the spherical conformation of the shell may be attributed to the low splay-to-surface-tension ratio $\mathpzc{K} < 0.01$ (Eq.~\eqref{eq:kappa_tactoid}, see Supplemental Fig.~S5) for short DNA-like filaments with $l_c=16\,\sigma$. Since $\mathpzc{K} \propto V^{-1/3}$ at fixed $\gamma$~\footnote{Note that in the case of polymerized membranes, the effective surface tension $\gamma_{\rm eff}$ is related to the radial stress imposed by the encapsulated chains, and is generally scale-dependent at given internal pressure $\Pi$ (Eq.~\eqref{eq:laplace}). Hence, Eqs.~\eqref{eq:kappa_tactoid} and~\eqref{eq:laplace} would instead lead to an even faster decay $\mathpzc{K} \propto V^{-2/3}$ under fixed osmotic conditions.}, this inequality is expected to be satisfied \textit{a fortiori} for all membrane dimensions $R$ larger than the typical radius $R_\eta \simeq \SI{45}{\nano\meter}$ corresponding to the bulk nematic density range in our simulations. For longer chains, it was reported in Ref.~\cite{prinsen2004parity} that shape anisotropy is generally destabilized by director twist --- with achiral, elongated morphologies being favored below a critical volume $V_{\rm crit}$, which is a complex function of the OF moduli, interfacial tension and anchoring strength~\cite{prinsen2004parity}. Such structures are not observed here, which implies that $V_{\rm crit}$ likely lies below the system sizes investigated. These results, combined with the previous discussion of orientational wetting, suggest that the spherical ground-state arrangements in Fig.~\ref{fig6}b should be widely valid for DNA-like chains confined within envelope-like elastic shells, provided that the membrane radius satisfies $R \gtrsim l_p\simeq\SI{50}{\nano\meter}$.
\par
These observations further point to a simple physical basis for the reported role of chromatin in the regulation of the shape and rigidity of eukaryotic nuclei~\cite{stephens2019chromatin}, whose typical dimensions are such that $R > \SI{1}{\micro\meter}$. In this case, the combination of Eqs.~\eqref{eq:swelling} and~\eqref{eq:surf_tens} yields a direct analog of the Young-Laplace relation for polymerized shells,
\begin{equation}
  \label{eq:laplace}
  \gamma_{\rm eff} \simeq \frac{\Pi R}{2} \propto V^{1/3},
\end{equation}
where we used $\Delta A/A \simeq 2 \Delta R/R$ in the limit of small deformations. Hence, for such large systems, a rise in the intra-nuclear osmotic pressure $\Pi$, as imposed (e.g.) by a local increase in chromatin stiffness due to histone modifications~\cite{stephens2018chromatin}, may lead to a significant upturn in the effective surface tension $\gamma_{\rm eff}$ of the membrane --- and thus help regulate nuclear morphology by inhibiting aberrant deviations from a healthy spherical shape, such as the pathological protrusions known as \textit{nuclear blebs}~\cite{stephens2019chromatin}.
\par
The relative independence of the phase diagram of DNA-like chains from the mean membrane radius $R$ at fixed packing fraction $\eta$ and contour length $l_c$ contrasts with the self-assembling behavior of short tubulin-like polymers ($l_p\simeq \SI{25}{\micro\meter}$), which generally depends on the full set of reduced parameters $(\eta, l_c/\sigma, R/l_c)$ at given $l_p/\sigma$~\cite{liang2019orientationally}. In weak confinement conditions ($R/l_c \gtrsim 1$), we find that orientational wetting is largely suppressed if $l_p/R \gg 1$, so that the system undergoes a transition from isotropic to full nematic order at densities close to the bulk binodal points $\big[\eta_{\rm iso}^{\rm bin},\eta_{\rm nem}^{\rm bin}\big]$ --- which are usually functions of both $l_c/\sigma$ and $l_p/\sigma$~\cite{egorov2016insight}, and asymptotically converge towards~Eq.~\eqref{eq:odijk_bin} in the limit $l_p/\sigma\gg l_c/\sigma \gg 1$. 
\par
For chains with $l_p \gg R\gtrsim l_c$, we report that the resulting nematic state for $\eta \gtrsim \eta_{\rm nem}^{\rm bin}$ bears a bipolar longitudinal symmetry, associated with a moderate elongation of the membrane along the polar axis --- as illustrated by the case $l_c=16\,\sigma \simeq\SI{0.4}{\micro\meter}$, for which $R_\eta \simeq \SI{0.65}{\micro\meter}$ (Fig.~\ref{fig7}a). The measured axial ratio of $\sim1.1$ is found to be consistent with the theory of Ref.~\cite{prinsen2003shape}, and is expected to decrease with increasing system sizes $R$ at fixed $l_c$ and $\eta$ based on the scaling behavior of $\mathpzc{K}$ discussed above. Larger, more isotropic shell conformations could similarly serve to stabilize twisted director patterns~\cite{prinsen2004parity}, although the considerably-stiffer bend modulus $K_{33}$ of rod-like filaments with $l_p \gg l_c$ (Figs.~\ref{fig3}d-f) --- relative to that of the more compliant, DNA-like chains --- is generally incompatible with the Williams criterion (Eq.~\eqref{eq:williams}) in the limit of strong alignment. Such structures could nonetheless be potentially observed in a limited density range close to the I-N transition, in which Eq.~\eqref{eq:williams} is found to hold (see Supplemental Fig.~S4). Finally, at higher concentrations, we predict the appearance of smectic phases, which may display either a smectic-A-like regular stacking of layers if $2R/l_c$ approaches an integer value (Fig.~\ref{fig7}b), or a spiral arrangement otherwise (Fig.~\ref{fig7}c).
\par
Conversely, the nematic phase of longer filaments with $l_p \gg l_c \gtrsim R$ is characterized by increasingly non-uniform density profiles within the shell, which precludes the use of simple continuum theories based on the assumption of homogeneous elastic moduli $K_{ii}$. For $l_c \gtrsim R \gtrsim l_c/\sqrt{2}$, the lateral depletion of polymers near the surface is found to induce a faceted tetrahedral membrane morphology, associated with a tennis-ball-like surface arrangement of chains bearing four $s=1/2$ topological defects localized at each vertex (Fig.~\ref{fig8}a). However, this local faceting is prohibited by geometry for $R \lesssim l_c/\sqrt{2}$ $(\ll l_p)$, in which case the underlying frustration must generally be resolved through global buckling transitions involving either the polymers or the confining membrane. 
\par
In this strong-confinement regime, the competition between filament and shell buckling leads to the introduction of a new lengthscale $L^*$, which may be phenomenologically derived as follows. Neglecting the effects of thermal fluctuations, the maximum total compressive force that can be sustained by the $N_c$ encapsulated chains in an extended conformation reads as (c.f.~Eq.~\eqref{eq:euler})
\begin{equation*}
F_{\rm poly}^{\rm buck} = N_c \frac{\pi^2l_p k_B T}{l_c^2},
\end{equation*}
while the maximum radial force that the membrane can exert while retaining its stable spherical shape may be crudely estimated from Eq.~\eqref{eq:p_buck},
\begin{equation*}
F_{\rm memb}^{\rm buck} \propto 4\pi R^2  P_{\rm buck} \propto \sqrt{\kappa_0Y_0}.
\end{equation*}
Balancing the two forces leads to the scaling relation $l_c \propto (N_cl_p)^{1/2} (\kappa_0Y_0)^{-1/4}$, and a more detailed analysis yields the corresponding numerical prefactor~\cite{cohen2003kinks},
\begin{equation}
  \label{eq:l_crit}
  l_c = \Bigg(\frac{\pi N_c l_p k_B T}{2\sqrt{2\kappa_0Y_0}}\Bigg)^{1/2} \equiv L^*.
\end{equation}
\par
Hence, for $F_{\rm poly}^{\rm buck} > F_{\rm memb}^{\rm buck}$ (i.e., $l_c < L^*$), the energetic cost of buckling the shell is lower than that of buckling the confined filaments --- so that the equilibrium structure of the system for $R \lesssim l_c/\sqrt{2}$ is expected to feature significant deformations of the membrane, but limited deviations of the polymers from an extended, linear state. This scenario is exemplified by the case $l_c=64\,\sigma$ ($N_c=512$, Fig.~\ref{fig9}a), for which $L^* \simeq 200\,\sigma$ in our simulations. Reciprocally, chains with $l_c > L^*$ display a higher relative compliance to buckling than the encapsulating shell ($F_{\rm poly}^{\rm buck} < F_{\rm memb}^{\rm buck}$), and may therefore bend smoothly to accommodate the spherical membrane --- as in the case $l_c=256\,\sigma$, for which $L^* \simeq 100\,\sigma$ ($N_c=128$, Fig.~\ref{fig10}c, center). In either case, the post-buckling structure may be subject to further buckling transitions with decreasing $R/l_c$, which give rise to prolate, twisted bundle-like filament arrangements for $l_c < L^*$ (Fig.~\ref{fig9}b), or to oblate, disk-like shell conformations for  $l_c > L^*$ (Fig.~\ref{fig10}c, right).
\par
Finally, for $l_c \simeq L^*$, we predict the emergence of intricate morphologies characterized by large, simultaneous deformations involving both the chains and the membrane (c.f.~Fig.~\ref{fig10}a for $l_c=128\,\sigma$ and $N_c=256$, for which $L^*\simeq150\,\sigma$). Interestingly, this regime should be particularly pertinent for sickled red blood cells, whose anisotropic shapes are generally stabilized by a number $N_c \gtrsim 15$ of intracellular hemoglobin fibers with persistence length $l_p \simeq \SI{240}{\micro\meter}$~\cite{wang2002micromechanics,lu2019quantitative} --- which leads to a critical length $L^* \gtrsim \SI{3}{\micro\meter}$ (Eq.~\eqref{eq:l_crit}), comparable in magnitude to both the typical contour length $l_c$ of the fibers and the mean radius $R$ of the erythrocyte membrane. This effect may thus provide a simple physical interpretation for the remarkable variety of structures observed in sickled red blood cells based on the detailed number and length distribution of the chains~\cite{lu2019quantitative}, and could be further compounded by the propensity of hemoglobin fibers to aggregate into cohesive bundles~\cite{wang2002micromechanics}, which is expected to yield a steeper dependence $L^* \propto N_c$ of $L^*$ on the number $N_c$ of encapsulated filaments~\cite{cohen2003kinks}.
\par
To summarize, for confined polymer solutions with fixed persistence length $l_p \lesssim R$, our results suggest orientational wetting as a powerful, density-dependent mechanism for the control of defect morphologies at given contour length $l_c/\sigma$ --- while in the opposite limit $l_p \gg R$, we predict a rich array of anisotropic assemblies with a complex dependence on density, $l_c/\sigma$ as well as the ratios $R/l_c$ and $l_c/L^*$. Besides their biological relevance, both regimes should be potentially accessible to \textit{in vitro} experiments, thanks to recent rapid progress in the combination of liposome and polymersome synthesis with various macromolecular encapsulation techniques~\cite{rideau2018liposomes}. On the theoretical front, further investigations would be required to explore the prospective roles of size polydispersity, and of specific polymer-polymer and polymer-membrane interactions beyond simple excluded volume. It is hoped that the current study will help motivate research efforts in these directions, to complement the vast body of work on low-molecular-weight LCs in microfluidic cavities~\cite{urbanski2017liquid}.

\begin{acknowledgments}
This work was supported by the Agence Nationale de la Recherche (ANR-18-CE12-0006-03, ANR-18-CE45-0022-01). We acknowledge PSMN (Pôle Scientifique de Modélisation Numérique) and Centre Blaise Pascal of the ENS de Lyon for computing resources.
\par
M.M.C.T. \& D.J. conceptualized the study. M.M.C.T designed the model, conducted the simulations, developed the theory, performed the analysis, interpreted the results and wrote the manuscript. D.J. acquired funding and commented on the final version of the manuscript.
\end{acknowledgments}


\bibliography{refs}


\end{document}



\title{Supplementary information: Morphogenesis and self-organization of persistent filaments confined within flexible biopolymeric shells}
\date{\today}
\author{Maxime M. C. Tortora}
\thanks{Correspondence to \href{mailto:maxime.tortora@ens-lyon.fr}{maxime.tortora@ens-lyon.fr}}
\author{Daniel Jost}
\affiliation{Université de Lyon, ENS de Lyon, Univ Claude Bernard, CNRS, Laboratoire de Biologie et Modélisation de la Cellule, Lyon, France}



\maketitle 


\section{Stretch response of bare membranes} \label{sec:force-stretch}

Let us consider the extension behavior of an empty polymerized membrane of reference radius $R$. We pick a random axis $\widehat{\mathbf{e}}_{\rm stretch}$, and identify two antipodal vertices based on their extremal projections onto $\widehat{\mathbf{e}}_{\rm stretch}$ in the reference elastic state. We measure the mean separation distance $L_0$ between the vertices in the absence of external forces by thermal averaging over $\mathcal{O}\big(10^8\big)$ MD steps. We then apply opposite outwards forces of increasing magnitude $F_{\rm stretch}$ along $\widehat{\mathbf{e}}_{\rm stretch}$ to each vertex through a ramp procedure, letting the system relax over $\mathcal{O}\big(10^7\big)$ MD steps after each force increment before similarly computing the mean inter-vertex distance $L$. 

\begin{figure}[htpb]
  \includegraphics[width=\columnwidth]{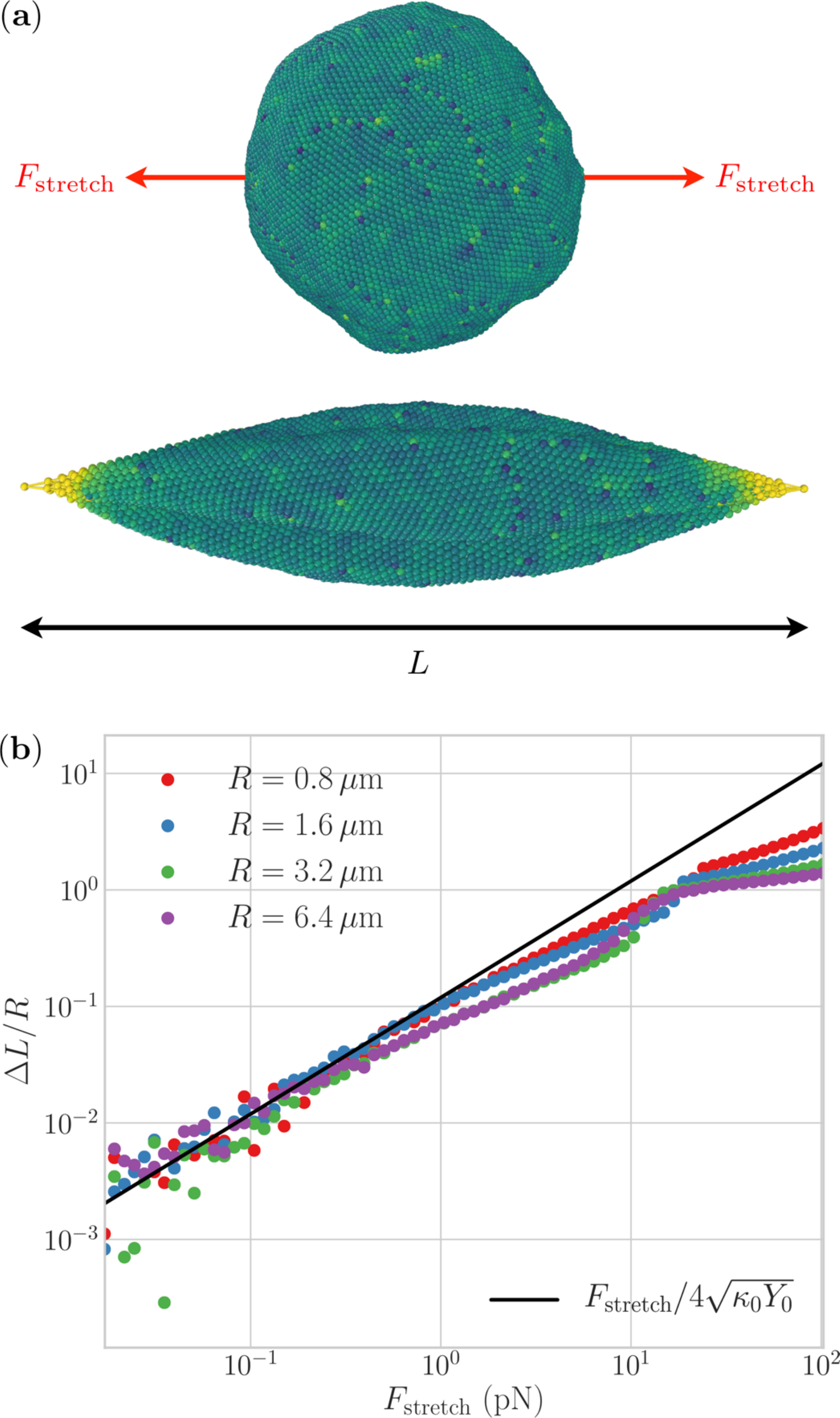}
  \caption{\label{figS1}Stretch behavior of empty thermalized membranes. (a) Sketch of the numerical setup and post-buckled state of the force pulling simulations. Vertices are colored according to local strain. (b) Force-extension relationship for erythrocyte-like spectrin shells ($Y_0 \simeq \SI{70}{\micro\newton\per\meter}$, $\kappa_0 \simeq \SI{4e-20}{\newton\meter}$) with $N_v=\num[group-separator={,},group-minimum-digits={3}]{9800}$ vertices and various reference radii $R$. The crossover to a non-linear regime at $\Delta L \simeq R$ corresponds to the buckling instability depicted in (a).}
\end{figure}

The resulting relationship between $\Delta L\equiv L-L_0$ and $F_{\rm stretch}$ is plotted in Fig.~\ref{figS1} for erythrocyte-like membranes of similar dimensions to those considered in the main text. The measurements are compared against the analytical solution of Reissner~[77] for thin uniform elastic shells subjected to radial point forces, 
\begin{equation}
  \label{eq:reissner}
  F_{\rm stretch} = \frac{4\sqrt{\kappa_0 Y_0}}{R} \Delta L.
\end{equation}
Eq.~\eqref{eq:reissner} is found to be well satisfied in all cases for small extensions ($\Delta L \lesssim 0.05\,R$), although larger deformations are associated with increasing deviations from a linear force response, and eventually lead to buckling transition to a collapsed state with significant lateral invaginations (Fig.~\ref{figS1}a) for $\Delta L \simeq R$.

\begin{figure}[htpb]
  \includegraphics[width=\columnwidth]{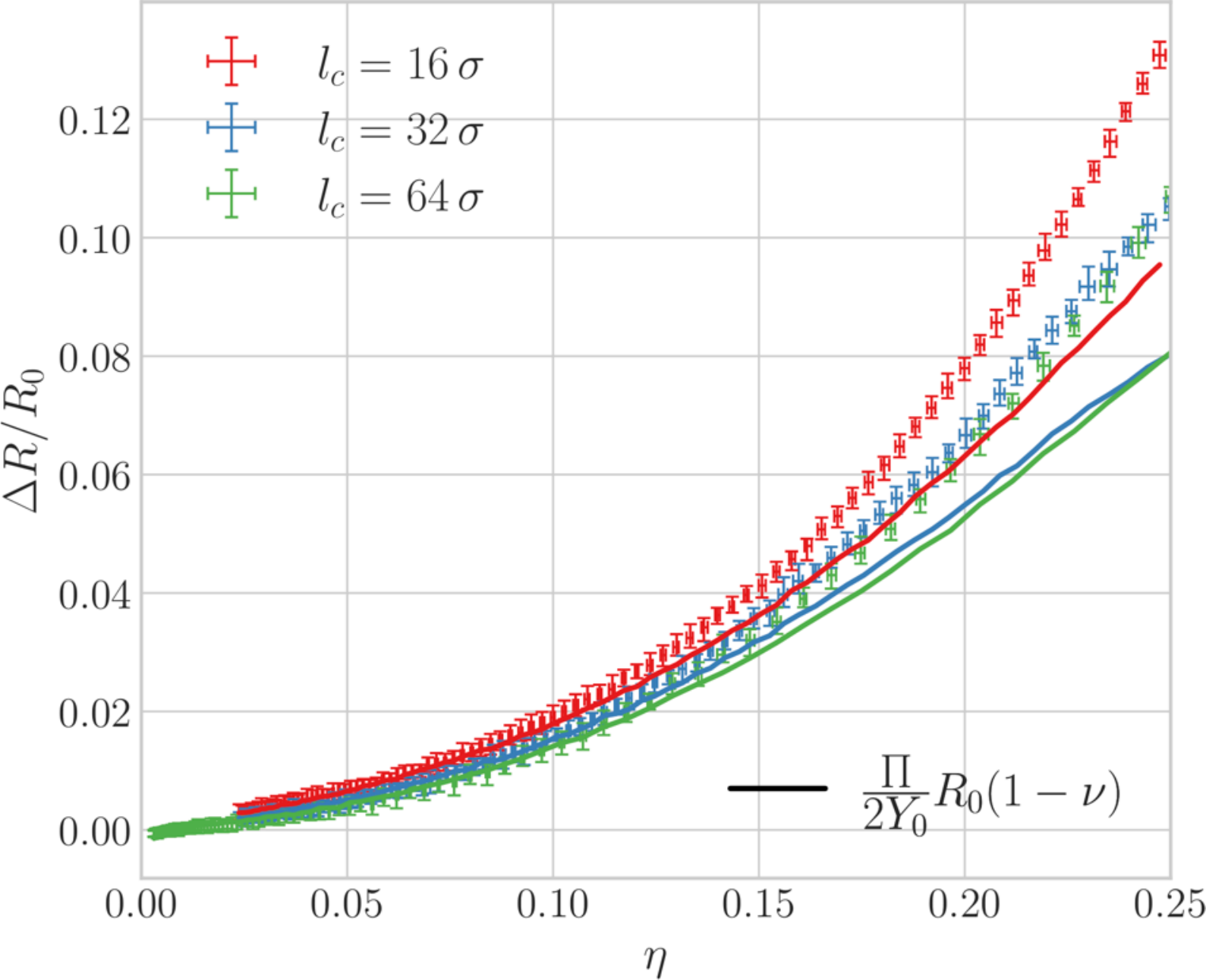}
  \caption{\label{figS2}Osmotic swelling of nuclear-like envelopes ($Y_0\simeq\SI{25}{\milli\newton\per\meter}$, $\kappa_0 \simeq \SI{3.5e-19}{\newton\meter}$) in the presence of DNA-like chains ($\sigma \simeq \SI{2}{\nano\meter}$, $l_p\simeq \SI{50}{\nano\meter}$). The mean radial expansion $\Delta R$ of the shell at density $\eta$ (markers), relative to the average radius $R_0$ of the empty thermalized membrane, may be quantitatively linked to the shell Poisson's ratio $\nu$ and osmotic pressure $\Pi$ of the enclosed polymer solution for $\Delta R/R_0\ll 1$ (solid lines, c.f.~Eq.~(11) of the main text).}
\end{figure}


\section{Averaged normal tensor from bond vector components} \label{sec:Q_tens}

\begin{lemma}
\label{lemma:dyadic}
Let $\mathbf{v} \equiv \begin{bmatrix} v_x\; v_y \;v_z \end{bmatrix}$ be an arbitrary unit vector, and $\begin{bmatrix} \mathbf{v} \end{bmatrix}_\times$ the skew-symmetric matrix such that
\begin{gather*}
   \mathbf{v} \times \mathbf{x} = \begin{bmatrix} \mathbf{v} \end{bmatrix}_\times \cdot \mathbf{x} \qquad \forall \mathbf{x} \in \mathbb{R}^3 \\ \big \Updownarrow \\ \begin{bmatrix} \mathbf{v} \end{bmatrix}_\times = 
  \begin{bmatrix} 
  0 & -v_z & v_y \\
  v_z & 0 & -v_x \\
  -v_y & v_x & 0
  \end{bmatrix},
\end{gather*}
where $\times$ and $\cdot$ denote the respective vector cross and dot products. Then,
\begin{equation}
  \label{eq:dyadic_cross}
  \mathbf{v} \otimes \mathbf{v} = \begin{bmatrix} \mathbf{v} \end{bmatrix}_\times^2 + \mathcal{I},
\end{equation}
with $\otimes$ the dyadic product and $\mathcal{I}$ the 3D identity matrix.
\end{lemma}

\begin{proof}
By definition of the dyadic product,
\begin{equation*}
    \mathbf{v} \otimes \mathbf{v} \equiv \begin{bmatrix} v_i v_j  \end{bmatrix} = \mathbf{v} \cdot \mathbf{v}^{\sf T}.
\end{equation*}
Furthermore, for any vector $\mathbf{x} \in \mathbb{R}^3$,
\begin{gather*}
  \big(\mathbf{v} \cdot \mathbf{v}^{\sf T}\big) \cdot \mathbf{x} = \mathbf{v} \big(\mathbf{v}^{\sf T} \cdot \mathbf{x}\big)  = (\mathbf{v} \cdot \mathbf{x}) \mathbf{v}, \\
  \begin{bmatrix} \mathbf{v} \end{bmatrix}_\times^2 \cdot \mathbf{x} = \mathbf{v} \times (\mathbf{v}\times \mathbf{x}) = (\mathbf{v}\cdot \mathbf{x}) \mathbf{v} - \mathbf{x},
\end{gather*}
from which Eq.~\eqref{eq:dyadic_cross} immediately follows.
\end{proof}

\begin{lemma}
\label{lemma:cross}
Let $\mathbf{v}$ and $\mathbf{w}$ be two arbitrary unit vectors such that $(\mathbf{v} \cdot \mathbf{w})^2 \neq 1$, and $\mathbf{n} \equiv \mathbf{v} \times \mathbf{w}$. Then,
\begin{multline}
  \label{eq:lemcross}
  \widehat{\mathbf{n}} \otimes \widehat{\mathbf{n}} = \mathcal{I} - \frac{1}{1+\mathbf{v} \cdot \mathbf{w}} \Big \{  \mathbf{v}\otimes \mathbf{w} +\mathbf{w}\otimes \mathbf{v} \\ + 2 \big(\widehat{\mathbf{v-w}}\big) \otimes \big(\widehat{\mathbf{v-w}}\big) \Big \},
\end{multline}
where the hat notation indicates normalized vectors,
\begin{equation*}
  \widehat{\mathbf{x}} \equiv \frac{\mathbf{x}}{\lVert \mathbf{x} \rVert} \qquad \forall \mathbf{x} \in \mathbb{R}^3,
\end{equation*}
with $\lVert \cdot \rVert$ the Euclidean norm.
\end{lemma}

\begin{proof}
Lemma~\ref{lemma:dyadic} immediately yields
\begin{equation}
  \label{eq:nxn1}
   \widehat{\mathbf{n}} \otimes \widehat{\mathbf{n}} = \mathcal{I} + \frac{1}{1-(\mathbf{v} \cdot \mathbf{w})^2} \begin{bmatrix} \mathbf{n} \end{bmatrix}_\times^2,
\end{equation}
where we used
\begin{equation*}
   \widehat{\mathbf{n}} = \frac{\mathbf{n}}{\sqrt{1-(\mathbf{v} \cdot \mathbf{w})^2}}.
\end{equation*}
Furthermore, for any vector $\mathbf{x} \in \mathbb{R}^3$,
\begin{align*}
  \begin{bmatrix} \mathbf{n} \end{bmatrix}_\times^2 \cdot \mathbf{x}
  &= (\mathbf{v} \times \mathbf{w}) \times \big\{ (\mathbf{v} \times \mathbf{w}) \times \mathbf{x} \big\} \\
  &= (\mathbf{v} \times \mathbf{w}) \times \big\{(\mathbf{v}\cdot \mathbf{x}) \mathbf{w} - ( \mathbf{w}\cdot\mathbf{x}) \mathbf{v}\big\}\\ 
  &= \big\{ (\mathbf{v}\cdot \mathbf{w}) (\mathbf{w}\cdot \mathbf{x}) - (\mathbf{v}\cdot \mathbf{x})\big\} \mathbf{v} \\
  & \qquad\qquad\qquad +\big\{ (\mathbf{v}\cdot \mathbf{w}) (\mathbf{v}\cdot \mathbf{x}) - (\mathbf{w}\cdot \mathbf{x})\big\} \mathbf{w} \\
  &= (\mathbf{v}\cdot \mathbf{w}) (\mathbf{v} \otimes \mathbf{w} + \mathbf{w} \otimes \mathbf{v}) \cdot \mathbf{x} \\
  &\qquad\qquad\qquad - (\mathbf{v}\otimes\mathbf{v}+\mathbf{w}\otimes\mathbf{w})\cdot \mathbf{x}.
\end{align*}
Thus,
\begin{align}
  \begin{bmatrix} \mathbf{n} \end{bmatrix}_\times^2 
  &=  (\mathbf{v}\cdot \mathbf{w}) (\mathbf{v} \otimes \mathbf{w} + \mathbf{w} \otimes \mathbf{v})-(\mathbf{v}\otimes\mathbf{v}+\mathbf{w}\otimes\mathbf{w}) \nonumber \\
  &= (\mathbf{v}\cdot \mathbf{w}-1)  (\mathbf{v} \otimes \mathbf{w} + \mathbf{w} \otimes \mathbf{v}) \nonumber \\
  \label{eq:n_x2}
  &\qquad\qquad\qquad - (\mathbf{v}-\mathbf{w})\otimes(\mathbf{v}-\mathbf{w}).
\end{align}
Plugging Eq.~\eqref{eq:n_x2} into Eq.~\eqref{eq:nxn1} leads to
\begin{multline}
  \label{eq:nxn2}
  \widehat{\mathbf{n}} \otimes \widehat{\mathbf{n}} = \mathcal{I} - \frac{1}{1+\mathbf{v} \cdot \mathbf{w}}(\mathbf{v} \otimes \mathbf{w} + \mathbf{w} \otimes \mathbf{v}) \\
 - \frac{1}{1-(\mathbf{v} \cdot \mathbf{w})^2}(\mathbf{v}-\mathbf{w})\otimes(\mathbf{v}-\mathbf{w}),
\end{multline}
and substituting $\lVert \mathbf{v}-\mathbf{w}\rVert^2 = 2\,(1-\mathbf{v} \cdot \mathbf{w})$ into Eq.~\eqref{eq:nxn2} directly yields Eq.~\eqref{eq:lemcross}.
\end{proof}

\begin{theorem}
\label{theorem:qi}
Let $\mathcal{N}_k$ be the normal tensor associated with the $i$-th pair of consecutive inter-monomer bonds,
\begin{equation*}
  \mathcal{N}_k \equiv \Big(\widehat{\mathbf{t}_k\times \mathbf{t}_{k+1}}\Big) \otimes \Big(\widehat{\mathbf{t}_k\times \mathbf{t}_{k+1}}\Big),
\end{equation*}
and $\mathbf{u}_k \equiv \mathbf{t}_{k+1}-\mathbf{t}_k$. Then,
\begin{equation}
  \label{eq:qi_exp}
  \mathcal{N}_k = \mathcal{I} - \frac{\mathbf{t}_k\otimes \mathbf{t}_k + \mathbf{t}_{k+1}\otimes \mathbf{t}_{k+1}}{2} - \widehat{\mathbf{u}}_k \otimes  \widehat{\mathbf{u}}_k  + \mathcal{O}\big(\mathbf{u}_k^2\big).
\end{equation}
\end{theorem}

\begin{proof}
Using Lemma~\ref{lemma:cross},
\begin{equation}
  \label{eq:qi_full}
  \mathcal{N}_k = \mathcal{I} - \frac{1}{1+\mathbf{t}_k \cdot \mathbf{t}_{k+1}}\big(\mathbf{t}_k \otimes \mathbf{t}_{k+1} + \mathbf{t}_{k+1} \otimes \mathbf{t}_k + 2\widehat{\mathbf{u}}_k \otimes  \widehat{\mathbf{u}}_k\big). 
\end{equation}
It is straightforward to check that
\begin{equation*}
  \mathbf{t}_k \otimes \mathbf{t}_{k+1} + \mathbf{t}_{k+1} \otimes \mathbf{t}_k = \mathbf{t}_k \otimes \mathbf{t}_k + \mathbf{t}_{k+1} \otimes \mathbf{t}_{k+1} + \mathcal{O}\big(\mathbf{u}_k^2\big),
\end{equation*}
along with 
\begin{equation*}
  \mathbf{t}_k \cdot \mathbf{t}_{k+1} = 1+ \mathcal{O}\big(\mathbf{u}_k^2\big).
\end{equation*}
The Taylor expansion of Eq.~\eqref{eq:qi_full} to first order in $\mathbf{u}_k$ thus immediately yields Eq.~\eqref{eq:qi_exp}.
\end{proof}

\begin{figure*}[htpb]
  \includegraphics[width=1.5\columnwidth]{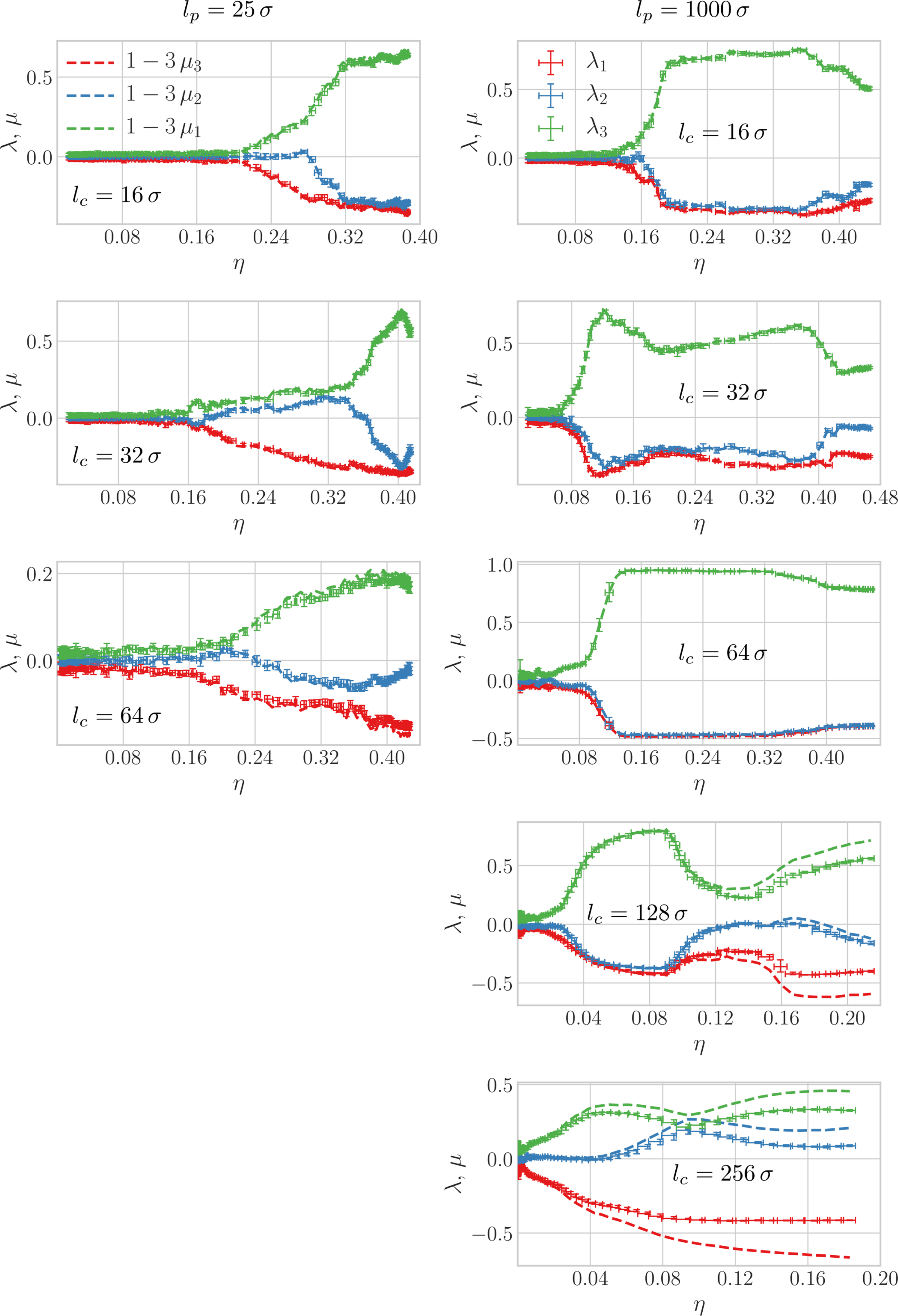}
  \caption{\label{figS3}Normal ($\mu_i$) and standard nematic tensor eigenvalues ($\lambda_i$) as a function of polymer volume fraction ($\eta$). Error bars are computed as described in the main text, and are only shown for the $\lambda_i$ for readability.}
\end{figure*}

\begin{corollary}
\label{corollary:contour}
Let $l_b$ be the fixed separation distance between two consecutive monomers along the chain and $\langle \cdot \rangle_c$ the (discrete) contour average,
\begin{equation*}
  \langle \cdot \rangle_c \equiv \frac{1}{N_m -2} \sum_{k=1}^{N_m-2} \; \cdot \;,
\end{equation*}
where $N_m$ is the total number of monomers per chain. Then, in the limit of long polymers ($N_m \gg 1$),
\begin{equation}
  \label{eq:q_contour}
  \big\langle \mathcal{N}_k \big\rangle_c = \mathcal{I} -\big \langle \mathbf{t}_k \otimes \mathbf{t}_k \big \rangle_c - \big\langle \widehat{\mathbf{u}}_k \otimes \widehat{\mathbf{u}}_k  \big\rangle_c + \mathcal{O}\big(l_b^2C_m^2\big),
\end{equation}
with $C_m \equiv \sqrt{\big\langle \mathbf{u}_k^2 \big\rangle_c}/l_b$ the root mean square curvature of the chains.
\end{corollary}

\begin{proof}
The Kremer-Grest (KG) definition of the discrete local curvature $C_k$ reads as
\begin{equation}
  \label{eq:curv}
  C_k \equiv \frac{\Vert \mathbf{t}_{k+1}-\mathbf{t}_{k} \rVert}{l_b} = \frac{\sqrt{\mathbf{u}_k^2}}{l_b}.
\end{equation}
Eq.~\eqref{eq:q_contour} then trivially follows from Eqs.~\eqref{eq:qi_exp} and~\eqref{eq:curv}, using $C_m = \sqrt{\big\langle C^2_k\big\rangle_c}$ and
\begin{equation*}
  \big\langle \mathbf{t}_k\otimes \mathbf{t}_k + \mathbf{t}_{k+1}\otimes \mathbf{t}_{k+1} \big\rangle_c = 2 \big\langle\mathbf{t}_k\otimes \mathbf{t}_k \big\rangle_c +\mathcal{O} \bigg ( \frac{1}{N_m} \bigg),
\end{equation*}
where the extremal terms of order $1/N_m$ may be neglected in the long-chain limit.
\end{proof}

\begin{theorem}
\label{theorem:qi_unconfined}
Let $\langle \cdot \rangle$ be the canonical thermodynamic average,
\begin{equation*}
  \langle \cdot \rangle \equiv \langle \langle \cdot \rangle_c \rangle_{\rho, T}
\end{equation*}
with $\langle \cdot \rangle_{\rho, T}$ the ensemble average over the accessible polymer configurational space at fixed temperature $T$ and density $\rho$. In the unconfined chain approximation (UCA)~[114], 
\begin{equation}
  \label{eq:q_uca}
  \big\langle \mathcal{N}_k\big\rangle = \frac{I-\big\langle \mathbf{t}_k \otimes \mathbf{t}_k \big\rangle}{2} + \mathcal{O}\bigg(\frac{l_b}{l_p}\bigg),
\end{equation}
with $l_p$ the polymer persistence length.
\end{theorem}

\begin{proof}
Let $\Theta_k$ be the angle between two consecutive bonds $\mathbf{t}_k$ and $\mathbf{t}_{k+1}$,
\begin{equation*}
  \cos\Theta_k \equiv \mathbf{t}_k \cdot \mathbf{t}_{k+1}.
\end{equation*}
Neglecting any spontaneous chain curvature potentially induced by the confining membrane ($R \gg l_b$), it follows from the local definition of the persistence length~[84] that, for sufficiently stiff filaments ($l_p \gg l_b$),
\begin{equation}
  \label{eq:theta_trans}
  \big\langle\cos\Theta_k\big\rangle_{\rho, T}  \xrightarrow[\rho \to 0]{}  \exp \bigg\{ - \frac{l_b}{l_p(T)} \bigg \} = 1 + \mathcal{O} \bigg( \frac{l_b}{l_p} \bigg).
\end{equation}
Note that in the case of finite densities within the nematic stability range, deflections of the chain by the surrounding polymers typically induce a further inhibition of transverse fluctuations~[102], so that Eq.~\eqref{eq:theta_trans} may generally provide an upper bound for the variations of $\Theta_k$. While similar considerations have been suggested to potentially favor the appearance of hairpin defects~[102], we find no evidence of hairpin formation in any of the systems investigated here, and hence neglect the probability of their occurrence in the following discussion. Thus, 
\begin{equation}
  \label{eq:curvature_ave}
  \big\langle C^2_m\big\rangle_{\rho, T} = \frac{\big\langle\big\langle\mathbf{u}_k^2\big\rangle_c\big\rangle_{\rho, T}}{l_b^2} = \frac{2\big\langle 1-\cos\Theta_k\big\rangle}{l_b^2} = \mathcal{O}\bigg(\frac{1}{l_b l_p}\bigg).
\end{equation}
Furthermore, let $\mathbf{t}_{\perp k}$ be an arbitrary unit vector such that $\mathbf{t}_k \cdot \mathbf{t}_{\perp k} = 0$. We may rewrite $\mathbf{u}_k$ in the form
\begin{equation}
  \label{eq:u_proj}
   \mathbf{u}_{k} = u_{\perp k} \, \mathbf{t}_{\perp k} + u_{\independent  k} \, \big(\mathbf{t}_k \times \mathbf{t}_{\perp k}\big) + \mathcal{O}\bigg( \frac{l_b}{l_p}\bigg),
\end{equation}
where we used
\begin{equation*}
     \big\langle\mathbf{u}_{k}\cdot\mathbf{t}_k \big\rangle_{\rho, T} = \big\langle1-\mathbf{t}_{k}\cdot\mathbf{t}_{k+1} \big\rangle_{\rho, T} =  \mathcal{O}\bigg( \frac{l_b}{l_p} \bigg).
\end{equation*}
The KG bending energy penalty then reads as
\begin{equation}
  \label{eq:kg_bend}
  U_{\rm bend} = \frac{\epsilon_b}{2} \sum_{k=1}^{N_m-2} \mathbf{u}_k^2 =  \frac{\epsilon_b}{2} \sum_{k=1}^{N_m-2} \big( u_{\perp k}^2 + u_{\independent k}^2\big) + \mathcal{O}(k_BT),
\end{equation}
with $\epsilon_b \equiv  k_B T l_p/ l_b$ the chain flexural modulus. Assimilating the transverse fluctuation components $u_{\perp k}$ and $u_{\independent k}$ to decoupled degrees of freedom,
\begin{gather} 
  \label{eq:u_ave}
  \big \langle u_{\perp k} u_{\independent k} \big \rangle_{\rho, T} = \big \langle u_{\perp k}\big \rangle_{\rho, T} \big \langle u_{\independent k}\big \rangle_{\rho, T} = 0,\\
    \label{eq:u2_ave}
  \big \langle u_{\perp k}^2\big \rangle_{\rho, T} = \big \langle u_{\independent k}^2\big \rangle_{\rho, T} = \frac{l_b}{l_p},
\end{gather}
where Eq.~\eqref{eq:u2_ave} results from the equipartition theorem. Note that Eq.~\eqref{eq:u_ave} assumes that the chains bear no local curvature at rest, consistently with Eq.~\eqref{eq:theta_trans}, while Eq.~\eqref{eq:u2_ave} further neglects additional Hamiltonian contributions beyond Eq.~\eqref{eq:kg_bend} which may arise from potential polymer-polymer and polymer-membrane interactions. Hence, Eqs.~\eqref{eq:u_ave} and~\eqref{eq:u2_ave} are only expected to hold in the so-called \textit{unconfined chain} regime~[114], in which local polymer conformations are largely unaffected by the presence of surrounding chains or membrane walls. In this case, Eqs.~\eqref{eq:u_proj},~\eqref{eq:u_ave} and~\eqref{eq:u2_ave} lead to
\begin{align}
  \big \langle \widehat{\mathbf{u}}_k \otimes \widehat{\mathbf{u}}_k \big\rangle_{\rho, T} &=  \frac{1}{2} \big \langle \big(\mathbf{t}_k \times \mathbf{t}_{\perp k}\big) \otimes \big(\mathbf{t}_k \times \mathbf{t}_{\perp k}\big) \big\rangle_{\rho, T} \nonumber \\
  &  \qquad\qquad\qquad +  \frac{1}{2} \big \langle \mathbf{t}_{\perp k} \otimes \mathbf{t}_{\perp k} \big\rangle_{\rho, T} +  \mathcal{O}\bigg( \frac{l_b}{l_p}\bigg) \nonumber \\
   \label{eq:uxu_ave}
  &= \frac{\mathcal{I} - \big\langle \mathbf{t}_k \otimes \mathbf{t}_k \big\rangle_{\rho, T}}{2} + \mathcal{O}\bigg( \frac{l_b}{l_p}\bigg),
\end{align}
where we used Lemma~\ref{lemma:cross}. Plugging Eqs.~\eqref{eq:curvature_ave} and~\eqref{eq:uxu_ave} into Eq.~\eqref{eq:q_contour} finally yields Eq.~\eqref{eq:q_uca}.
\end{proof}

\begin{corollary}
\label{corollary:qi_eig}
Let $\mathcal{Q}_k$ be the standard nematic order parameter tensor,
\begin{equation*}
  \mathcal{Q}_k \equiv \frac{3 \big(\mathbf{t}_k \otimes \mathbf{t}_k\big) - \mathcal{I}}{2}.
\end{equation*}
Then, for long polymers ($N_m \gg 1$) in the UCA,
\begin{equation}
  \label{eq:q_rel}
  \big\langle \mathcal{Q}_k\big\rangle  = \mathcal{I} - 3\big\langle \mathcal{N}_k\big\rangle + \mathcal{O}\bigg(\frac{l_b}{l_p}\bigg).
\end{equation}
Hence, the tensors $\mathcal{Q} \equiv \big\langle \mathcal{Q}_k\big\rangle$ and $\mathcal{N}\equiv\big\langle \mathcal{N}_k\big\rangle$ generally share the same eigenvectors, with their respective ascending eigenvalues $\big\{\lambda_{ j}\big\}$  and $\big\{\mu_{ j}\big\}$ related through
\begin{equation}
\label{eq:eig_perp}
\begin{gathered}
    \lambda_{ 1} = 1-3\,\mu_3, \\
    \lambda_{ 2} = 1-3\,\mu_2, \\
    \lambda_{ 3} = 1-3\,\mu_1.
\end{gathered}
\end{equation}
\end{corollary}

\begin{figure*}[htpb]
  \includegraphics[width=1.5\columnwidth]{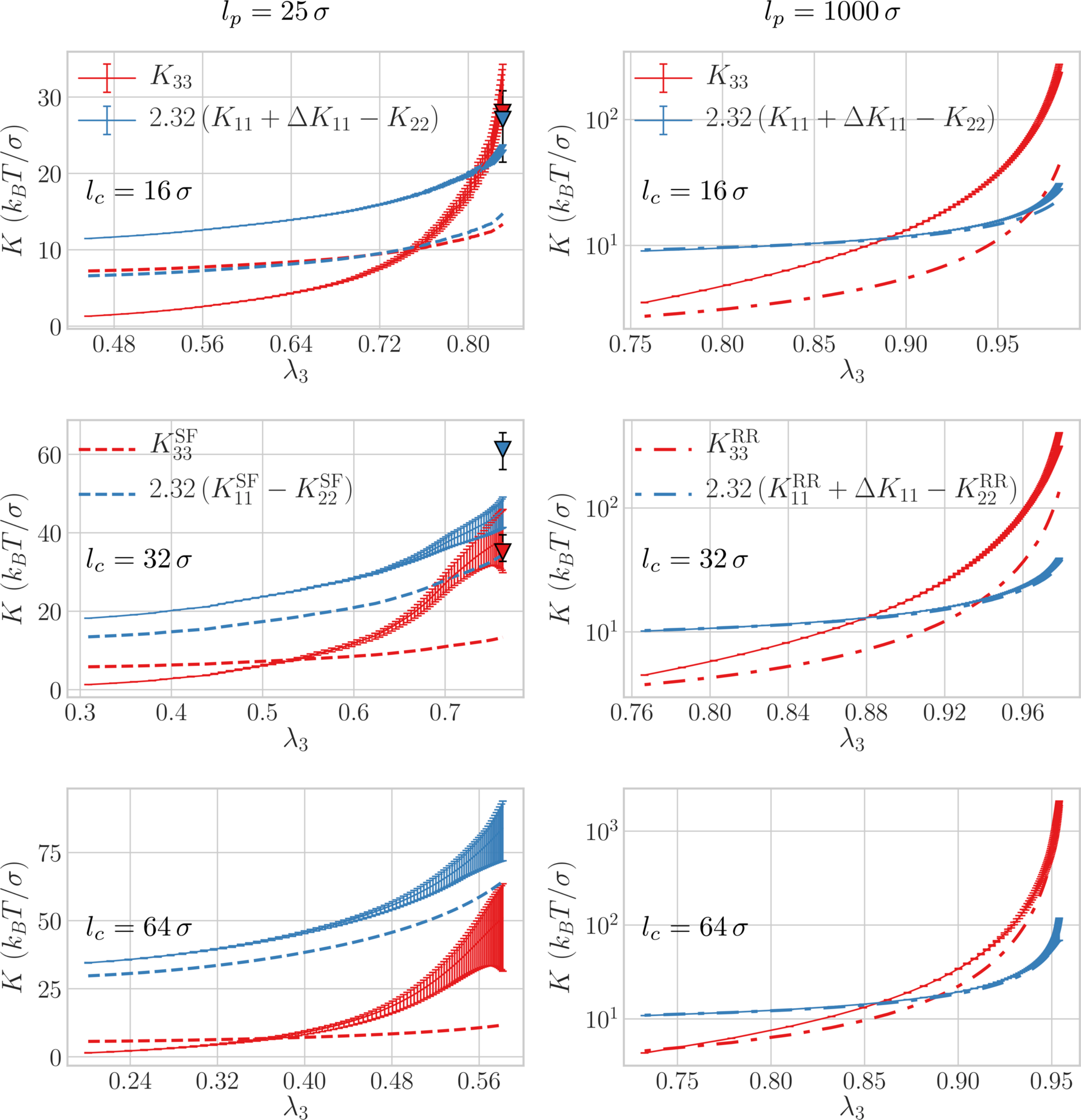}
  \caption{\label{figS4}Oseen-Frank elastic anisotropies of DNA- ($l_p=25\,\sigma$) and tubulin-like ($l_p=1000\,\sigma$) chains as a function of the degree $\lambda_3$ of nematic order. Symbols are as in Fig.~3 of the main text. Note that the Williams inequality (Eq.~(26)) is satisfied at all densities for $l_p=25\,\sigma$ and $l_c \gtrsim 32 \,\sigma$, but violated for $l_c=16\,\sigma$ in the strong alignment regime characterizing the bipolar state ($\lambda_3\gtrsim 0.80$, c.f.~Fig.~1d of the main text).}
\end{figure*}

\begin{proof}
Eqs.~\eqref{eq:q_rel}-\eqref{eq:eig_perp} follow directly from Eq.~\eqref{eq:q_uca}. 
\end{proof}
Note that while Theorem~\ref{theorem:qi} and Corollary~\ref{corollary:contour} may be derived based solely on geometrical considerations, and are therefore quite generally valid for $N_m \gg 1$, the additional thermodynamic assumptions underlying Theorem~\ref{theorem:qi_unconfined} and Corollary~\ref{corollary:qi_eig} restrict their applicability to phases in which the UCA may reasonably hold --- i.e., in which local chain fluctuations do not significantly deviate from those expected in the dilute regime. However, it is shown in Fig.~\ref{figS3} that Eq.~\eqref{eq:eig_perp} is remarkably well satisfied for all systems considered, with relative discrepancies of the order of $30\,\%$ in the corresponding eigenvalues being observed only in the limit of extreme confinement ($l_p = 1000\,\sigma$, $l_c/R_\eta \gg 1$). This observation further evidences the suitability of the UCA in our case, which provides the basis of the Fynewever-Yethiraj density functional theory employed in the main text~[114].

\begin{figure}[htpb]
  \includegraphics[width=\columnwidth]{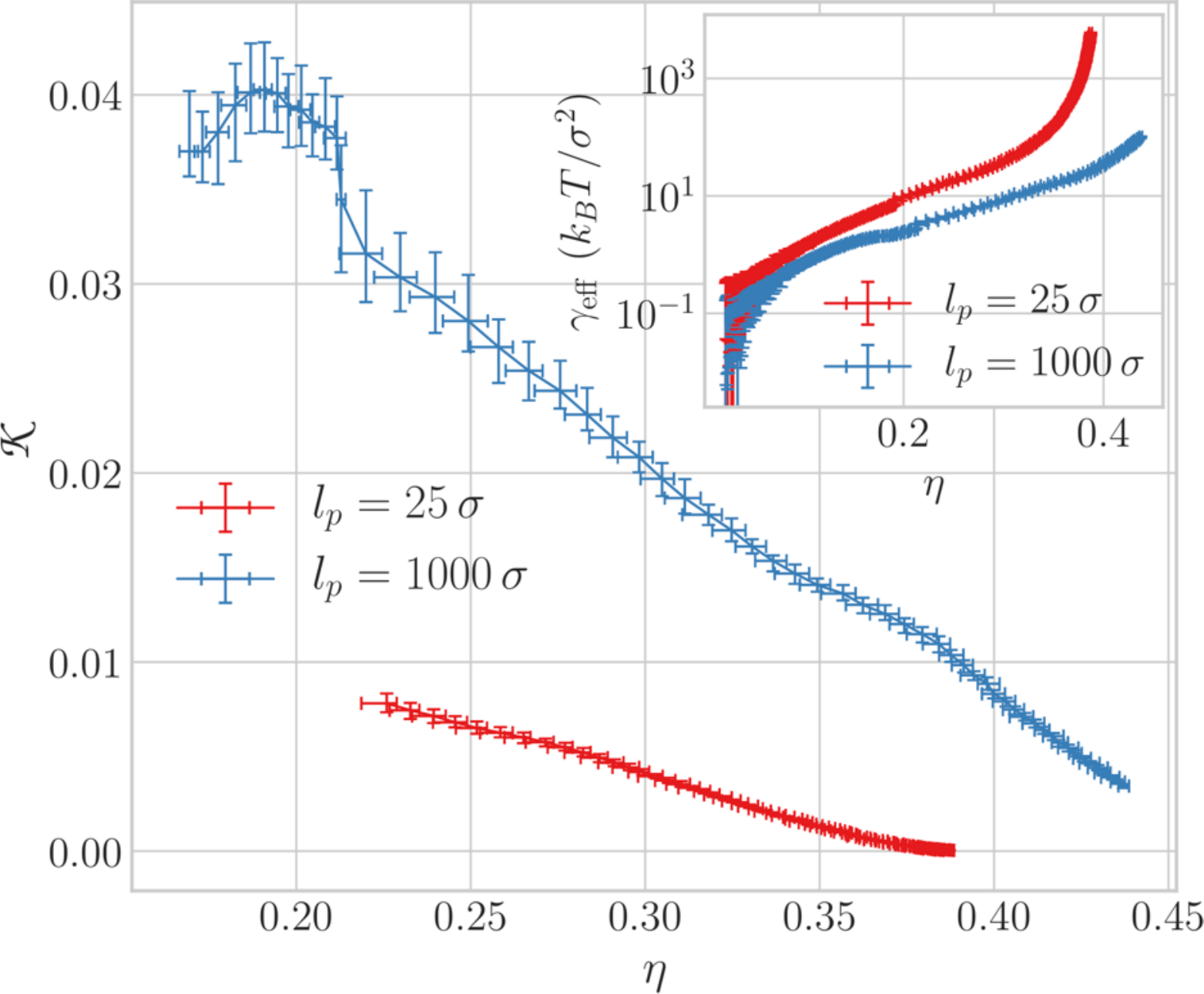}
  \caption{\label{figS5}Reduced modulus ($\mathpzc{K}$) and effective surface tension ($\gamma_{\rm eff}$) as a function of $\eta$ for filaments with $l_c =16\,\sigma$ (Eqs.~(27)-(28) of the main text). Red markers correspond to the case of DNA-like chains ($\sigma \simeq \SI{2}{\nano\meter}$, $l_p=25\,\sigma$) confined within a nuclear-like envelope ($Y_0\simeq\SI{25}{\milli\newton\per\meter}$, $\kappa_0 \simeq \SI{3.5e-19}{\newton\meter}$), and blue markers that of tubulin-like polymers ($\sigma \simeq \SI{25}{\nano\meter}$, $l_p=1000\,\sigma$) within an erythrocyte-like spectrin shell ($Y_0 \simeq \SI{70}{\micro\newton\per\meter}$, $\kappa_0 \simeq \SI{4e-20}{\newton\meter}$).}
\end{figure}


